%% file: main.tex
\documentclass[acmtog, nonacm = true]{acmart}
\acmSubmissionID{papers\_435s2}

\usepackage{booktabs} 
\usepackage{xcolor}

\usepackage{tcolorbox} 

\usepackage{subcaption}
\usepackage{cleveref}

\usepackage{svg}
\usepackage{float}
\usepackage[group-separator={,}]{siunitx}

\usepackage[ruled]{algorithm2e} 

\SetAlFnt{\small}
\SetAlCapFnt{\small}
\SetAlCapNameFnt{\small}
\SetAlCapHSkip{0pt}

\usepackage{makecell}
\setcopyright{rightsretained} 
\acmJournal{TOG}
\acmYear{2021}\acmVolume{40}\acmNumber{6}\acmArticle{236}\acmMonth{12} \acmDOI{10.1145/3478513.3480569}

\DeclareMathOperator*{\argmin}{arg\,min}

\begin{document}
\title{Neural Radiosity}

\author{Saeed Hadadan}
\orcid{1234-5678-9012-3456}
\affiliation{%
  \institution{University of Maryland, College Park}
  \city{College Park}
  \state{MD}
  \postcode{20740}
  \country{USA}}
\email{saeedhd@umd.edu}
\author{Shuhong Chen}
\affiliation{%
  \institution{University of Maryland, College Park}
  \city{College Park}
  \country{USA}
}
\email{shuhong@terpmail.umd.edu}
\author{Matthias Zwicker}
\affiliation{%
 \institution{University of Maryland, College Park}
 \city{College Park}
 \state{MD}
 \country{USA}}
\email{zwicker@cs.umd.edu}

\renewcommand\shortauthors{Hadadan, S. et al}

\begin{abstract}

We introduce Neural Radiosity, an algorithm to solve the rendering equation by minimizing the norm of its residual, similar as in classical radiosity techniques. Traditional basis functions used in radiosity, such as piecewise polynomials or meshless basis functions are typically limited to representing isotropic scattering from diffuse surfaces. Instead, we propose to leverage neural networks to represent the full four-dimensional radiance distribution, directly optimizing network parameters to minimize the norm of the residual. Our approach decouples solving the rendering equation from rendering (perspective) images similar as in traditional radiosity techniques, and allows us to efficiently synthesize arbitrary views of a scene. In addition, we propose a network architecture using geometric learnable features that improves convergence of our solver compared to previous techniques. Our approach leads to an algorithm that is simple to implement, and we demonstrate its effectiveness on a variety of scenes with diffuse and non-diffuse surfaces.

\end{abstract}

%
%

\ccsdesc[500]{Computing methodologies~Ray tracing}
\ccsdesc[300]{Computing methodologies~Supervised learning by regression}
\ccsdesc[300]{Computing methodologies~Neural networks}%
%

\keywords{Neural Rendering, Neural Radiance Field}

\begin{teaserfigure}
  \centering
  \subcaptionbox{Solving the rendering equation using a neural network by minimizing $\mathcal{L}(\theta)$\label{fig:rendering_equation}}{%
    \includegraphics[width=0.5\textwidth]{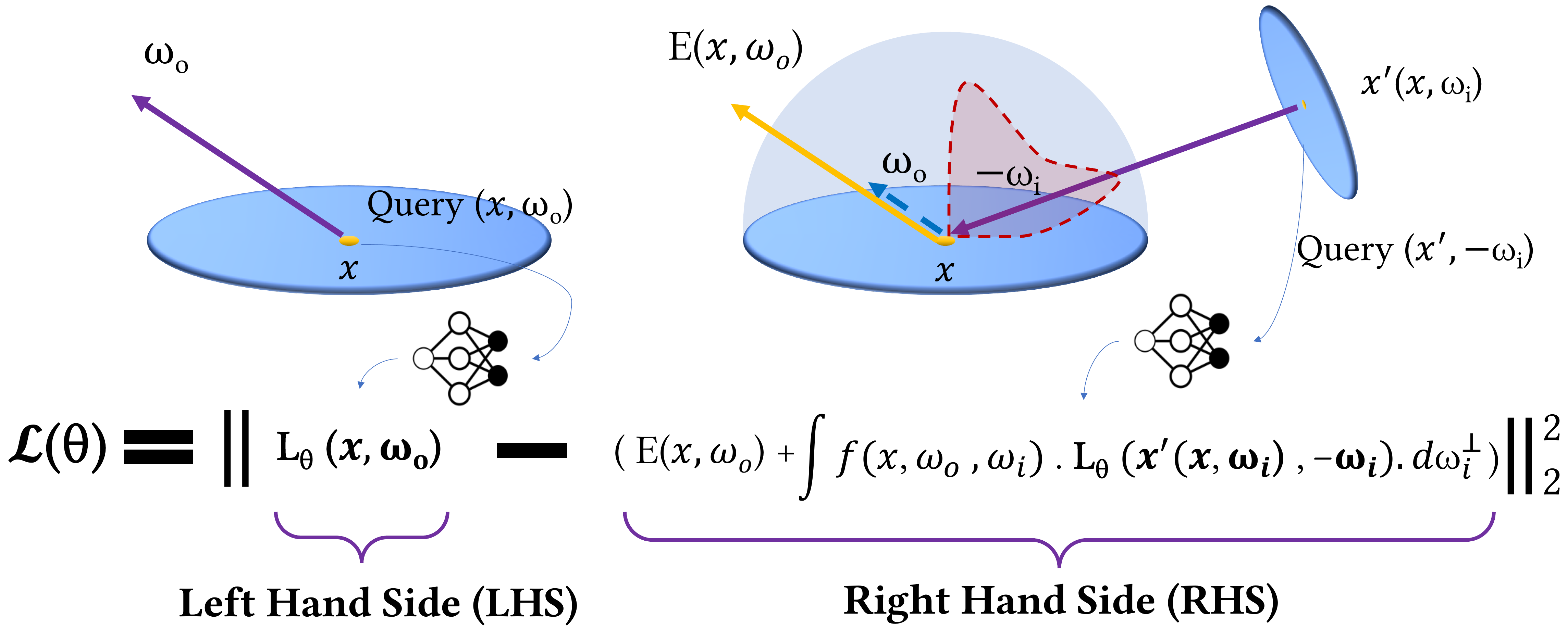}
  }
  \hspace{0.5cm}
  \subcaptionbox{Multiple views of a solution of (a)\label{Chair_MultiView}}{%
    \includegraphics[width=0.4\textwidth]{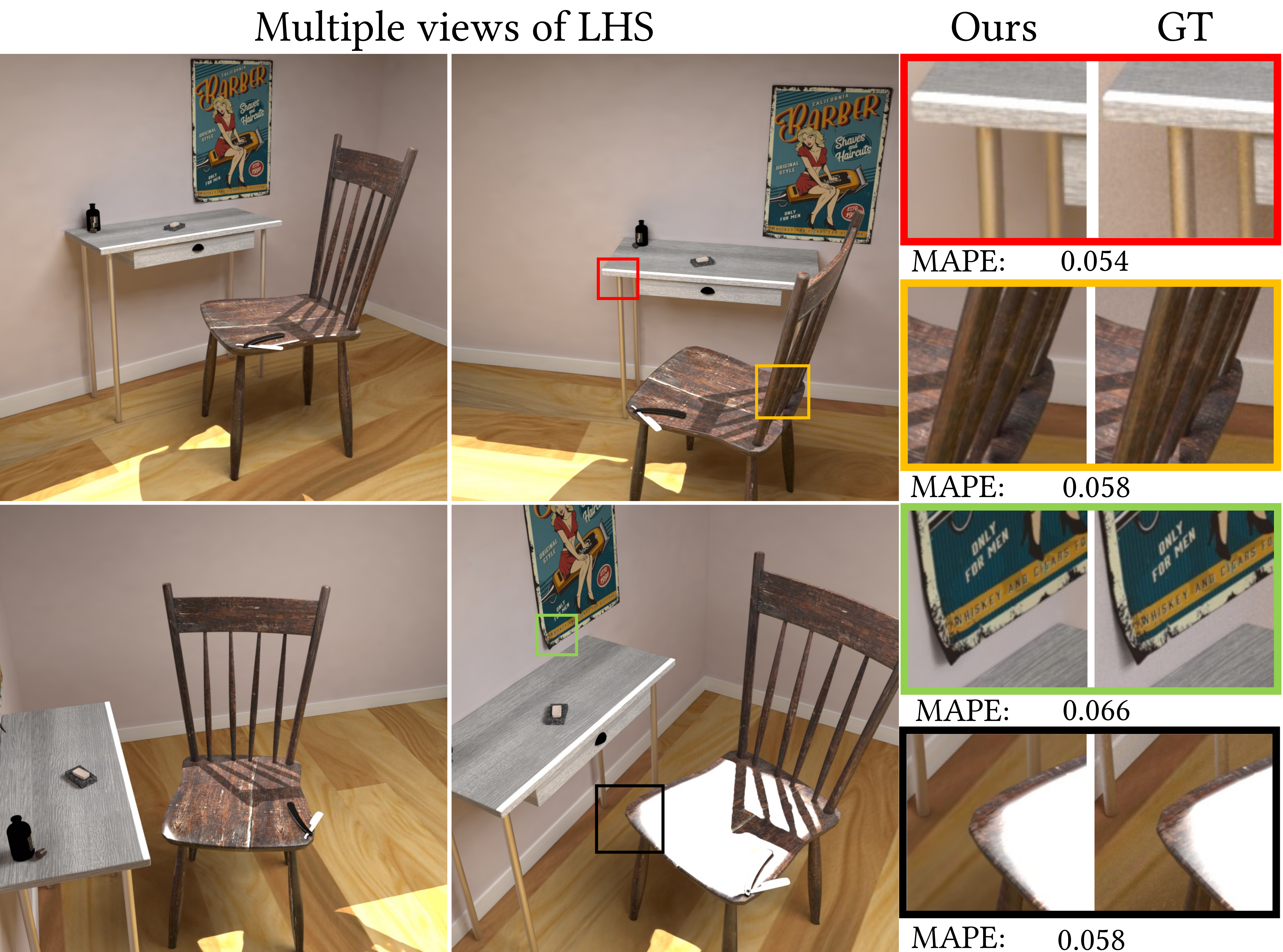}
  }
  \caption{(a) Neural Radiosity directly solves the rendering equation by minimizing the norm of its residual $\mathcal{L}(\theta)$, using a single neural network with parameters $\theta$ to represent the radiance distribution $L_{\theta}(x,\omega_o)$.  
  (b) 
  After solving for the radiance  $L_{\theta}(x,\omega)$, images from arbitrary viewpoints can be computed efficiently. Here, the left hand side (LHS) of the rendering equation represented by the network $L_{\theta}(x,\omega_o)$ is visually indistinguishable from the ground truth (GT).
  }
  \label{fig:teaser}
\end{teaserfigure}

\maketitle

\input{body}

\appendix

\section{Supplementary materials}

\begin{figure*}[htbp]
\centering
\subcaptionbox{Before training}{
\includegraphics[width = 0.485 \textwidth]{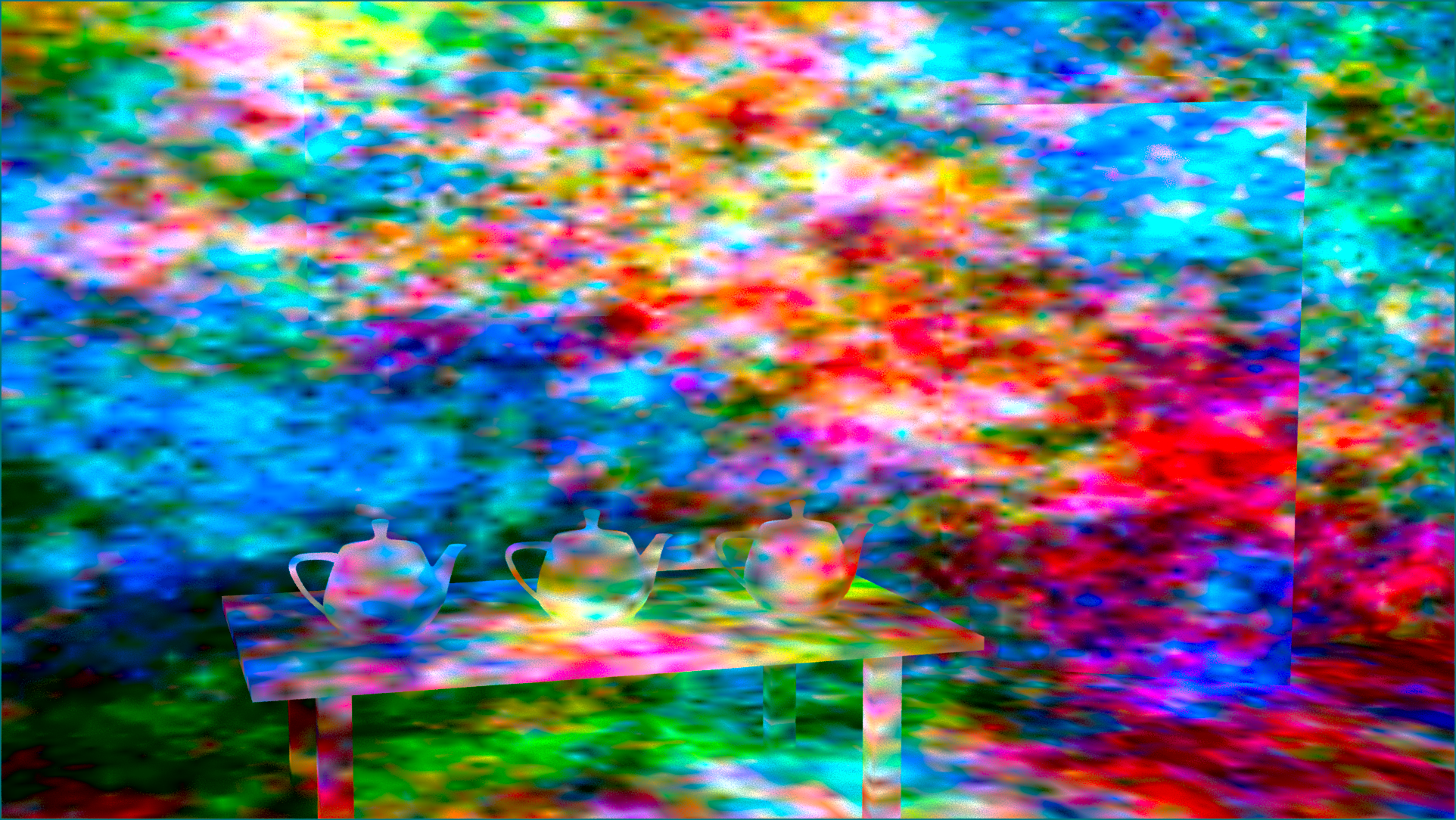}
}
\subcaptionbox{After training}{
\includegraphics[width = 0.485 \textwidth]{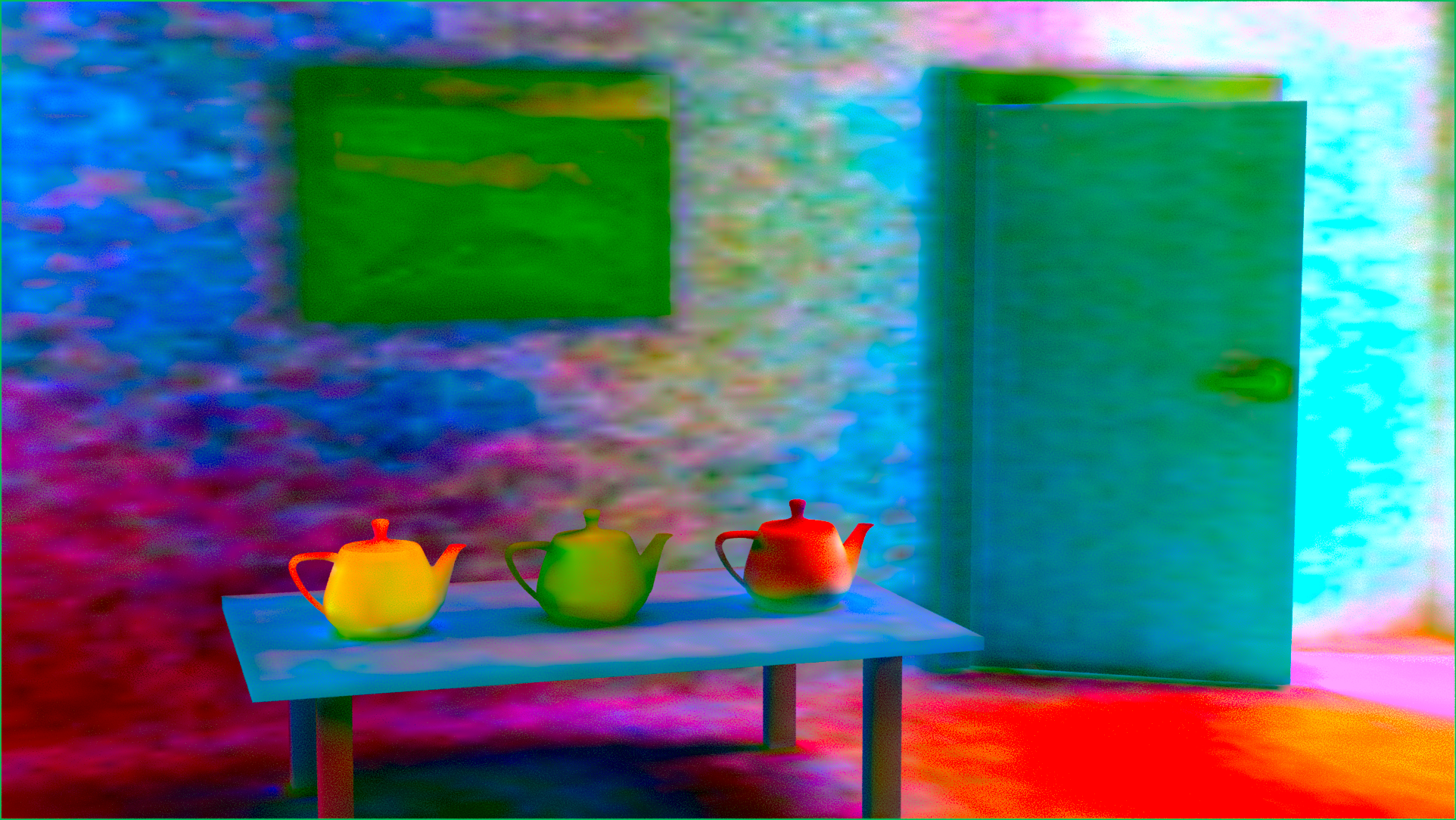}
}
  \caption{Visualization of the grids in the \emph{Veach Door} scene. The interpolated features are rendered into 16-D pixels, which then are reduced by PCA to 3-D \emph{Lab} colors.}
  \label{fig:grid_vis}
\end{figure*}

\subsection{Visualization of Feature Grids}

We provide a visualization of the learned feature grids in Fig. \ref{fig:grid_vis} on the \emph{Veach Door} scene.  To generate each visualization, we first find camera ray intersections, and collect the interpolated multi-level grid features (levels 4-128) at each hit point, resulting in a 16-channel feature image (one channel per feature dimension). We then perform PCA across the 16-D pixels, and reduce the dimensionality to the three top principal components.  The transformed 3-channel pixels are then normalized to span the \emph{ Lab} color space for visualization.

From the visualizations, we see that although the network started with random initializations, network training encourages locations with similar radiance properties to have similar weights.  In particular, note how each of the teapots has different PCA colors, indicating that the model has learned different features for handling various material types.  Another interesting observation is that the network learns different features for areas with shadows; see the alternative PCA colors for shadows of the table and teapots.

\subsection{Residual Normalization}

\begin{figure*}[hbt]
  \centering
\subcaptionbox{No normalization, MSE = 0.013, MAPE= 0.208}{
\includegraphics[width = 0.48\textwidth]{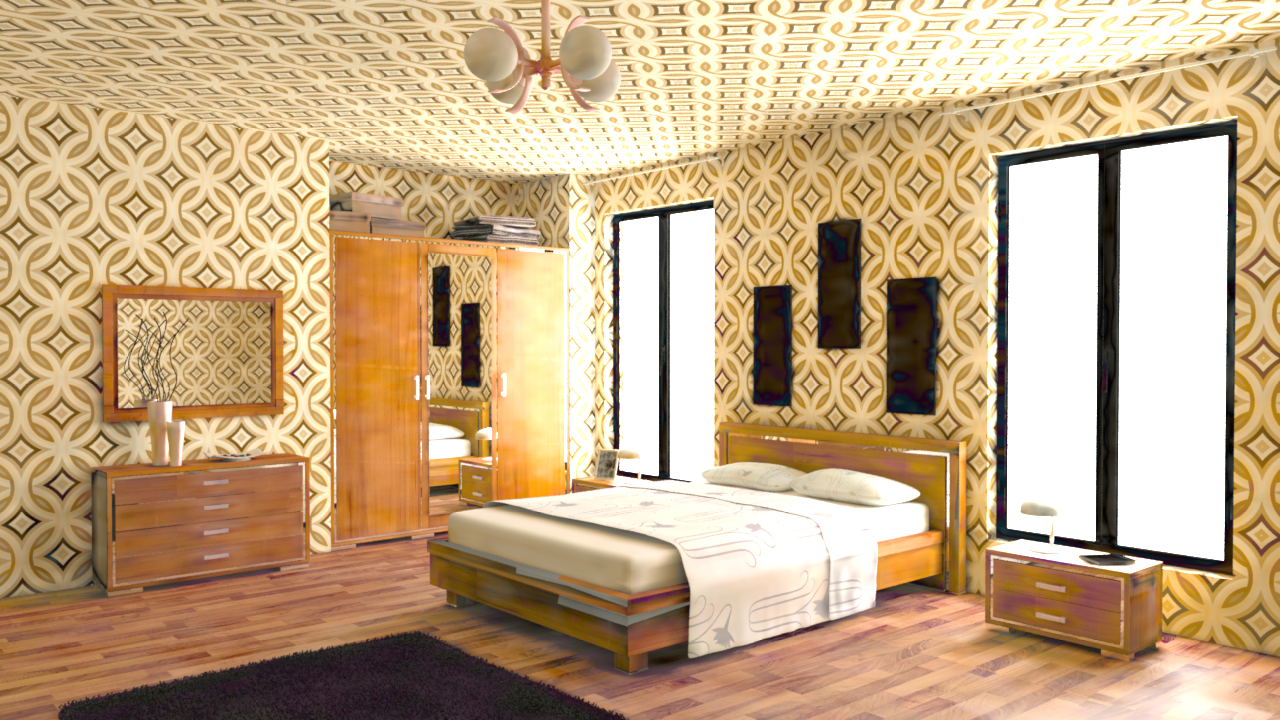}
}
\subcaptionbox{Our normalization, MSE = 0.018, MAPE: 0.111}{
\includegraphics[width = 0.48\textwidth]{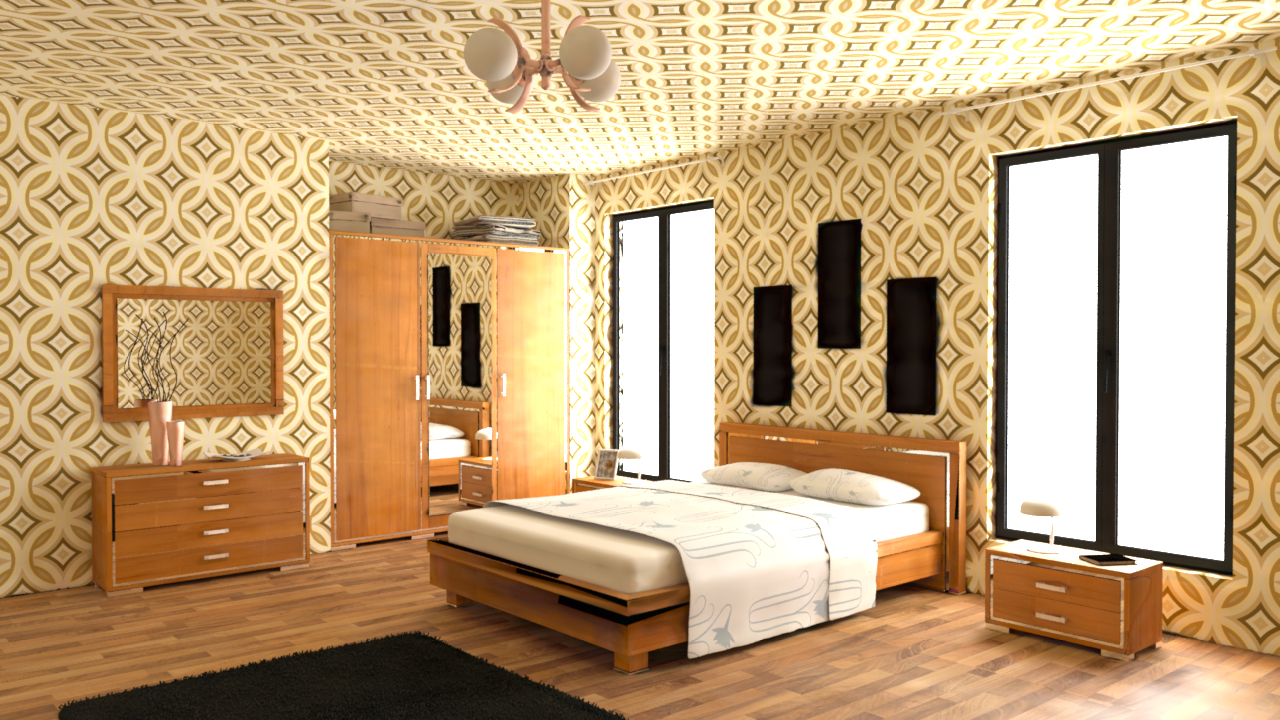}
}
\subcaptionbox{Normalize only by LHS, MSE = 0.030, MAPE: 0.147}{
\includegraphics[width = 0.48\textwidth]{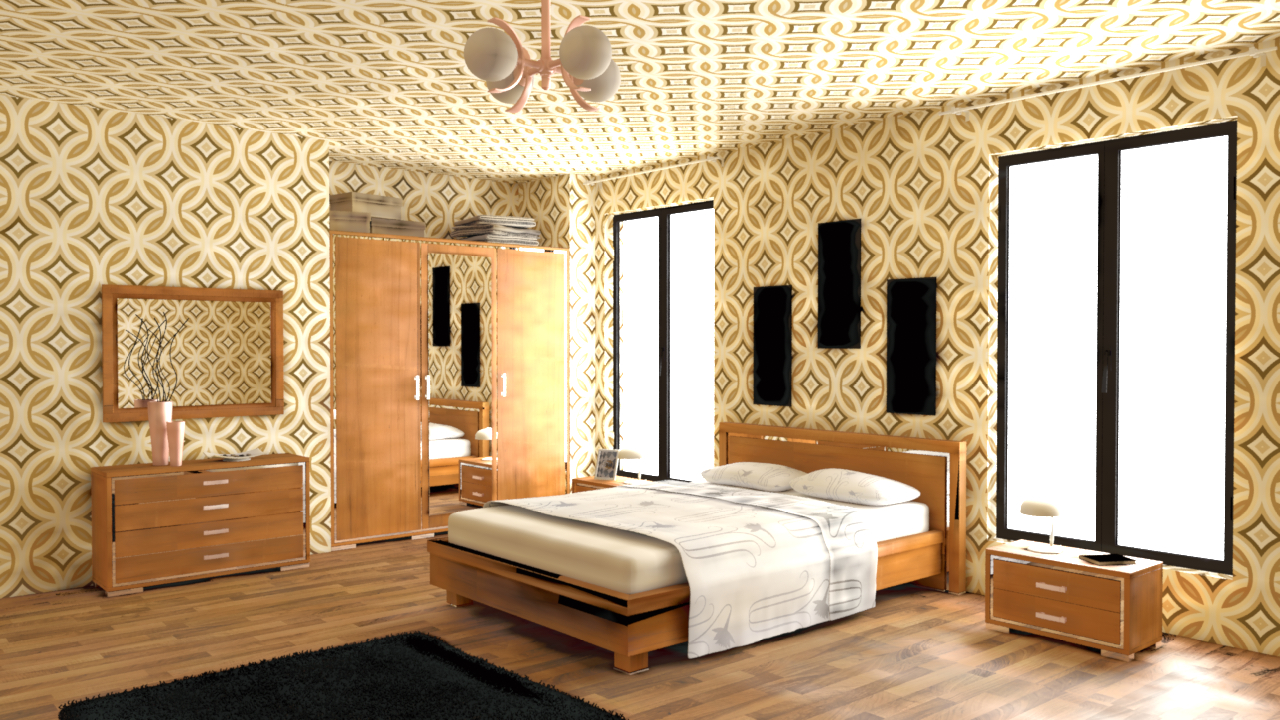}
}
\subcaptionbox{Normalize only by RHS, MSE = 0.061, MAPE: 0.173}{
\includegraphics[width = 0.48\textwidth]{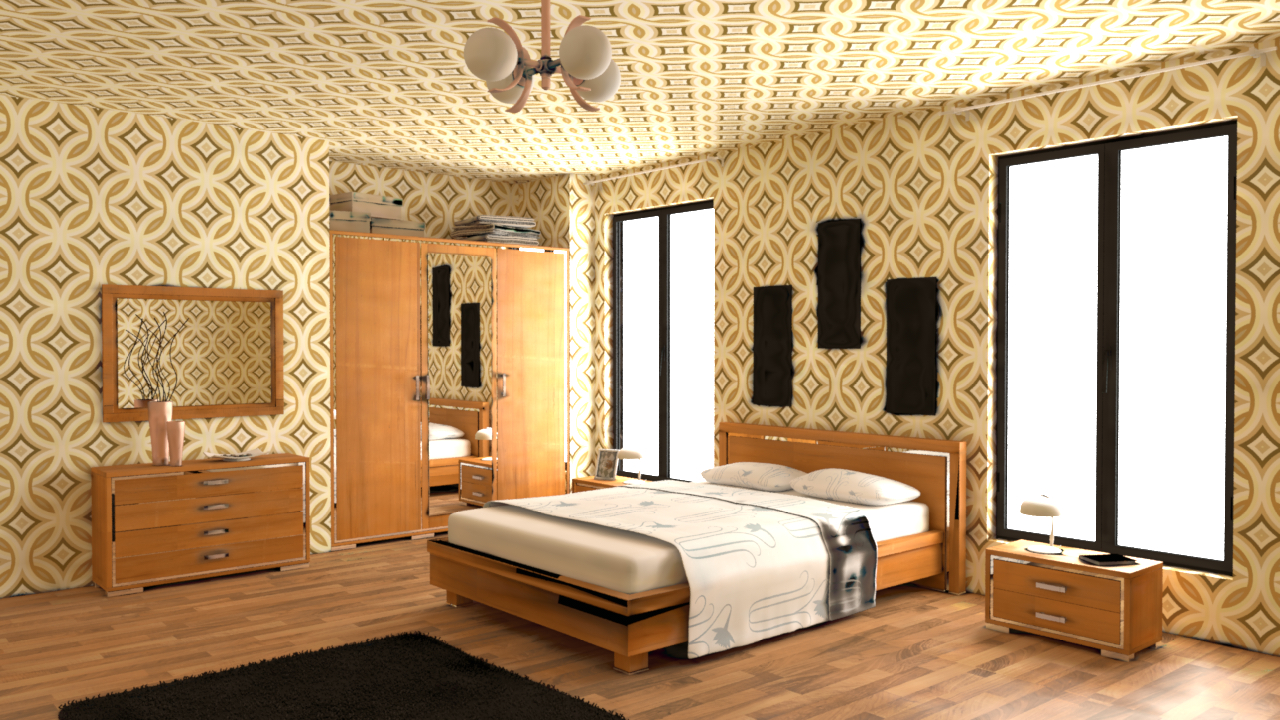}
}

  \caption{Comparing results using residuals with different normalizations.}
  \label{fig:normalizer}
\end{figure*}

We compare the use of different residuals as loss functions, and conclude that the relative residual proposed in the main paper performs the best. The comparison is made by optimizing four different loss functions: relative residual normalized by average of LHS and RHS as proposed in the paper, by LHS only, by RHS only, or with no normalization.  Figure~\ref{fig:normalizer} illustrates different renderings of the \emph{Bedroom} scene after optimizing the above loss terms separately. We found that our approach using normalization of the average of the LHS and RHS leads to the most accurate solutions. We observed that normalizing by the LHS often creates numerical instabilities and exploding gradients during training. 

\subsection{Network Inputs}

We input the normalized position $x\in R^3$ along with the vectors obtained from our multi-level feature grid.  As mentioned in the main paper, we also provide normalized directional inputs including the outgoing direction as $\omega_o$ and surface normals. The other inputs are surface roughness in $R^2$, and diffuse or specular reflectance values in $R^3$.
Table \ref{tab:network inputs} lists all the network inputs used in our method.

\begin{table}[htb]
    \centering
    \caption{Network Inputs}
    \begin{tabular}{c|c |c}
        Input & Symbol & Range \\
        \hline
        Position & $ x \in \mathbb{R}^3$ & $[0,1]$\\
        Direction & $ \omega \in \mathbb{R}^3$ & $[-1,1]$\\
        Surface Normal &  $n \in \mathbb{R}^3$ & $[-1,1]$\\
        Diffuse Reflectance &  $f_{d} \in \mathbb{R}^3$ & $[-1,1]$\\
        Specular Reflectance & $f_{r} \in \mathbb{R}^3$ & $[-1,1]$\\
        Interpolated features & $G(x) \in \mathbb{R}^{16}$ & -\\
    \end{tabular}
    \label{tab:network inputs}
\end{table}

\subsection{Implementation of Sparse Feature Grids}

Storing dense feature vectors incurs cubic storage costs in the grid resolution; the space inefficiency comes from the fact that many voxels do not contain any scene surfaces. By using a fast pre-processing step, we find and store only the voxels with some surface inside them. The algorithm begins by finding voxels that contain at least one triangle; next we store a hash value for each relevant voxel in a sorted \emph{array of hashes}. An array of feature vectors with the same size as the \emph{array of hashes} is then allocated. With this scheme, the hash of the queried voxel is computed in $O(1)$, and look-up of the feature vector array can be done efficiently with PyTorch's GPU-accelerated sorted search. In our implementation, the hash value we used is simply $x + y\times{grid size} + z\times{grid size}^2$, where $(x,y,z)$ is the voxel coordinate. 



\end{document}

%% file: body.tex
\section{Introduction}

In this paper\footnote{The published version available at  https://doi.org/10.1145/3478513.3480569}, we introduce a novel approach to solve the rendering equation~\cite{Kajiya86REq} using deep learning. We represent the radiance function, constituting the solution to the rendering equation, using a neural network. Our key idea is to optimize the parameters of this radiance function directly to minimize the residual of the rendering equation. This approach is conceptually similar to finite element methods, which are known as radiosity algorithms in computer graphics. Therefore, we call our approach Neural Radiosity.

Neural networks are most commonly associated with applications in artificial intelligence, such as computer vision, natural language processing, or robotics. At the core of these applications, neural network training involves very high-dimensional optimization problems, which are addressed using gradient-based techniques with automatic differentiation (i.e., error backpropagation). Neural networks are attractive for these applications because they can represent complex high-dimensional functions with millions of parameters, while supporting effective gradient-based optimization. Yet, the powerful combination of a flexible, high-dimensional function representation and automatic differentiation has many potential applications outside of artificial intelligence. For example, neural network architectures have recently been proposed to represent radiance fields~\cite{mildenhall2020nerf}, or to represent solutions of partial differential equations~\cite{sitzmann2019siren}.

Here, we propose using neural networks as a function representation to solve the rendering equation. Similarly as in radiosity techniques, we solve the rendering equation by minimizing the norm of its residual, which we achieve by optimizing over the parameters of the function representation as illustrated in Fig.~\ref{fig:teaser}(a). Our approach decouples solving for the radiance distribution in the scene from rendering perspective images, similar to traditional radiosity techniques. After obtaining the radiance distribution, arbitrary views can be rendered efficiently as shown in Fig.~\ref{fig:teaser}(b). A major challenge in traditional radiosity approaches is to represent solutions of particular scenes efficiently, by providing sufficient accuracy while avoiding storage and computation of unnecessary degrees of freedom. This can be achieved with methods that adapt the function representation on the fly during the optimization process. Adaptive techniques, which include progressive
meshing~\cite{Cohen88PRR,Lischinski92DMR}, wavelet radiosity~\cite{Wavelet}, or hierarchical meshless basis functions~\cite{Lehtinen08MHR}, are often cumbersome to implement, however. In addition, traditional radiosity approaches are typically limited to purely diffuse scenes in practice, although extensions to non-diffuse scenes have been proposed ~\cite{Immel86RND}.

In contrast, our neural network-based approach represents the full four-dimensional radiance field and is not limited to diffuse scenes. Our approach also does not require meshing or adaptive techniques. These advantages are due to the non-linear nature of neural networks, which also implies that minimizing the residual norm as a function of the network parameters is a non-linear optimization problem. Yet this challenge can be solved robustly using mini-batch stochastic gradient descent, taking advantage of neural network frameworks that provide automatic differentiation (error backpropagation). Minimizing the norm of the residual using mini-batch stochastic gradient descent corresponds to Monte Carlo estimation of the gradient of the residual in each step. To improve convergence, we propose a network architecture including multi-resolution learnable features.

In summary, this paper makes the following contributions:
\begin{itemize}
    \item We propose Neural Radiosity, a method to solve the rendering equation using neural networks. As in radiosity techniques, we minimize the norm of the residual of the rendering equation over the parameters of the function representation.
    \item We demonstrate a practical approach to solve the resulting non-linear optimization using mini-batch stochastic gradient descent. To improve convergence, we propose a network architecture that includes multi-resolution learnable features.
    \item We illustrate the feasibility of this approach experimentally by showing successful solutions for several scenes. In addition, we provide the source code of our approach.
\end{itemize}

\section{Related Work}

\paragraph{Radiosity Techniques.} The radiosity algorithm was proposed by Goral et al.~\shortcite{Goral84Radiosity} to model indirect illumination in diffuse scenes. Similar to our approach, the key idea is to use a function space to represent the solution of the rendering equation; the original radiosity approach used a linear function space consisting of piecewise constant functions. Zatz~\shortcite{Zatz1993GRd} generalized this approach by reformulating it using Galerkin finite element methods and introducing higher-order polynomial basis functions. Even with higher-order polynomials, however, it is challenging to represent solutions of complex scenes accurately and compactly. Hence, various techniques have been proposed to adapt the basis functions to the solution of a scene~\cite{Cohen88PRR,Lischinski92DMR,Wavelet,Lehtinen08MHR}. 
While it is possible to extend such techniques to non-diffuse environments~\cite{Immel86RND}, high storage requirements make them unattractive for complex scenes. More recently, Dahm and Keller~\shortcite{dahm2017learning} used reinforcement learning to solve the rendering equation, which leads to a similar optimization problem as in our technique. They did not use neural networks in their implementation, however. 

\paragraph{Neural Representations of Radiance Fields.} Using neural networks to represent radiance fields has been popularized by Mildenhall et al.'s work on Neural Radiance Fields (NeRF)~\shortcite{mildenhall2020nerf}. They address the problem of novel view synthesis given a set of input photographs of a static scene. Their approach leverages a multilayer perceptron (MLP) as a regression function to interpolate the radiance samples given by the pixels of the input images. 
Our approach similarly represents radiance fields using a neural network, but we optimize the network parameters to minimize the norm of the residual of the rendering equation. This stands in contrast to solving a regression problem based on input samples that already contain global illumination.

\paragraph{Neural Network Techniques for Realistic Rendering.} Neural network techniques have been very successful for post-processing of rendered images, including denoising, antialiasing, and super-resolution~\cite{Vogels18DKP}. In these techniques, neural networks operate on images, and training is typically performed in a supervised manner using a dataset of high-quality ground truth images. Neural networks have also been leveraged for importance sampling in Monte Carlo rendering~\cite{Importance_sampling,Muller2019NIS}. These techniques compute sets of Monte Carlo samples as training data to learn probability densities represented by neural networks. More generally, the rendering equation is a Fredholm equation of the second kind, and previous work has proposed neural networks to solve such equations~\cite{Fredholm}.

Recently, M\"uller et al.~\shortcite{NCV} proposed neural networks as control variates in Monte Carlo integration (neural control variates, or NCV). This work has similarities with our approach, although we start from a different perspective. Control variates are a well-known variance reduction technique in Monte Carlo integration. In the basic formulation, the desired integral of an original function is expressed as the known integral of a ''simpler'' function (the control variate) plus the Monte Carlo estimate of the difference to the original integrand. Intuitively, if the difference between the original integrand and the control variate is closer to a constant, then the variance of this estimate is lower. The application of control variates to the rendering equation requires the computation of parameterized integrals; in other words, separate integrals representing the radiance for each pixel, or the outgoing radiance for each location and direction on a surface. M\"uller et al. therefore formulate parameterized control variates, which are factored into a scalar approximating the integral at each surface location and outgoing direction (the integral of the CV), and a normalized shape function (the shape of the CV). They then train separate neural networks to represent the integral of the CV, the shape of the CV, a sampling density proportional to the difference between the shape of the CV and the true integrand, and an additional weight for the control variate.

In contrast, we take a perspective akin to traditional radiosity techniques. We only use a single neural network representation of the outgoing radiance, and we directly optimize the network parameters to minimize the norm of the rendering equation residual. Our approach completely decouples solving the rendering equation from rendering (perspective) images using a given radiance field, as in classical radiosity techniques. Finally, our approach is not based on the series expansion of the rendering equation, which forms the basis of path tracing and Metropolis light transport. We do not compute any path integrals, but instead solve for the outgoing radiance field directly. Our approach is similar to NCV in that we both use neural networks to represent the outgoing radiance distribution; however, we only train one neural network with a simple residual loss. Despite the simplicity of our method, we demonstrate successful results on a variety of scenes. Concurrent work by M\"uller et al.~\shortcite{Mueller2021NRC} can be seen as a simplified version of NCV that is more similar to our approach since it also uses a single neural network, but it is geared towards real-time rendering and does not use our approach of minimizing the residual of the rendering equation.
\paragraph{Neural Network Architectures.} Neural network architecture often has a significant influence on the convergence properties of network training. For example, NeRF~\cite{mildenhall2020nerf} proposed a positional encoding technique to embed the spatial input parameters into a higher dimensional vector,
which led to improved results in practice. Tancik et al.~\cite{tancik2020fourier} analyzed this strategy using neural tangent kernels~\cite{jacot2020neural}, showing that passing the input parameters through a Fourier transform allows an MLP network to capture high frequency details more effectively. As an alternative, Sitzman et al.~\cite{sitzmann2019siren} introduce periodic activation functions. Using a principled initialization scheme, they show that their approach can better represent high-frequency detail compared to vanilla MLPs with ReLU activations.
Our approach is inspired by the Neural Geometric Level of Detail technique by Takikawa et al.~\shortcite{nlod} and similar techniques~\cite{liu2020neural,Local_Implicit_Grid_CVPR20,Peng2020ECCV,chabra2020deep,chibane2020implicit}. They propose MLP enhancements using a hierarchy of local learnable features, and demonstrate how this can improve the accuracy of implicit representations of 3D geometry.




\section{Problem Formulation}

Our goal is to compute solutions of the rendering equation, which can be expressed as
\begin{align}
\label{eq:rendering-equation}
L(x,\omega_o) = E(x,\omega_o) + \int_{\mathcal{H}^2} f(x,\omega_i, \omega_o) L(x'(x,\omega_i),-\omega_i) d\omega_i^{\perp},
\end{align}
where $L(x,\omega_o)$ represents the unknown radiance distribution, which is a function of 3D locations $x$ on surfaces and outgoing directions $\omega_o$. Light sources or emitters are represented by the emitted radiance distribution $E(x,\omega_o)$, and light scattering on surfaces is described by the integral over the incident directions $\omega_i$ on the hemisphere $\mathcal{H}^2$ using the projected solid angle measure $d\omega_i^{\perp}$. Here, $f(x,\omega_i, \omega_o)$ is the bidirectional reflectance distribution function (BRDF).
In addition, $L(x'(x,\omega_i),-\omega_i)$ represents the incident radiance at $x$ from incident direction $\omega_i$. Under the assumption that only surface scattering is considered, this is equivalent to the outgoing radiance in the direction $-\omega_i$ at the surface location intersected by a ray starting at $x$ in direction $\omega_i$, which is denoted by $x'(x,\omega_i)$.

\paragraph{Neural network-based radiance fields.} In our approach, the unknown solution, that is the radiance distribution $L(x,\omega_o)$, is represented using a neural network. The variables in our problem are the set of neural network weights and bias values denoted by $\theta$, which are nonlinearly related to the output values of the network. In addition, the network is a monolithic function representation that does not consist of locally supported basis functions. Therefore, the typical finite element approach is not applicable.

Let us denote a radiance distribution given by a set of network parameters $L_{\theta}(x,\omega_o) \in \mathcal{F}$. Here $\mathcal{F}$ denotes 
the space of functions that can be represented by the chosen network architecture. In addition, the residual of the rendering equation is 
\begin{align}
\label{eq:renderingresidual}
    r_{\theta}(x,\omega_o) = & L_{\theta}(x,\omega_o) - E(x,\omega_o) \nonumber \\
    & - \int_{\mathcal{H}^2} f(x,\omega_i, \omega_o) L_{\theta}(x'(x,\omega_i),-\omega_i) d\omega_i^{\perp},
\end{align}
where the notation $r_{\theta}$ indicates that the residual depends on the parameters $\theta$ of the radiance function $L_{\theta}$. In our approach, we express a suitable norm of the residual (here the squared $L_2$ norm for example) 
as a nonlinear function of the network parameters $\theta$,
\begin{align}
\label{eq:renderingloss}
\mathcal{L}(\theta) &= \left\| r_{\theta}(x,\omega_o) \right\|_2^2 \\
&= \int_{\mathcal{M}}\int_{\mathcal{H}^2} r_{\theta}(x,\omega_o)^2 dxd\omega_o, \nonumber
\end{align}
where integration is over all scene surfaces $\mathcal{M}$ and the hemisphere $\mathcal{H}^2$. Our desired solution is the radiance $L_{\theta^*}(x,\omega_o)$, where 
\begin{align}
    \label{eq:solution}
    \theta^* = \argmin_{\theta} \mathcal{L}(\theta).
\end{align}

We propose to minimize the residual norm $\mathcal{L}(\theta)$ using minibatch stochastic gradient descent, where the norm of the residual is repeatedly estimated using Monte Carlo integration, and gradients of the radiance field are computed using automatic differentiation.

\paragraph{Monte Carlo Estimate.} The MC estimate of the residual norm is
\begin{align}
    \mathcal{L}(\theta) \approx \frac{1}{N}\sum_{j=1}^{N} \frac{r_{\theta}(x_j,\omega_{o,j})^2}{p(x_j,\omega_{o,j})},
\end{align}
where $N$ is the number of samples, and samples of surface locations $x_j$ and outgoing directions $\omega_{o,j}$ are distributed according to probability density $p(x,\omega)$.

\paragraph{Gradients.} The MC approximation of the gradient $\nabla_{\theta} \mathcal{L}(\theta)$ is
\begin{align}
\label{eq:residualsampling}
\nabla_{\theta} \mathcal{L}(\theta) \approx& \frac{1}{N}\sum_{j=1}^{N} \frac{2 r_{\theta}(x_j,\omega_{o,j}) \nabla_{\theta} r_{\theta}(x_j,\omega_{o,j})}{p(x_j,\omega_{o,j})},
\end{align}
and the gradient of the residual is
\begin{align}
\nabla_{\theta} r_{\theta}(x,\omega_o) = & \nabla_{\theta} L_{\theta}(x,\omega_o) - \nonumber \\
&\int_{\mathcal{H}^2} f(x,\omega_i, \omega_o) \nabla_{\theta} L_{\theta}(x'(x,\omega_i),-\omega_i) d\omega_i^{\perp},
\end{align}
%
This gradient will also be approximated by evaluating the hemispherical integral using Monte Carlo integration,
\begin{align}
\label{eq:incidentintegral-MCestimation}
&\nabla_{\theta} r_{\theta}(x_j,\omega_{o,j}) = \nabla_{\theta} L_{\theta}(x_j,\omega_{o,j}) \nonumber \\
&-\frac{1}{M} \sum_{k=1}^{M} \frac{f(x_j,\omega_{i,j,k}, \omega_{o,j}) \nabla_{\theta} L_{\theta}(x'(x_j,\omega_{i,j,k}),-\omega_{i,j,k})}{p(\omega_{i,j,k})}.
\end{align}
Here $\omega_{i,j,k}, k \in \{1,\dots, M\}$ is a set of $M$ samples of incident directions for each sample $x_j, \omega_{o,j}$ ($i$ stands for ''incident'', not an index). 

\paragraph{Minibatch Stochastic Gradient Descent.}

We minimize the residual norm using minibatch stochastic gradient descent, as described by the pseudocode in Algorithm~\ref{alg:sgd}. Because the network is used to evaluate both the left and right hand side of the rendering equation as shown in Equation~\ref{eq:renderingresidual}, we could call this a ``self-training'' approach. Since the Monte Carlo gradient estimates are unbiased, this stochastic gradient descent is guaranteed to converge to a local minimum. If the network capacity is unlimited, it is guaranteed to converge to the exact solution where the residual norm vanishes. In practice, we improve convergence using adaptive momentum methods such as Adam~\cite{Kingma2015Adam}.

\begin{algorithm}
\SetAlgoLined
 initialize network parameters $\theta$\;
 \While{not converged}{
  sample a set of surface points $\{x_j | j=1\dots N\}$ and outgoing directions $\{\omega_{o,j} | j=1\dots N\}$\;
  for each $(x_j, \omega_{o,j})$, sample a set of incident directions $\{\omega_{i,j,k} | k=1\dots M\}$\;
  use the samples to evaluate the Monte Carlo estimate of $\nabla_{\theta} \mathcal{L}(\theta)$ using Equations~\ref{eq:residualsampling} and~\ref{eq:incidentintegral-MCestimation}\;
  $\theta = \theta - \eta \nabla_{\theta} \mathcal{L}(\theta)$\;
 }
 return $\theta$\;
 \caption{\label{alg:sgd} Minibatch stochastic gradient descent, learning rate $\eta$.}
\end{algorithm}

\paragraph{Image Synthesis.}

The solution $L_{\theta^*}(x,\omega_o)$ obtained from~\eqref{eq:solution} represents the radiance field over the surfaces of the entire 3D scene. This allows us to synthesize 2D images rapidly using perspective projection or ray tracing, and by evaluating the left hand side (LHS) of the rendering equation $L_{\theta^*}(x,\omega_o)$ for each surface location $x$ corresponding to an image pixel, and the direction $\omega_o$ that points from $x$ to the center of projection (the virtual camera location). Alternatively, we can render images by evaluating the right hand side (RHS) using Monte Carlo integration of the hemispherical integral. A comparison video of LHS/RHS and path-traced renderings is presented as supplemental material.

\paragraph{Evaluation Metric.}

Throughout this paper, we use Mean Squared Error (MSE) and Mean Absolute Percentage Error (MAPE), defined as $|\mathrm{img}-\mathrm{ref}|/(\mathrm{ref}+0.01)$, to compare image-based error with respect to the path-traced ground truth.

\begin{figure}[ht]
\centering
\includegraphics[width = \columnwidth]{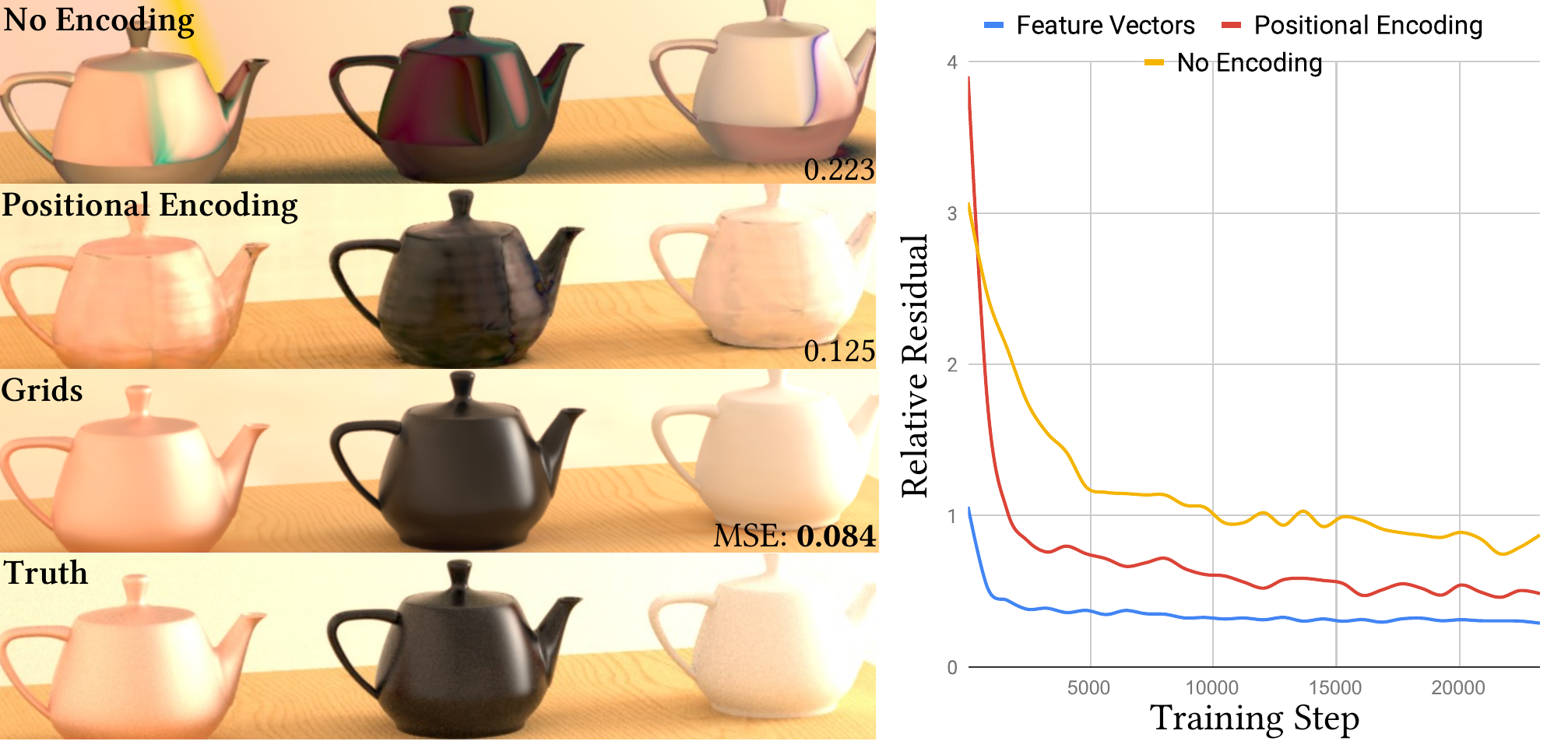}
\hspace{-5pt}
\caption{Comparison of LHS renderings of \emph{Veach Door} with different encoding techniques. For large scenes, the positional encoding approach produces artifacts in regions with more detailed geometry. LHS renderings are most faithful to the ground truth when multi-resolution feature grid vectors are used. From the training curves, we observe that multi-resolution learnable feature grids show superior convergence and training stability. 
}
\label{Comparison}
\end{figure}

\section{Implementation}

In this section we discuss several design choices of our implementation that are important to achieve accurate results in practice.

\subsection{Relative Residual}
The highly dynamic range of radiance values can introduce numerical instability to the training process, in particular with the $L_2$ norm described in Equation \ref{eq:renderingloss}.
As observed in previous work~\cite{NCV}, normalizing the loss values leads to a more stable training process, and facilitates learning darker parts of the scene. Thus, we modify our loss function in Equation~\ref{eq:renderingloss} to 
\begin{align}
\label{eq:renderingloss_normal}
\mathcal{L}(\theta) &= \left\| \frac{ r_{\theta}(x,\omega_o)  }{sg( m_{\theta}(x,\omega_o) ) + \epsilon} \right\|_2^2 \\
m_{\theta}(x,\omega_o) &= \frac{L_{\theta}(x,\omega_o) + E(x,\omega_o) + \mathrm{T}\{L_{\theta}\}(x,\omega_o)}{2} \\
\label{eq:taointegral}
\mathrm{T}\{L_{\theta}\}(x,\omega_o) &= \int_{\mathcal{H}^2} f(x,\omega_i, \omega_o) L_{\theta}(x'(x,\omega_i),-\omega_i) d\omega_i^{\perp},
\end{align}
where $\epsilon$ is a constant. The stop-gradient $sg(.)$ operator excludes the mean $m_\theta$ from contributing to the gradient during optimization. We compare additional normalizations in our supplemental material.  

\subsection{Emission Reparameterization}

The outgoing radiance in Equation~\ref{eq:rendering-equation} consists of the sum of the emitted and scattered radiance.  In our implementation, we treat the emittance as a term given by the scene, and only learn to approximate the scattered radiance term. 
Hence we bypass artifacts that arise when trying to model high-contrast discontinuities at the boundaries of light sources. More formally, we reparameterize as follows:
\begin{align}
L_{\theta}(x,\omega_o) &= N_{\theta}(x,\omega_o) + E(x,\omega_o), \\
r_{\theta}(x,\omega_o) &= L_{\theta}(x,\omega_o) - E(x,\omega_o) -\mathrm{T}\{L_{\theta}\}(x,\omega_o)\\
&= N_{\theta}(x,\omega_o) -\mathrm{T}\{N_{\theta} + E \}(x,\omega_o), \label{eq:emissionreparam}
\end{align}
%
%
where $\mathrm{T}\{L_{\theta}\}$ is the scattered radiance as defined in Eq.~\ref{eq:taointegral}, and $N_{\theta}(x,\omega_o)$ is now the network being optimized.  


\begin{figure*}[t]
    \centering
    \includegraphics[width = \textwidth]{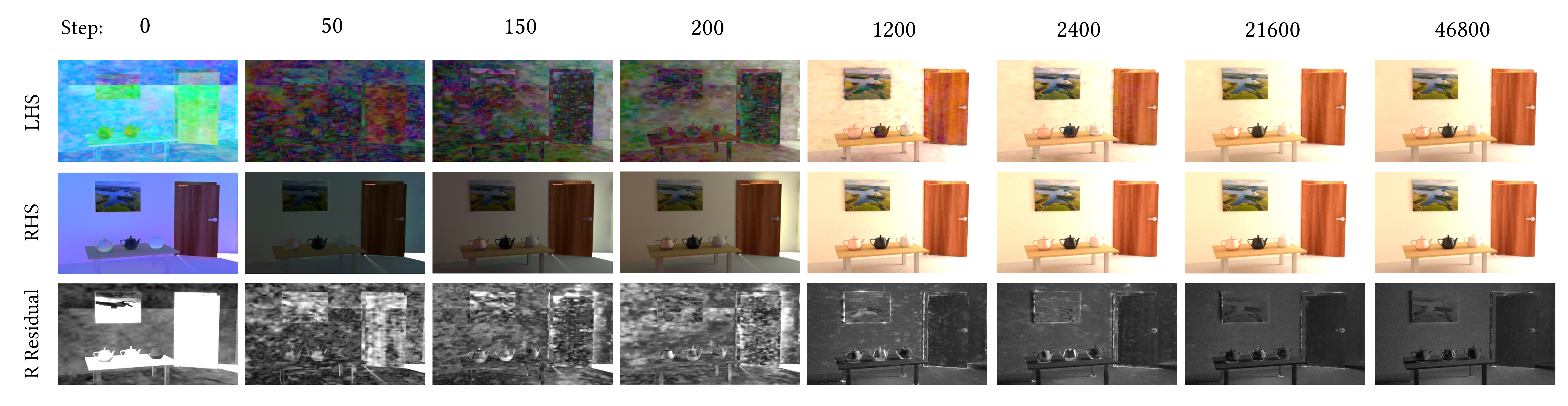}
    \caption{Training milestones for \emph{Veach Door}. RHS, LHS, and the relative residual for certain training steps are visualized. As training proceeds, the LHS gradually learns the radiance distribution, and the light coming from the room behind the door propagates to the RHS renderings. The residual visualisations show the relative residual. We use batch size $N=8192$, $M=32$, and 48K steps $S$. Thus, the total number of samples is $N \times M \times S = \num[group-separator={,}]{12580}$ million.  }
    \label{fig:veach_training_milestones}
\end{figure*}

\subsection{Multi-resolution Feature Grid}


Our learned radiance field function $L_{\theta}$ takes in a 3D point position $x\in \mathbf{R}^3$ and direction $\omega_o\in \mathcal{H}^2$.  A naive implementation may be a simple multi-layer perceptron with a 5D input $[x,\omega_o]$.
%
%
However, it has been observed that the input encoding may play a large role in determining model performance.  A popular mapping in previous work~\cite{mildenhall2020nerf} is sinusoidal positional encoding.
%
%

%
%
Instead, we saw significant performance gains from encoding the input using learned multi-resolution feature vectors, which has been observed similarly in other applications~\cite{nlod,liu2020neural,Local_Implicit_Grid_CVPR20,Peng2020ECCV,chabra2020deep,chibane2020implicit}.
Specifically, we define $n$ regular 3D lattice grids containing the entire scene, each at a different resolution.  At each point on each lattice, we initialize a feature vector of length $l$.  To query the model at a given point, we trilinearly interpolate the features stored at the enclosing voxel's eight corners. This is done for each lattice, and the resultant interpolations from each level of detail are averaged into a final feature vector.  We finally feed this embedding to the downstream network in addition to the raw position coordinates. More formally,
\begin{align}
    L_{\theta}(x,\omega_o) = MLP\left(\begin{bmatrix}
        x\\
        G(x)\\
        \omega_o
    \end{bmatrix}\right), \quad
    G(x) = \frac{1}{n}\sum_{i=0}^{n-1} trilinear(x, V_i[x]),
\end{align}
where $G: \mathbf{R}^3 \rightarrow \mathbf{R}^l$ is the multi-resolution grid embedding function, and $V_i[x]$ represents the eight $l$-dimensional features at the corners of the voxel enclosing $x$ on lattice $i$. Since the $trilinear$ interpolation is differentiable, we may easily train the features. We compare this approach to positional encoding~\cite{mildenhall2020nerf} in Fig.~\ref{Comparison}.

\subsection{Scene Information as Additional Input}
\label{sec:addscene}

The BSDF $f(x,\omega_i, \omega_o)$ plays a critical role when integrating for the residual loss; it is directly multiplied with incoming radiances $L_{\theta}(x'(x,\omega_i),-\omega_i)$ sampled from the model in training.  However, for complex materials such as micro-facets, the BSDF may be highly non-linear in $(x, \omega_o)$, significantly destabilizing training.  To mitigate the effects of BSDF behavior on training, we follow practices in previous work \cite{NCV,Global_illumination_regression}, providing the network with given scene information such as surface normal, diffuse reflectance, and specular reflectance in addition to the position and angle $(x, \omega_o)$.  Given this extra information, the network is successful at dealing with various materials with complex BSDFs. We report the network inputs in detail in our supplemental material.

\subsection{Specular BSDFs and Caustics}
\begin{figure}[bt]
  \centering
\subcaptionbox{LHS, spp: 512}{  
\includegraphics[width = 0.47\columnwidth]{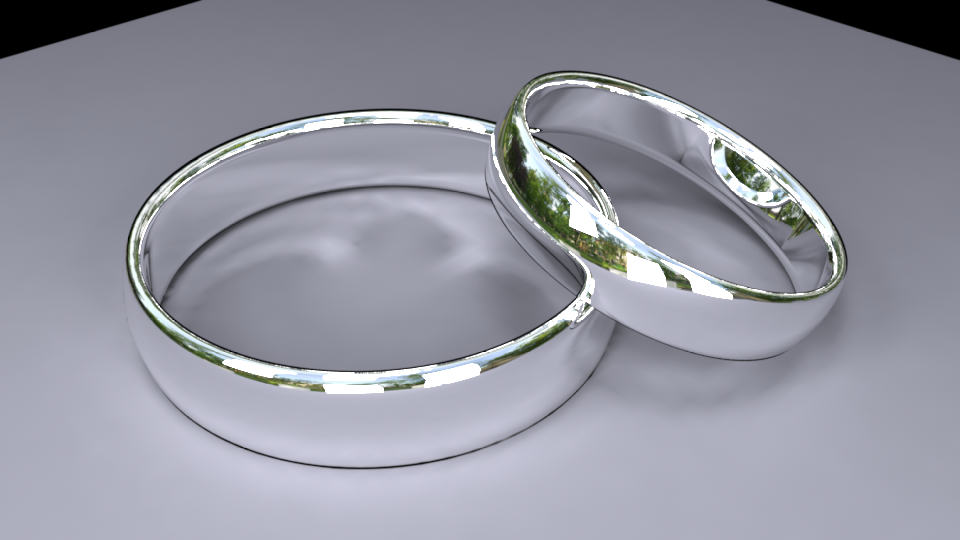}
}
\subcaptionbox{RHS, spp: 512$\times$4}{  
\includegraphics[width = 0.47\columnwidth]{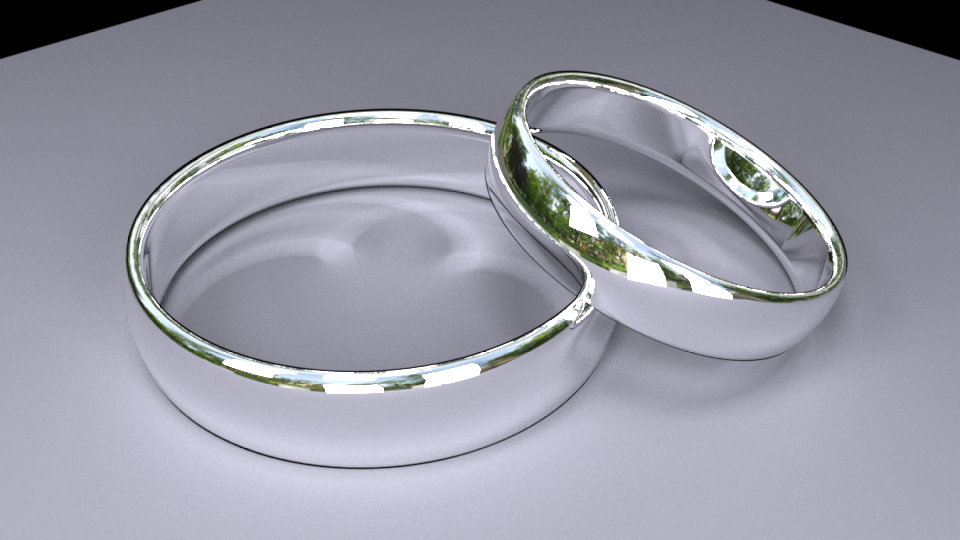}
}
  \caption{\emph{Rings} scene with caustics}
  \label{fig:Cautiscs}
\end{figure}

Perfect mirror BSDFs follow a delta-dirac distribution, which is not square-integrable~\cite{NCV}
and cannot be learned effectively by the model. Instead, 
we trace each ray through (potentially many) specular reflections until it hits a non-specular surface where we can query radiance from our model. Since this removes the need to evaluate our model on specular surfaces, we do not sample points on specular surfaces during training. 
Fig.~\ref{fig:Cautiscs} shows that we can handle caustics formed by specular surfaces, and we demonstrate our model trained on the \emph{Chair} scene with a specular chair in Fig.~\ref{fig:examplescenes}.

\subsection{Training}

\begin{figure}[t]
    \centering
    \subcaptionbox{}{
        \includegraphics[width = 0.47\columnwidth]{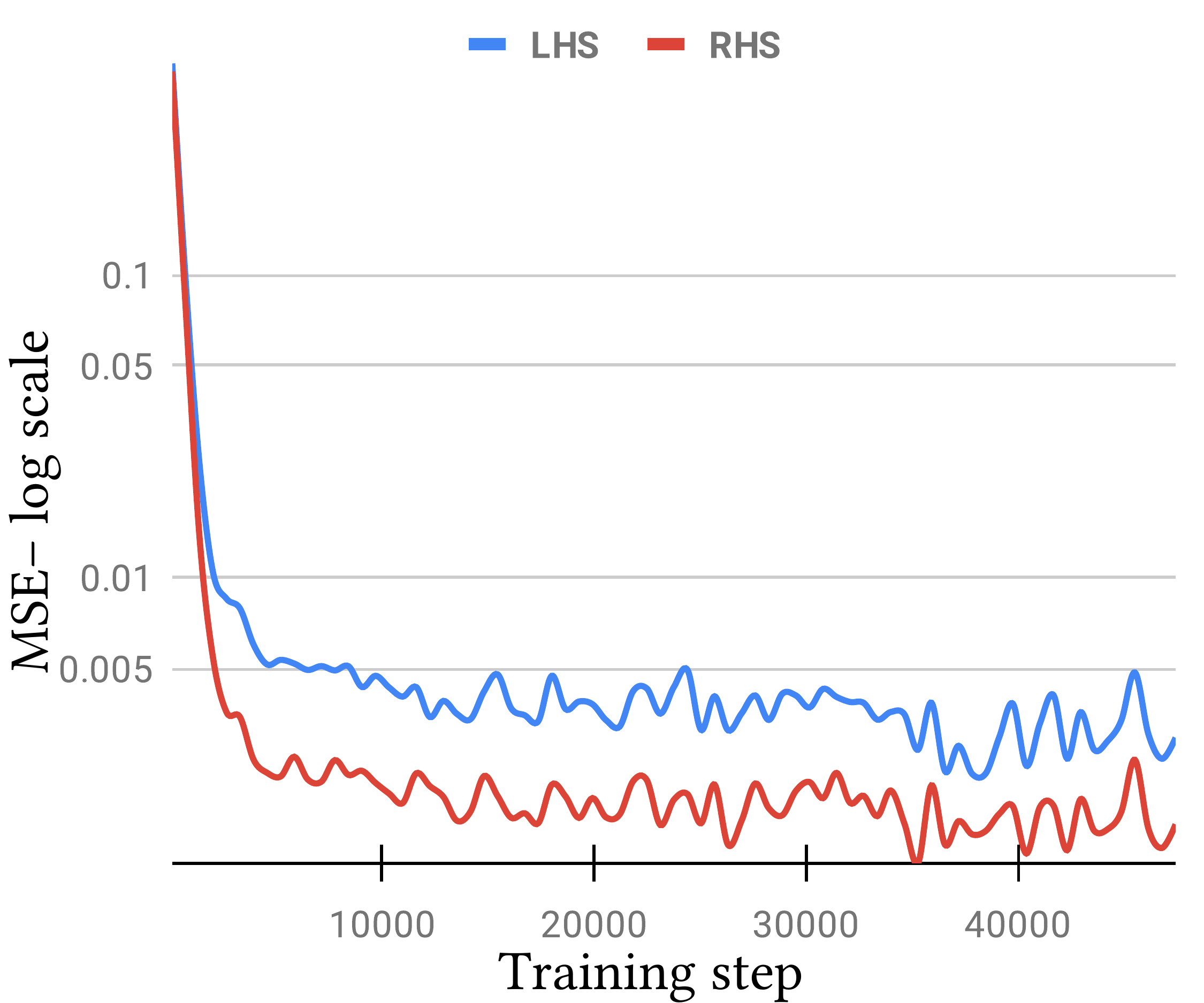}
    }
    \subcaptionbox{}{
        \includegraphics[width = 0.48\columnwidth]{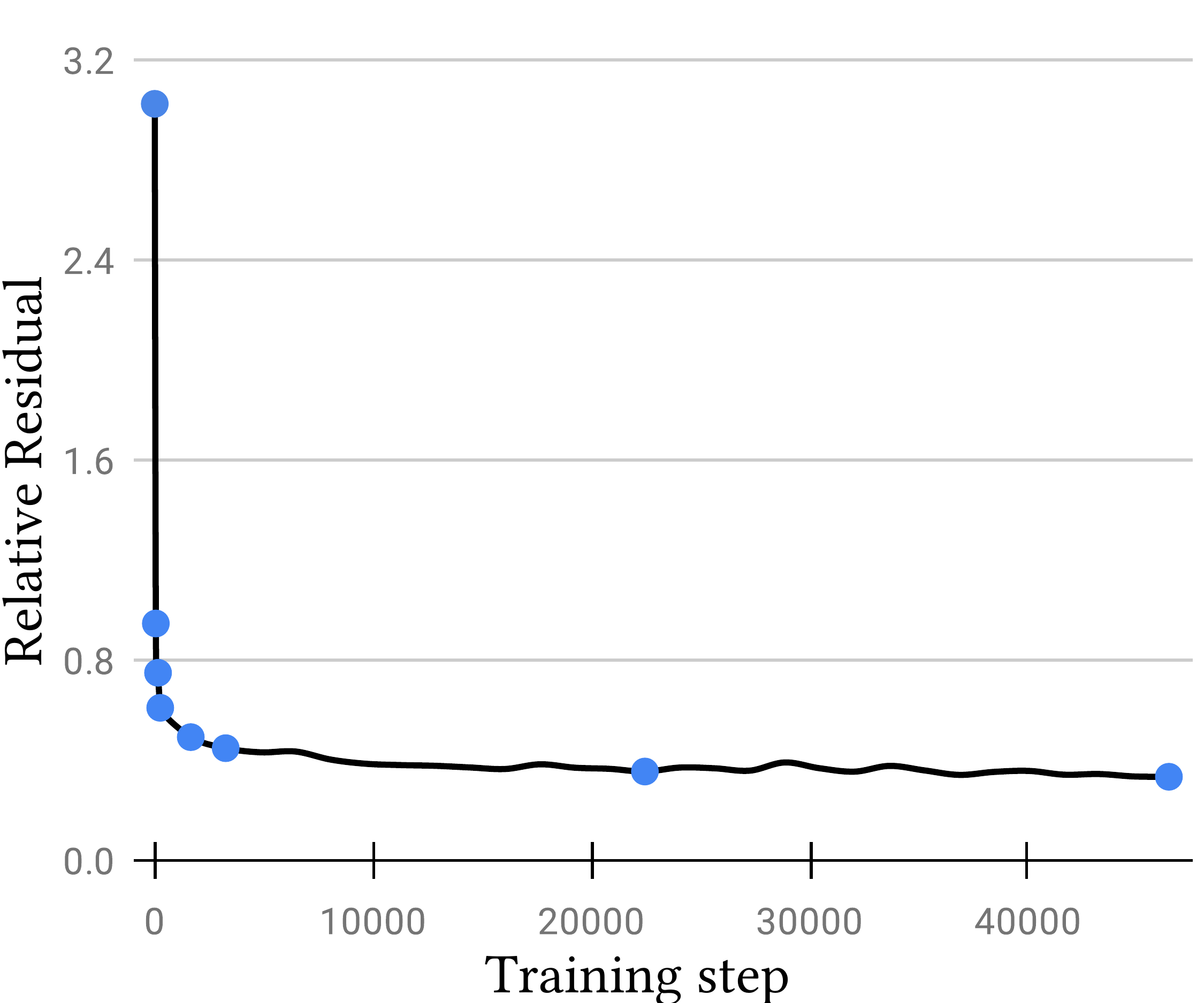}
    }

    \caption{(a) Image-based MSE of LHS/RHS compared to ground truth of the \emph{Veach Door} scene. Note that the vertical axis is log-scale to show the two curves distinctively. As the chart illustrates, the RHS is always more faithful to the ground truth compared to the LHS. This is because RHS involves one more step of ray tracing that integrates multiple network outputs, as opposed to the LHS which computes one network output per sample. (b) Relative residual loss curve during training. The milestones marked on the line chart correspond to the training steps visualized in Fig.~\ref{fig:veach_training_milestones}.\label{fig:veach_training_charts}}
\end{figure}

\input{ablation_fig}
\input{noisyest_fig}

\begin{table*}[t]
    \centering
    \caption{Training costs for the examples in Fig.~\ref{fig:examplescenes}. 
    In practice, training times can be reduced significantly at the cost of small decreases in accuracy. *The large number of objects in the \emph{Bedroom} scene causes Mitsuba 2~\cite{Mitsuba2} to generate large CUDA kernels that require offloading of GPU registers to memory. This leads to slower training and memory leakage in the current implementation of Mitsuba 2. Hence we needed to use a smaller batch size, but still ran out of memory periodically due to memory leakage. We worked around this problem by resuming training three times.
    }    
\begin{tabular*}{\textwidth}{c|c|c|c|c|c|c|c}
    Scene & \makecell[c]{$N$ \\ (Batch Size)} & $M$ & \makecell[c]{Grid \\ Resolution} & \makecell[c]{$S$ \\ (Training Steps)} &\makecell[c]{N $\times$ M $\times$ S \\ (Total \# of Samples)}  & Training Time & Peak Memory GiB \\
    \hline
    Living Room & 16384 & 32 & 64 & 38 K & 20128 M & 5 h 22 m & 9.43 \\
    Bedroom & 8192 & 32 & 64 & 72 K & 18880 M & 7 h 1 m & 10.69* \\
    Dining Room & 16384 & 32 & 512 & 36 K & 18880 M & 5h 23m & 10.58\\
    Veach Door & 16384 & 32 & 128 & 24 K & 12576 M & 3h 1m & 8.7\\
    Chair & 16384 & 32 & 64 & 24 K & 12576 M & 1h 38m & 7.20\\
    Mirror Chair & 16384 & 32 & 64 & 22 K & 11744 M & 1h 47m & 7.46\\
    Rings & 16384 & 32 & 128 & 40 K & 20960 M & 4h 50m  & 10.43\\
    Copper Man - Rest & 16384 & 32 & 32 & 8 k & 4192 M & 52m & 10.10\\
    Cornell Box & 16384 & 32 & 32 & 4 K & 2080 M & 22 m & 9.54
    \end{tabular*}

    \label{tab:training_time}
\end{table*}

Figs.~\ref{fig:veach_training_milestones} and~\ref{fig:veach_training_charts} illustrate the optimization process for the \emph{Veach Door} scene. We use the Adam optimizer~\cite{Kingma2015Adam} with a learning rate of $5e-4$, decayed by a factor of $0.33$ at every one-third of the training process. 
The experiments in this paper use from $n=5$ to $n=9$ resolution levels of feature grids, the first level always being $2\times2\times2$.  Empirically, we found that model performance is stable with feature dimension $l>8$, and so we fix $l=16$ in all our experiments.

\subsection{Sampling}

Our sampling approach is different for the left hand and right hand sides. For the left hand side, the samples $(x_j,\omega_{o,j})$ in Equation~\ref{eq:incidentintegral-MCestimation} are drawn from a uniform distribution. More specifically, $x_j$ is sampled from a uniform distribution over all the surfaces in the scene; $\omega_{o,j}$ is sampled uniformly either on the local unit hemisphere (for one-sided BSDFs) or on the local unit sphere (for two-sided BSDFs).

To sample the transport operator in Equation~\ref{eq:emissionreparam}, the directions $\omega_{i,j,k}$ are sampled 
using BSDF and emitter sampling. Note that sampling $\omega_{i,j,k}$ is repeated $M$ times and averaged to achieve less noisy gradients.

\subsection{Architecture}
The neural network used in our experiments consists of 6 linear layers of width 512 with ReLU activation functions. Thanks to the representational power of multi-level feature grids, we found that the MLP prediction head still delivered satisfying performance despite being relatively shallow. The model parameter initialization that worked best in our experiments was a uniform $[-1,1]$ initialization scheme, scaled by inverse root of weight size (default PyTorch~\cite{PyTorch} linear layer initialization). Our network inputs are discussed in detail in our supplemental document.

\subsection{Renderer}

Neural Radiosity requires calling several rendering methods during training.  In particular, evaluating the right hand side of the rendering equation residual requires one ray-tracing step per sampled point to gather the predicted scattered emittance at intersected points; i.e. to calculate $\mathrm{T}\{L_{\theta}\}$, we must use the model to predict many $L_{\theta}(x'(x,\omega_i),-\omega_i)$.  We display RHS renderings for several scenes in Fig. \ref{fig:examplescenes}, along with the relative residual loss objective. 
The left hand side is simpler to calculate, as we only need to feed sampled points to $L_{\theta}$ once.  Due to our reparameterization of emittance, both RHS and LHS evaluations require retrieving emitted radiance from the given scene properties.  

We use Mitsuba 2~\cite{Mitsuba2} as our rendering engine, and implement the deep learning components with PyTorch \cite{PyTorch}. Mitsuba 2 provides Python bindings that facilitate communication between C++ data structures and PyTorch tensors. Both the rendering and training occur on a single GPU (Nvidia RTX 2080).

\begin{figure*}[t]
  \centering
  \subcaptionbox{Equal LHS rendering time of 14 seconds. \label{fig:LHS_time}}{%
\includegraphics[width = 0.48\textwidth]{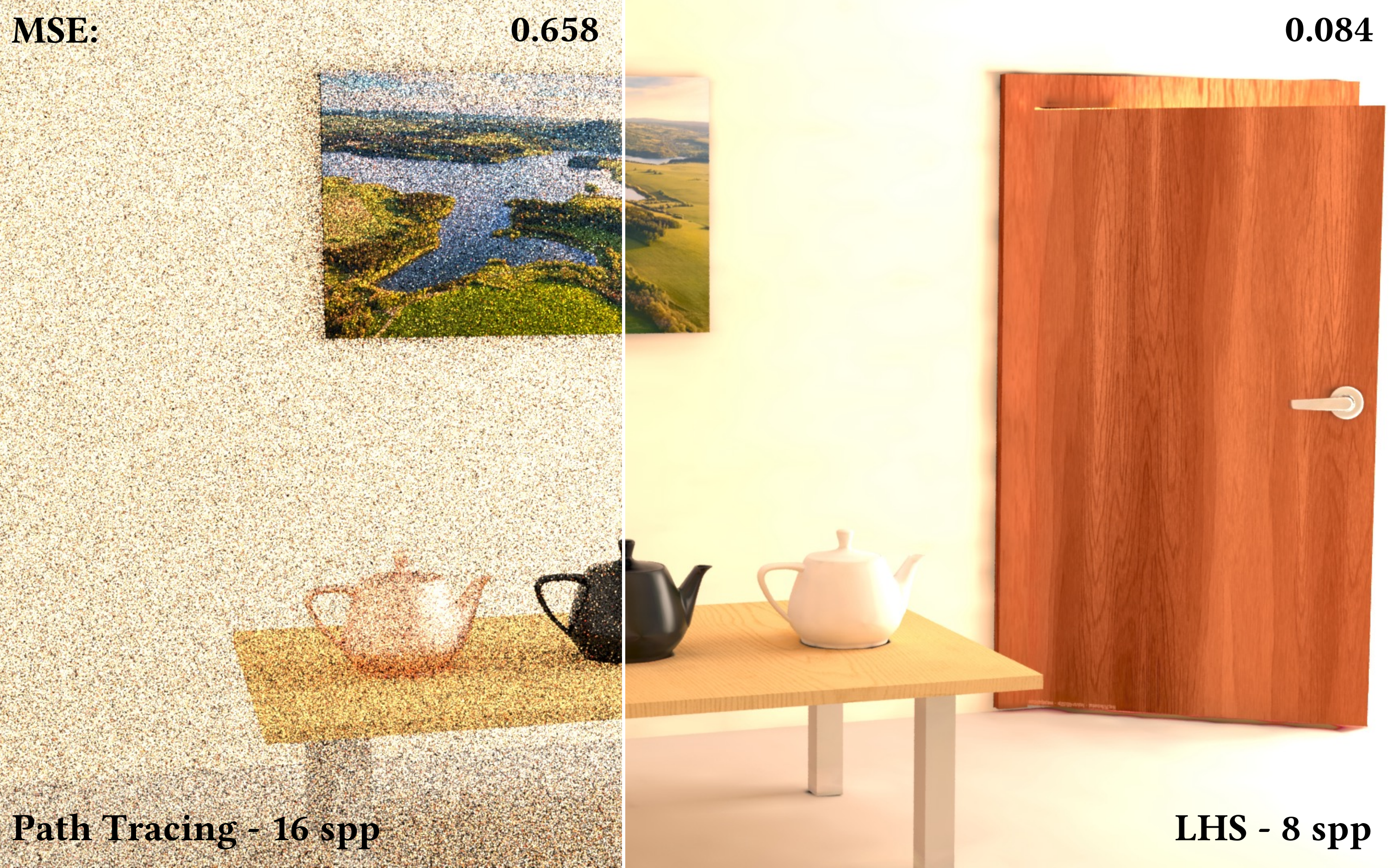}
  }
  \subcaptionbox{Equal RHS ($M=16$) rendering time of 3.5 minutes \label{fig:RHS_time}}{%
    \includegraphics[width = 0.48\textwidth]{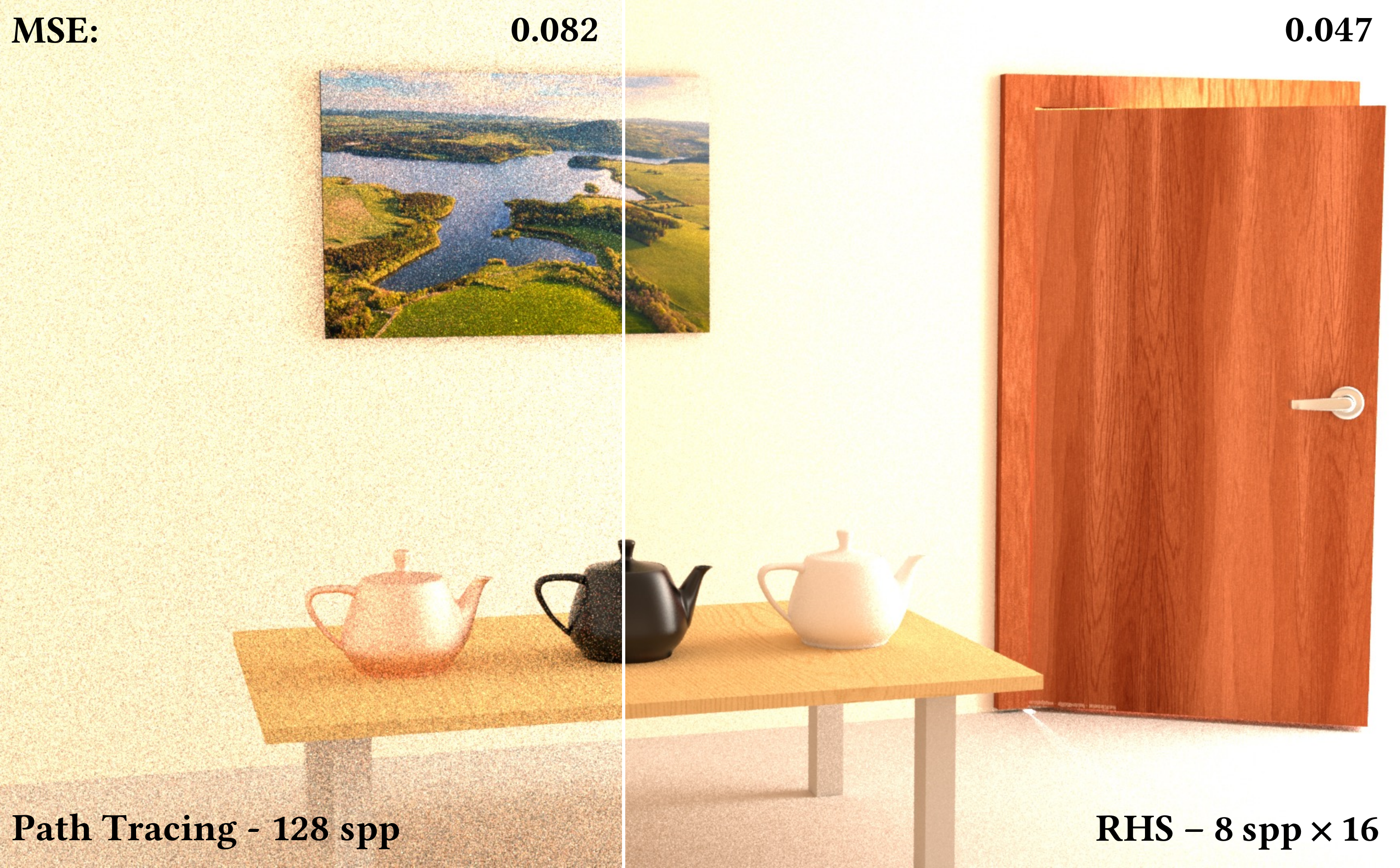}
  }
  \caption{Equal rendering time comparison between our pre-trained model and path-tracing. The training time for the model took about 3 hours (Table \ref{tab:training_time}). As an implementation detail, note that the rendering time depends on parameters such as the GPU thread block size. To use larger spp, we must accommodate GPU memory availability by dividing the computation into smaller blocks. Thus, 128 spp path-tracing is not exactly 8 times as costly as 16 spp.}
  \label{fig:timing_study}
\end{figure*}

\section{Results and Analysis}

\subsection{Training and Rendering Results}

We demonstrate rendering results for several scenes in Fig. ~\ref{fig:examplescenes}, including diffuse, rough specular, plastic, and perfect mirror materials, and report training costs for all scenes in Table~ \ref{tab:training_time}. As our model represents the entire radiance distribution, we may efficiently perform multi-view renderings of static scenes at inference time.  Given a novel view, rendering the LHS consists of simply finding intersections of all rays from the camera pixels, and passing them to our learned neural network. 
Since our neural architecture is rather shallow, the cost of querying the network is small, and is already efficiently parallelized for GPUs by mini-batching. Multiple views of the chair scene are shown in Fig. ~\ref{Chair_MultiView}. 

Fig.~\ref{fig:timing_study} compares LHS/RHS rendering based on a trained model using our method against path-traced results on the \emph{Veach Door} scene given the same amount of render time.  As expected, LHS rendering completes relatively quickly; we only need to ray-trace camera rays with few samples per pixel. On the other hand, path-tracing incurs the typical Monte Carlo noise artifacts. In equal rendering time, our LHS with only 8 spp produces images with much lower error than path-tracing with 16 spp.  Similarly, our RHS with one additional ray-tracing step also produces better quality renderings than path-tracing with equal time and spp. In summary, our approach is competitive with path tracing for novel view synthesis using a trained model, but training takes on the order of minutes or hours.

\subsection{Ablation Study}

We conducted an ablation study on various system components to evaluate their effectiveness, and we show results using the \emph{Living Room} scene (Fig. \ref{fig:Ablation_Living}). We trained our model without feature grids, local property inputs, or multi-resolution grids, and also adjusted network width and depth. The study shows how every component contributes to a better LHS rendering with lower error and faster convergence. As the ablations' differences are less visible with RHS renderings, we only show LHS in the figure. Grids play an important role in learning shadows, and in general any high frequency changes in radiance along the geometry. If no local properties are inputted to the network, the textures are blurry, the glossy reflections are not learned properly, and sharp edges/corners (large surface normals changes) are blurred. We see that a narrower network with width 128 instead of 512 is insufficient to model the complexity of reflections, and converges much more slowly. The table also shows that our architecture is less affected by network depth, with our complete 6-layer model performing marginally better than a 4-layer network.




\input{retraining_fig}

\subsection{Dynamic Scenes}

Neural Radiosity can be extended to dynamic scenes, in which object movements cause changes in the radiance distribution.  Instead of retraining the entire network from scratch at every new scene configuration, we use transfer learning to fine-tune weights of one scene to those of another.  We demonstrate this in Figs.~\ref{fig:retraining_charts} and~\ref{fig:Animation}, where a \emph{Copper Man} is reposed from a \emph{Rest} into a \emph{Speak} pose.  While directly reusing the pretrained \emph{Rest} pose network on the modified pose performs poorly (Fig.~\ref{fig:Animation}) as expected, the network is able to quickly adapt after only a few steps of fine-tuning (Fig.~\ref{fig:retraining_charts}).


\subsection{Training with Noisy Unbiased Data}

As an alternative to optimization by ``self-training'' (using the model to evaluate the RHS), we compare results to training on noisy unbiased estimates of the RHS.  That is, we compare a loss calculated with respect to path-traced samples of the RHS (``noisy ground truth'') to our loss, which queries our model.  In Fig. \ref{fig:noisy_estimate}, we show an LHS rendering of our method self-trained with $M=32$ model queries per sample, next to LHS renderings of our model trained with path-traced samples of varying noise levels (including $M=32$ for direct comparison).  As we see from Fig. \ref{fig:noisy_estimate} and the training curves in Fig. \ref{fig:noisy_train_curves}, self-training outperforms noisy estimate training. The path-traced estimates suffer from a fixed amount of noise that destabilizes training, whereas radiance estimates queried from self-training improve across iterations, leading to better convergence.



\subsection{Study of Multi-resolution Feature Grids}

\begin{figure*}[t]
  \centering
  \subcaptionbox{Increasing multi-resolution levels \label{fig:multi_res}}{%
\includegraphics[width = 0.28\textwidth]{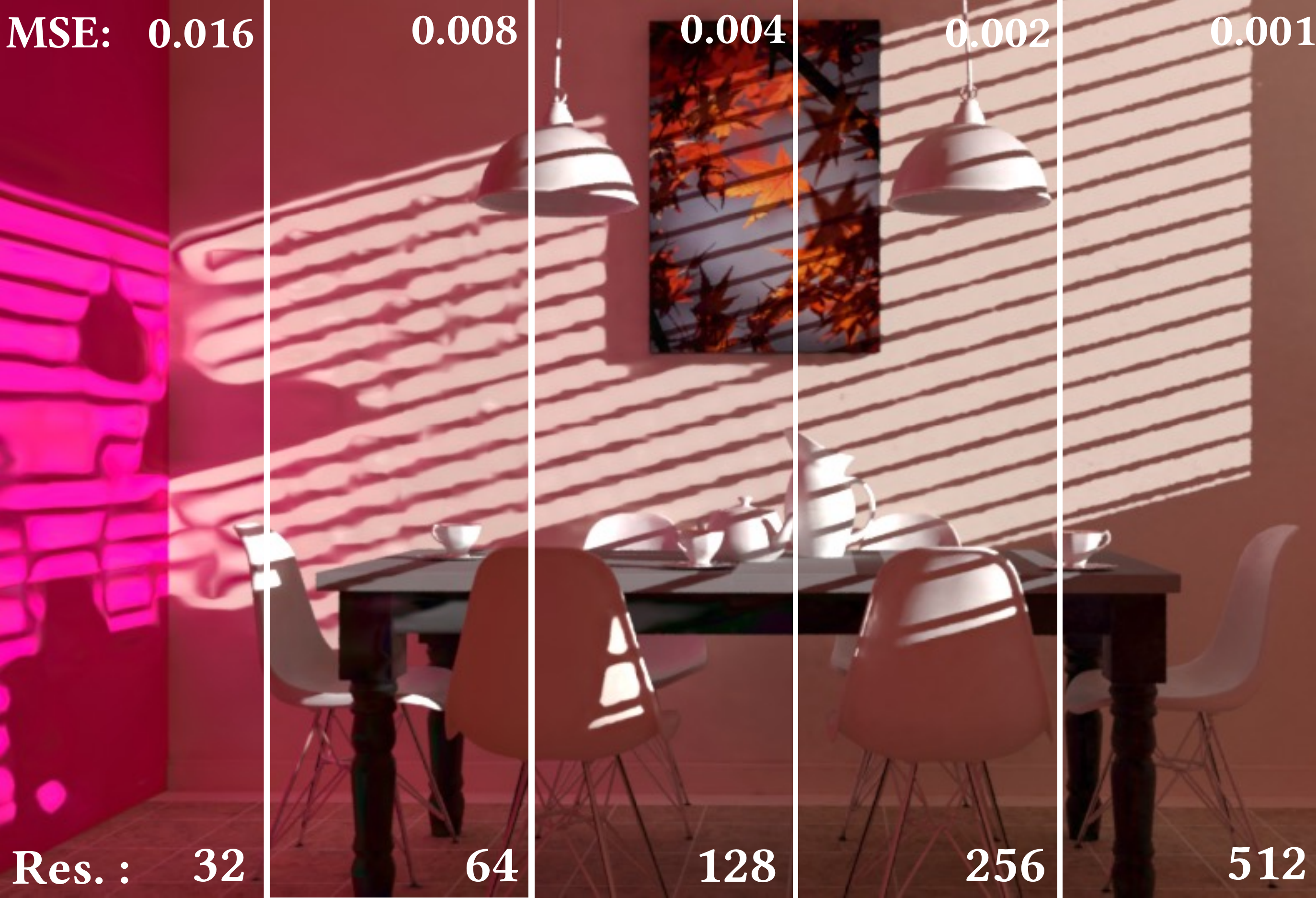}
  }
  \subcaptionbox{Single- vs. multi-resolution \label{fig:single_res}}{%
    \includegraphics[width = 0.28\textwidth]{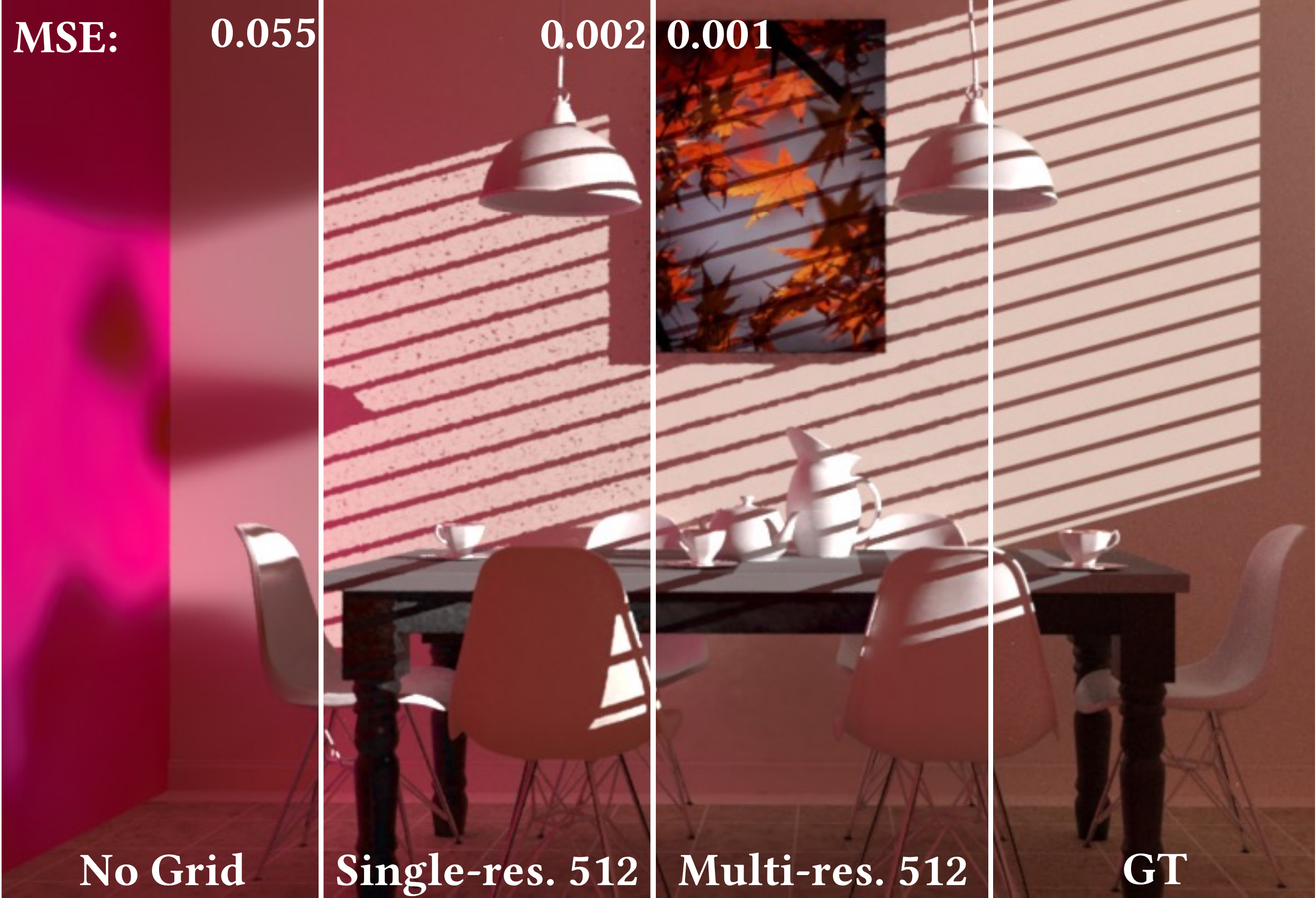}
  }
  \subcaptionbox{Training of (a)
  \label{fig:multi_res_curve}}{%
    \includegraphics[width = 0.20\textwidth]{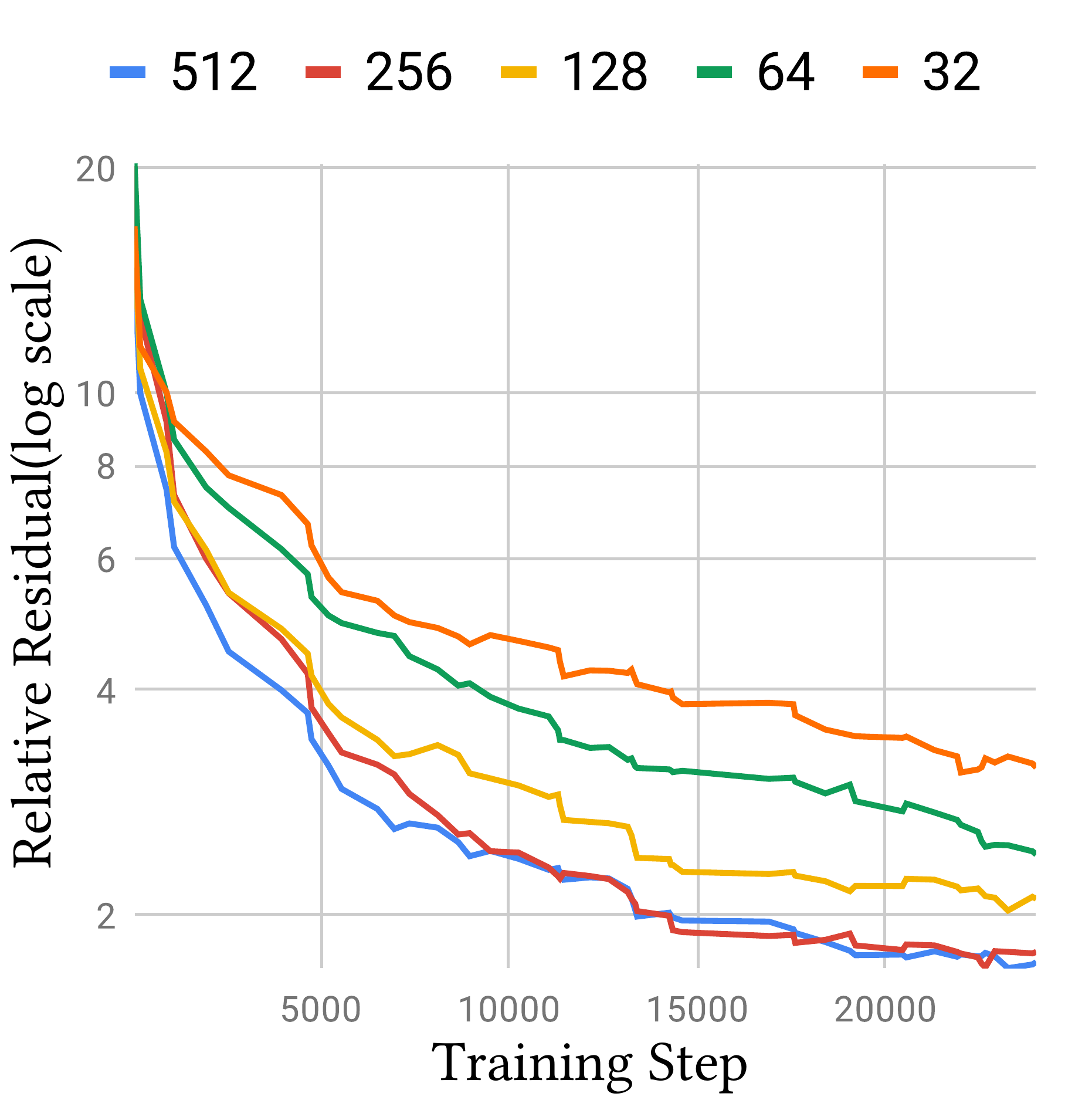}
  }
  \subcaptionbox{Training of (b) \label{fig:single_res_curve}
  }{%
    \includegraphics[width = 0.20\textwidth]{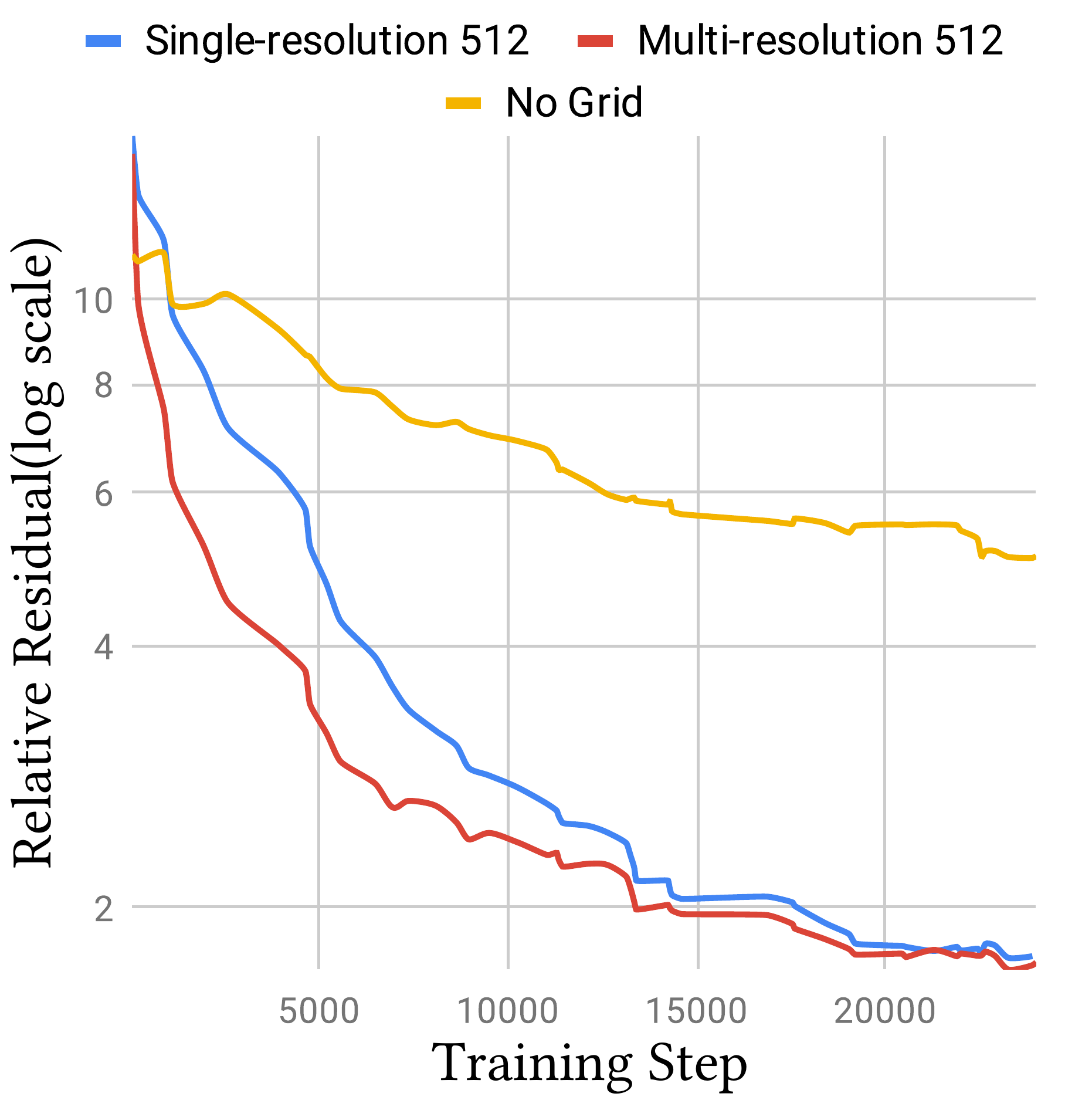}
  }
  \caption{Impact of grid resolution and our multi-resolution approach on LHS renderings. (a) Our multi-resolution approach leads to high quality results. (b) Omitting our feature grid produces blurry artifacts. Using a single-resolution grid results in dotty renderings and higher MSE. Our approach is most faithful to the ground truth. (c,d) Faster and better convergence occurs when using multi-resolution with higher levels. }
  \label{fig:Res_study}
\end{figure*}

We study the effect of multi-resolution feature grids on convergence and rendering quality in Fig.~\ref{fig:Res_study}. We use the \emph{Dining Room} scene with various levels of single- and multi-resolution grids. Fig. \ref{fig:Res_study} shows that increasing the multi-resolution level reduces error, and that the absence of multi-resolution grids causes dotty artifacts. The training curves in Fig. \ref{fig:Res_study} further quantify the superior convergence and accuracy using multi-resolution. We use an efficient sparse grid implementation with quadratic space complexity by only storing those grid vertices that contribute to the trilinear interpolation at surface samples. We report memory requirements in Table \ref{tab:res_storage}.

\begin{table}[t]
    \centering
    \caption{Space and training time costs for the resolution experiments in Fig. \ref{fig:Res_study}. The voxel count and density columns indicate the number and percentage of allocated grid vertices, and the feature size column shows the storage needed for each resolution level. Training time for a given resolution in each row involves all levels up to that row. That is, for 512 multi-resolution training, we train using all the levels from 4 to 512, and we provide the total feature vector storage in the last row. The training runs use 12,576M samples. }
    \begin{tabular}{|c|c|c|c|c|c}
        \hline
    Res. & Voxel Count & Density \% & Features Size & Train Time \\
    \hline
    32 & 3045 & 9.3 & 190 KB &  2h 52m \\
    64 & 11 K & 4.4 & 719 KB & 3h 17m \\
    128 & 51 K & 2.4 & 3.11 MB & 2h 59m \\
    256 & 214 K & 1.3 & 13.1 MB & 3h 8m \\
    512 & 863 K & 0.64 & 52.7 MB & 3h 11m \\
    \hline
    Total &  &  & \textbf{69.9 MB} &  \\
    \hline
    \end{tabular}

    \label{tab:res_storage}
\end{table}

\subsection{Limitations and Future Work}

A main limitation of our current implementation is that the training times are typically not competitive with state of the art Monte Carlo integration techniques. However, we can amortize initial training time when rendering multiple images in animations. A promising direction of future work is improved adaptation for dynamic scenes.  Our \emph{Copper Man} demonstration (Fig. \ref{fig:retraining_charts}) constitutes a form of transfer learning, whereby we transfer learned weights from a pretrained network to train on a novel domain, in this case a different scene.  Other methods of handling dynamic scenes include meta-learning, in which we would optimize a model initialization that can quickly adapt to any general scene; similar meta-learning formulations have already been proposed \cite{tancik2020learned}.  Other directions include few-shot learning, which would entail fine-tuning to a novel scene configuration given a very limited sample budget; in the context of our problem, we may design specialized sampling methods targeted at regions with changed geometry. For real-time rendering applications, we envision systems where the neural network training process to solve the rendering equation runs continuously and concurrently with a real-time image synthesis process. 

Additionally, Neural Radiosity could be coupled with an adaptive sampling approach for faster convergence while maintaining the unbiased nature of the gradient Monte Carlo estimates.
More future work may consist of network designs to accommodate scenes containing participating media, and improvements to positional encoding techniques such as multi-level feature octrees.

\section{Conclusions}

We propose Neural Radiosity, a novel method leveraging deep neural networks to solve the rendering equation. Inspired by traditional radiosity techniques, our approach formulates the full radiance distribution as a learnable network architecture, and is optimized by minimizing the norm of the rendering equation residual. By taking advantage of the representational capacity of neural models, we are able to solve for global illumination on scenes with complex lighting and non-diffuse surfaces. Our implementation incorporates key architectural design decisions, such as multi-resolution feature grids, into an accurate and compact scene representation.  Finally, we demonstrate that our system is capable of efficient multi-view rendering, and can be extended to handle complex dynamic scenes.

\begin{acks}
We would like to thank Wenzel Jakob for insightful initial discussions and providing the Mitsuba 2 system. We also thank Sebastien Speierer for helping us troubleshoot and modify Mitsuba 2. Baptiste Nicolet's script to convert blender files to Mitsuba is highly appreciated. Finally, we want to thank the scene authors Benedikt Bitterli~\shortcite{rendering_resources} and Blendswap user ``barcin''~\shortcite{chair}.
\end{acks}

\input{all_scenes}

\bibliographystyle{ACM-Reference-Format}
\bibliography{bibliography}



%% file: ablation_fig.tex
\begin{figure*}[hbt]
  \centering
\subcaptionbox{No grid}{
\includegraphics[width = 0.235\textwidth]{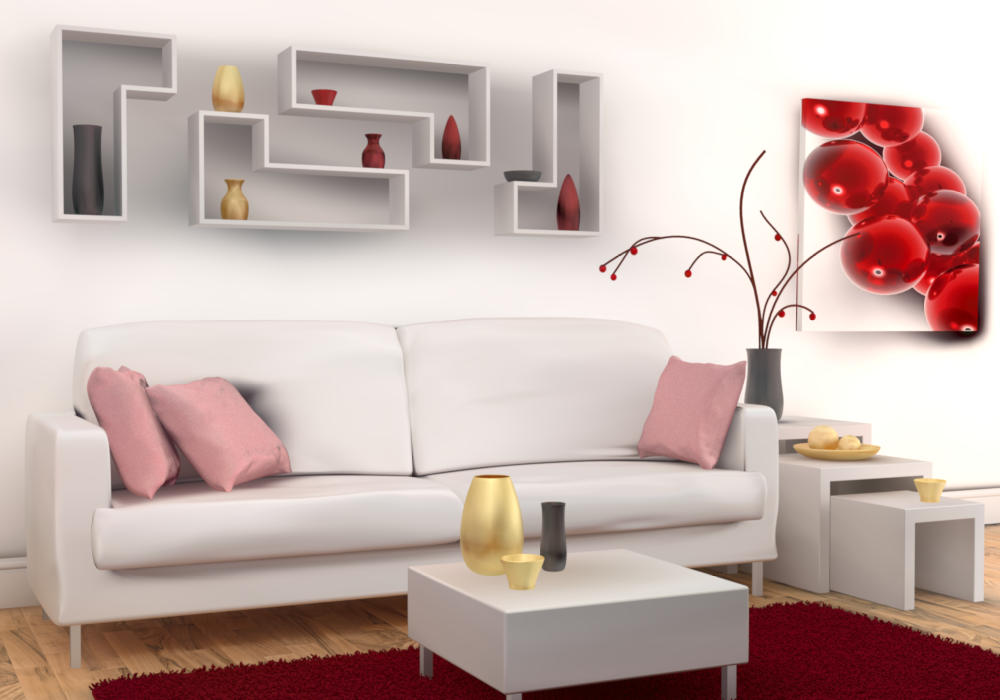}
}
\subcaptionbox{No local properties}{
\includegraphics[width = 0.235\textwidth]{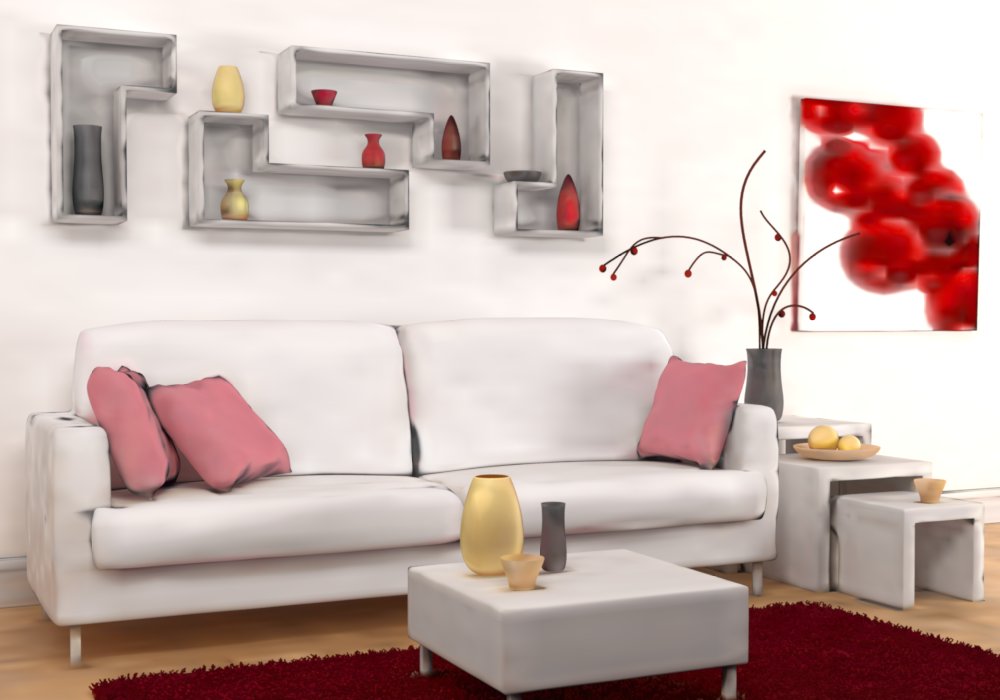}
}
\subcaptionbox{Single-resolution-64}{
\includegraphics[width = 0.235\textwidth]{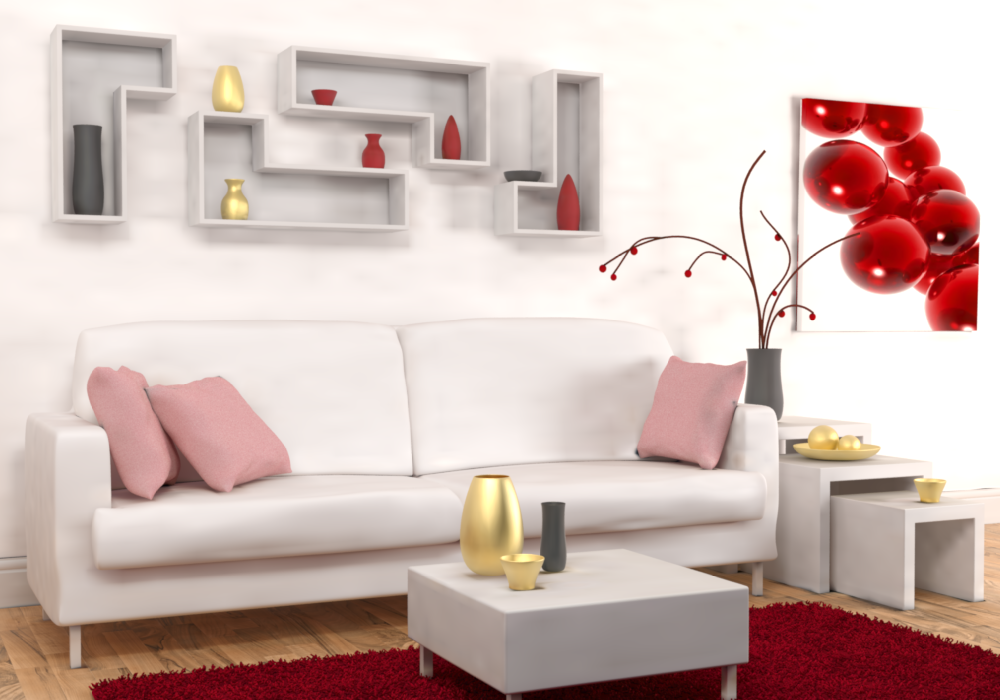}
}
\subcaptionbox{Shallow network, 4 linear layers}{
\includegraphics[width = 0.235\textwidth]{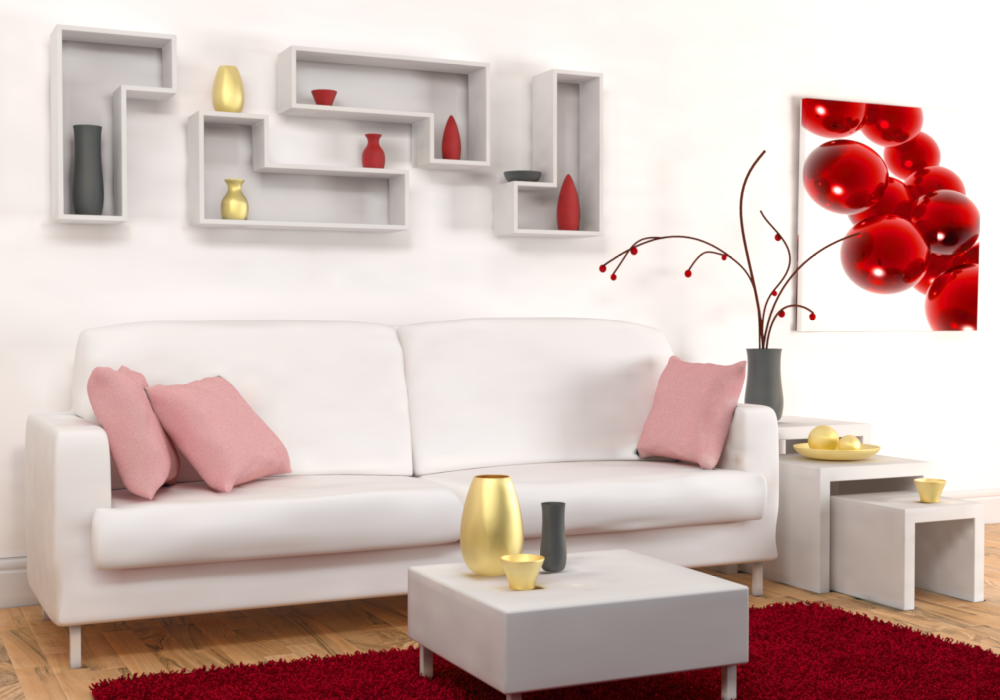}
}

\subcaptionbox{Narrow network, 128 neurons}{
\includegraphics[width = 0.235\textwidth]{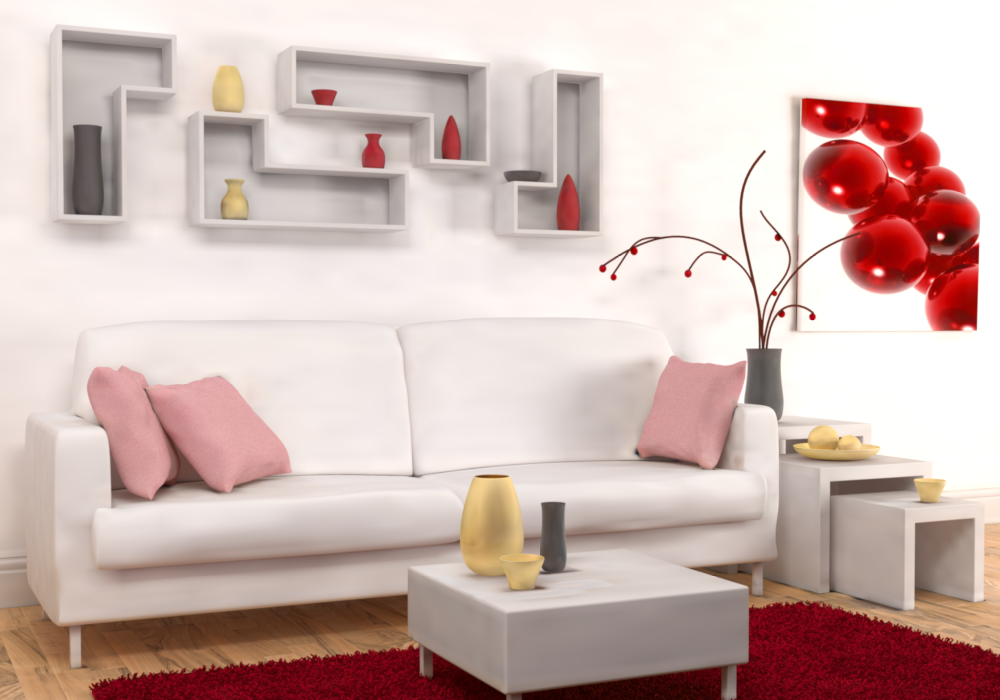}
}
\subcaptionbox{Complete method}{
\includegraphics[width = 0.235\textwidth]{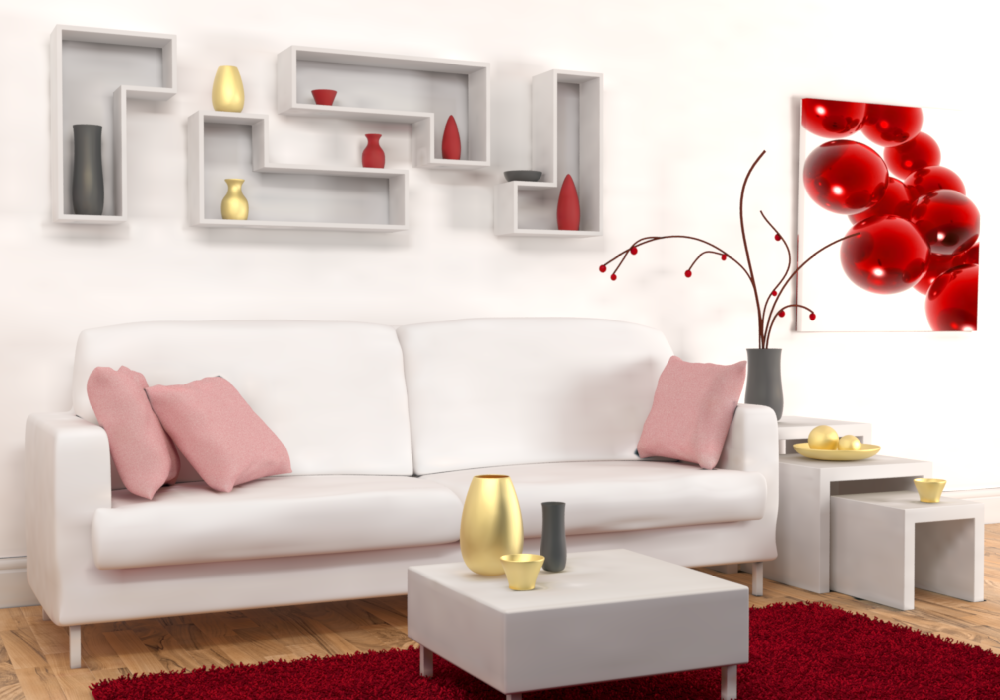}
}
\subcaptionbox{Complete method - 2.4x samples}{
\includegraphics[width = 0.235\textwidth]{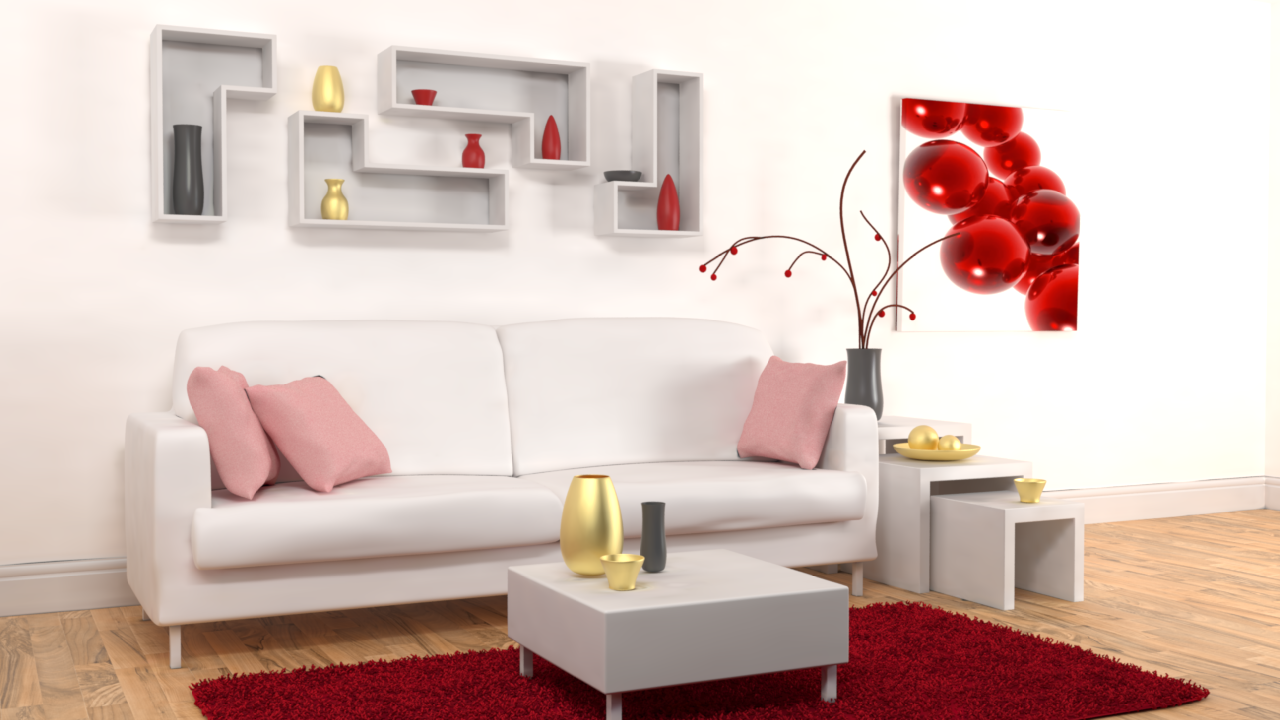}
}
\subcaptionbox{GT}{
\includegraphics[width = 0.235\textwidth]{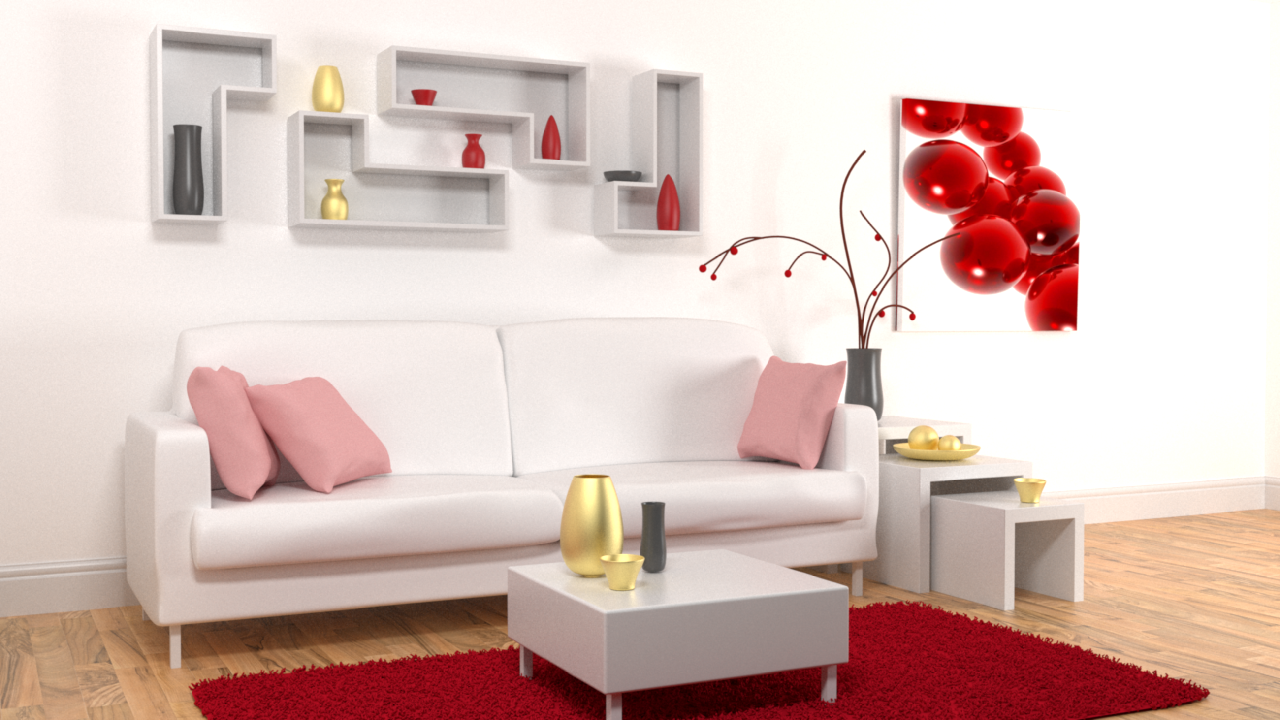}
}

\subcaptionbox{Error to path traced GT}{
\includegraphics[width = 0.44\textwidth]{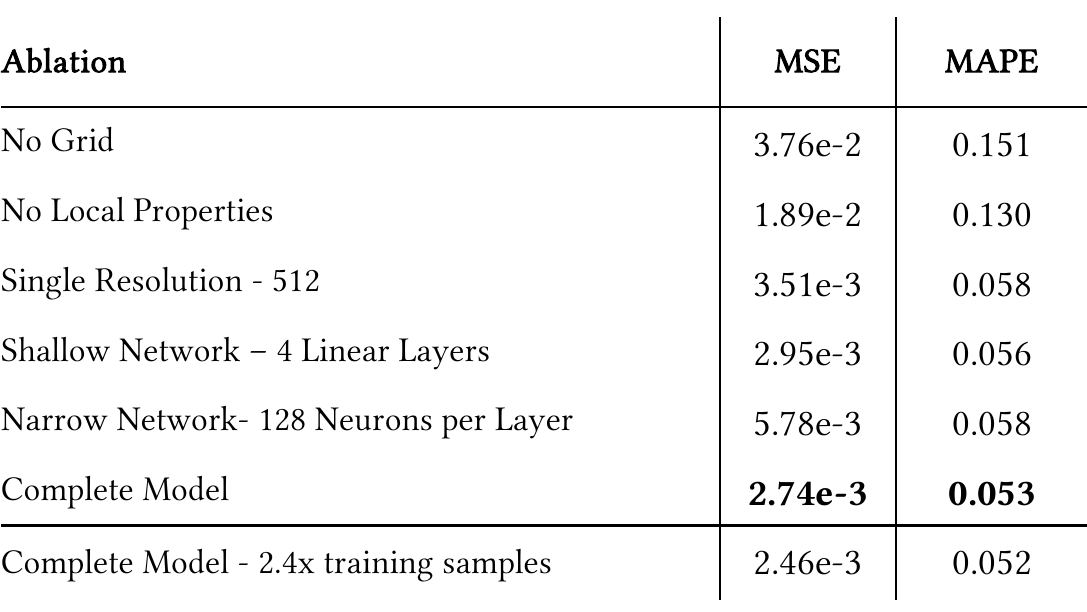}
}
\subcaptionbox{Training curves}{
\includegraphics[width = 0.44\textwidth]{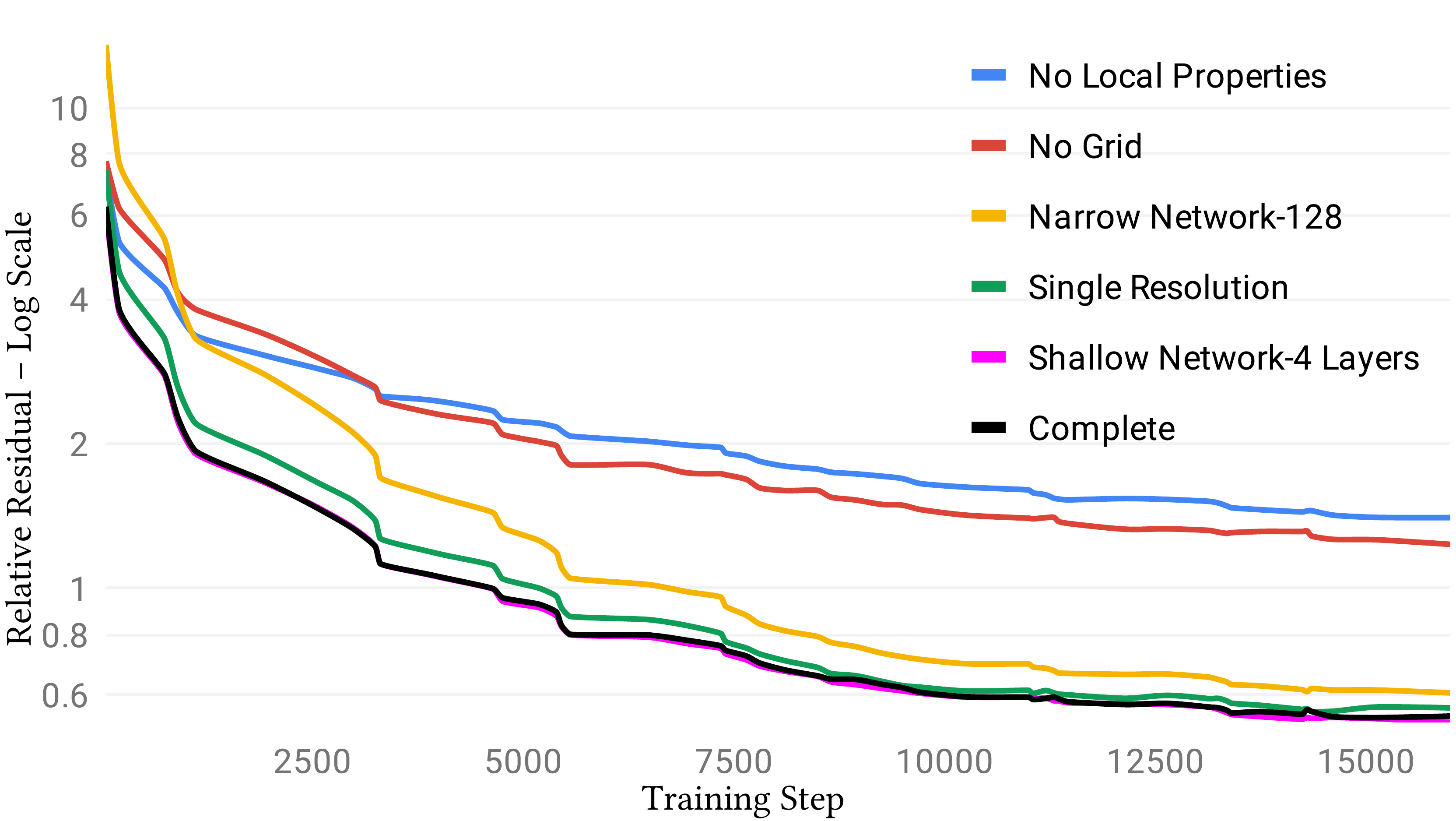}
}

\caption{Ablation Study. All renderings show the LHS. The complete model uses multi-resolution of 64, and an MLP with 6 layers and 512 width. We use a batch size of $N=16,384$, $M=32$, and 16K steps of training for a total of 8,384M samples for all the experiments (except the one mentioned that has 2.4x more samples). The complete models are the ones with the best convergence and least error.}
\label{fig:Ablation_Living}
\end{figure*}

%% file: noisyest_fig.tex
\newcommand{\expnumber}[2]{{#1}\mathrm{e}{#2}}

\begin{figure}[hbt]
  \centering
\subcaptionbox{Our LHS, M=32, MSE=\textbf{2.46e-3}}{
\includegraphics[width = 0.475\columnwidth]{NNMitsuba Paper/RenderingsV2/LivingRoom/livingroom-4800mlp-spp128-m1.png}
}
\subcaptionbox{Noisy Est., M=32, MSE=3.16e-3}{
\includegraphics[width = 0.475\columnwidth]{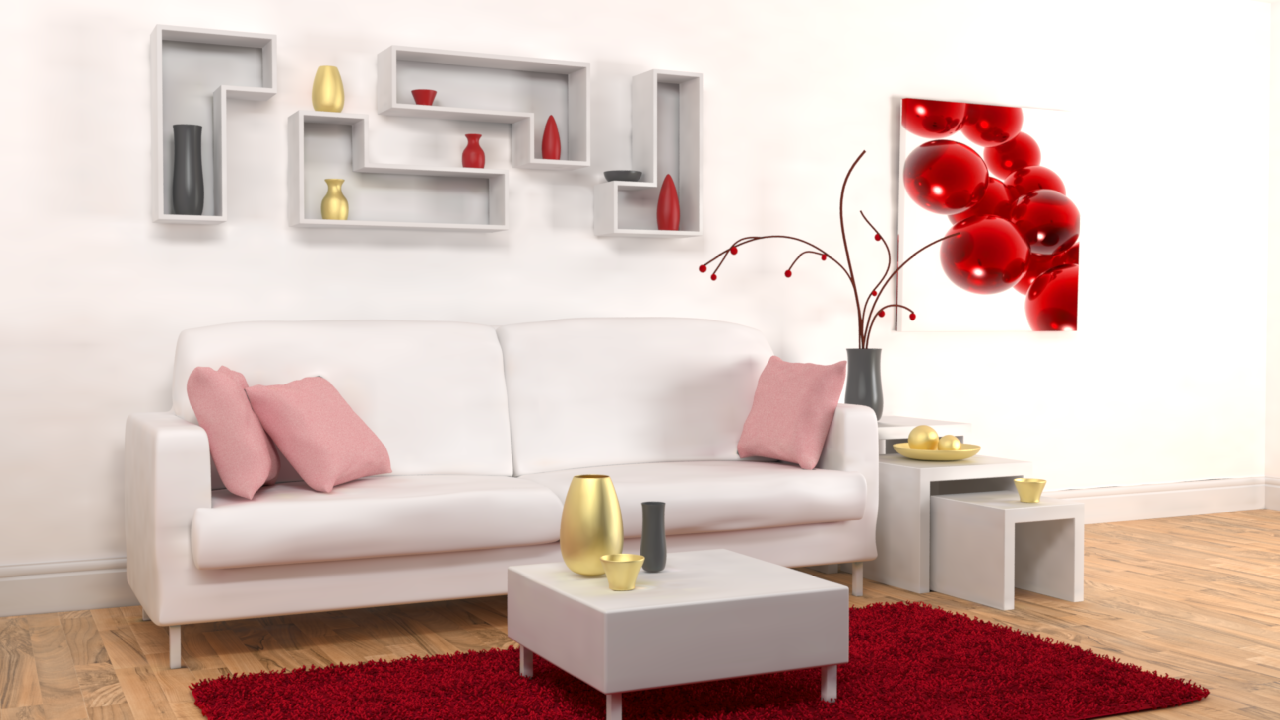}
}

  \caption{Comparing our approach (a) to fitting the model to a noisy estimate of scattered radiance (b), where $M$ is the number of paths cast per position sample to compute the scattering integral. The noisy estimate uses an equal number of samples $M=32$ as ours.}
  \label{fig:noisy_estimate}
\end{figure}

%% file: retraining_fig.tex
\begin{figure}[tb]
    \centering
    \subcaptionbox{ \label{fig:retraining_charts} Fine-tuning to scene changes
    }{\includegraphics[width = 0.485\columnwidth]{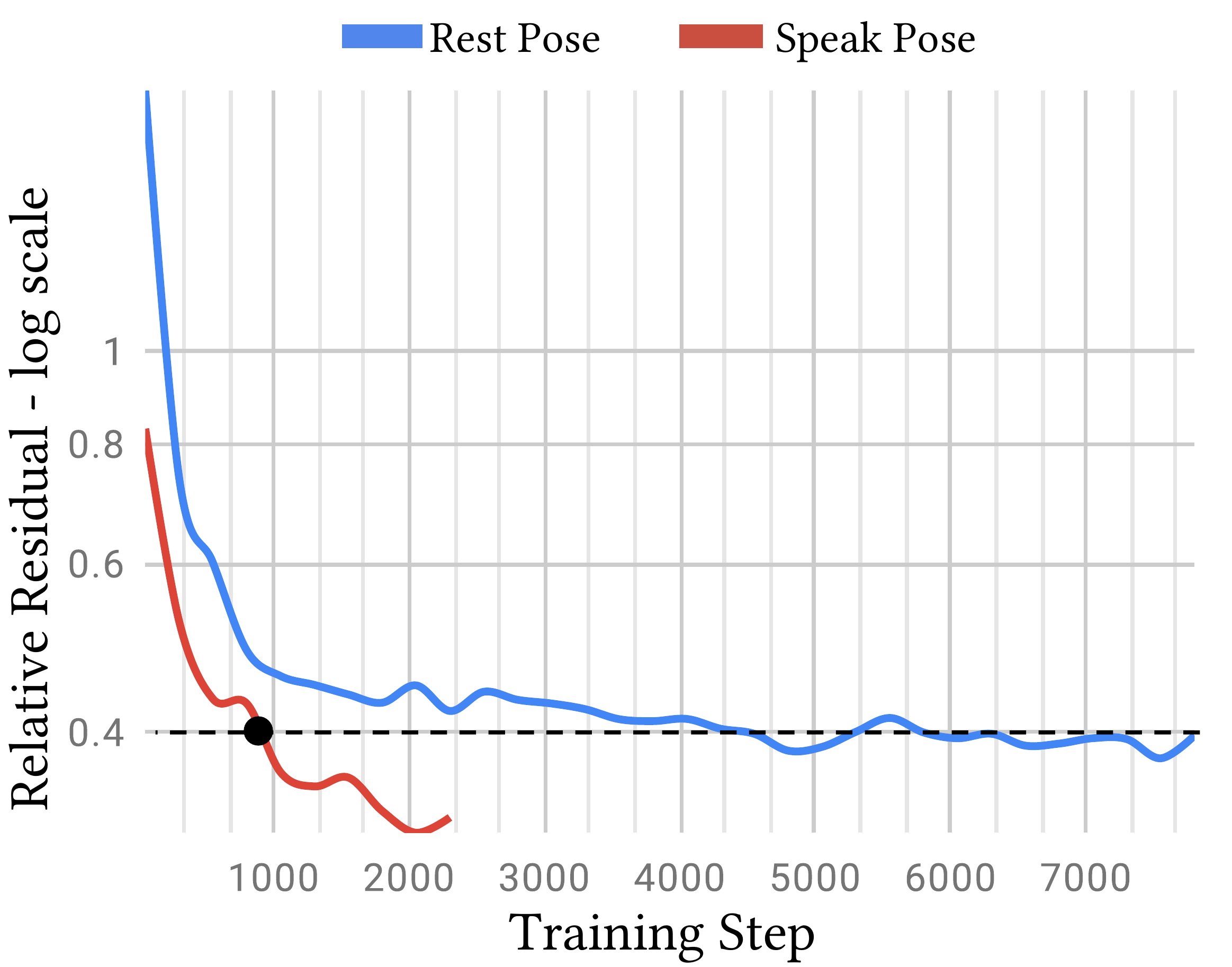}
    }
\subcaptionbox{\label{fig:noisy_train_curves} Training with noisy data}
{\includegraphics[width = 0.485 \columnwidth]{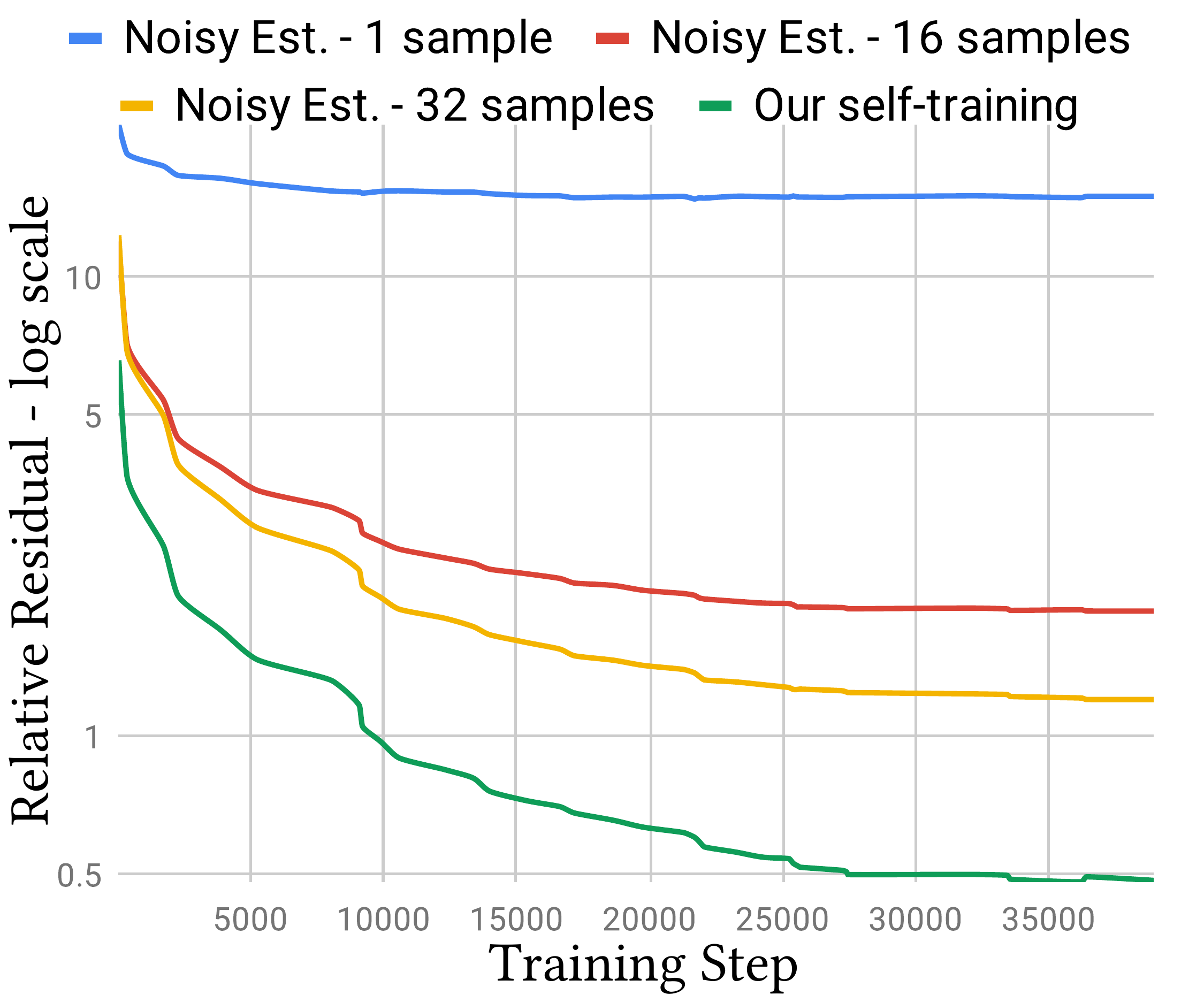}  }  

\caption{(a) Training and fine-tuning the \emph{Copper Man} scene. Our model trained on the \emph{Rest} pose can be rapidly fine-tuned to the scene changes in \emph{Speak}. 
(b) Comparing our self-training to fitting to noisy estimates. Our method ($M=32$) converges better than training on an equal number of path-traced estimates. Increasing $M$ in the noisy estimate brings the training curve closer to ours (we cannot increase $M$ above $32$ due to memory limitations). }
\end{figure}

\newcommand\adamakwidth{0.13}
\newcommand\adamakDoubleWidth{0.26}
\newcommand\adamakTripleWidth{0.40}
\newcommand\adamakBigWidth{0.18}
\newcommand\adamakBigHieght{0.22}

\newcommand\mapeoffsethere{14pt}
\newcommand\mapedeoffset{-16pt}

\begin{figure*}[htb]
    \centering
    
     \makebox[0pt]{\rotatebox{90}{\hspace{35pt} Rest}}
     \includegraphics[width=\adamakwidth\textwidth]{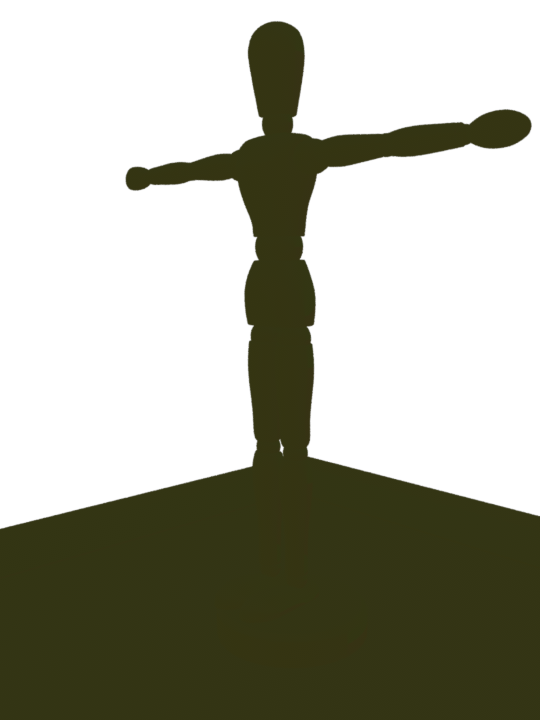}
     \includegraphics[width=\adamakwidth\textwidth]{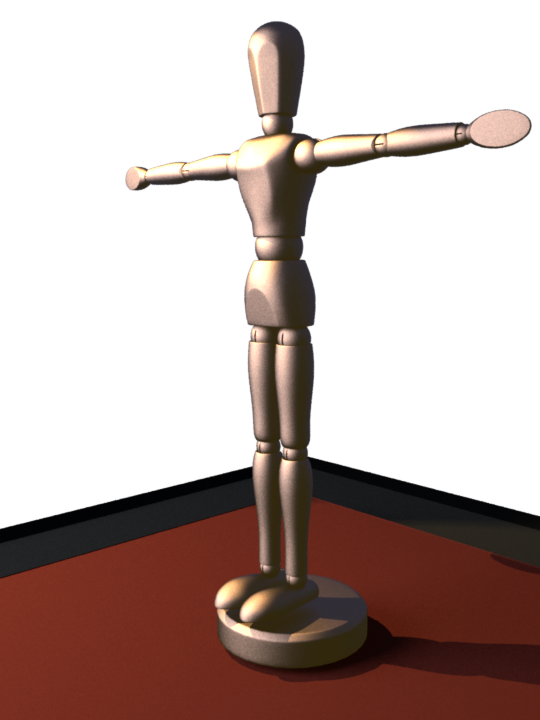}
     \includegraphics[width=\adamakwidth\textwidth]{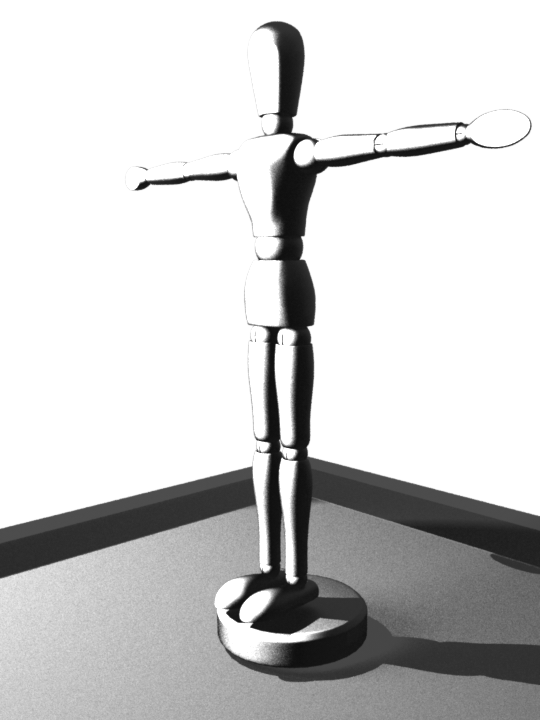}
     \includegraphics[width=\adamakwidth\textwidth]{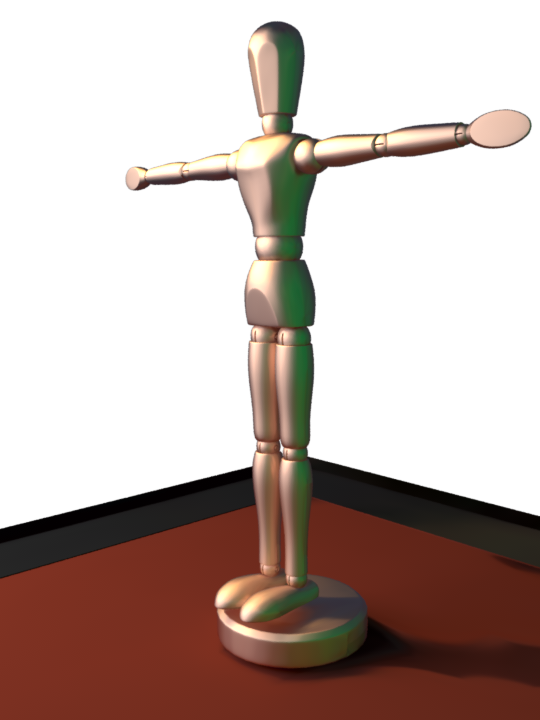}
     \includegraphics[width=\adamakwidth\textwidth]{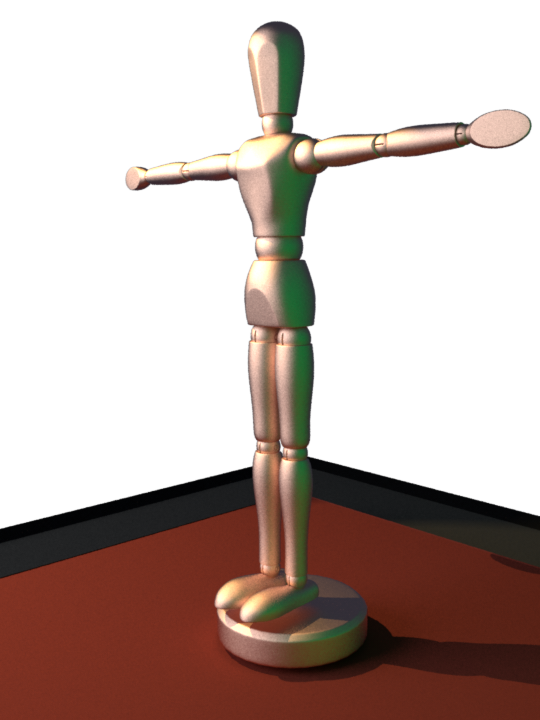}
     \includegraphics[width=\adamakwidth\textwidth]{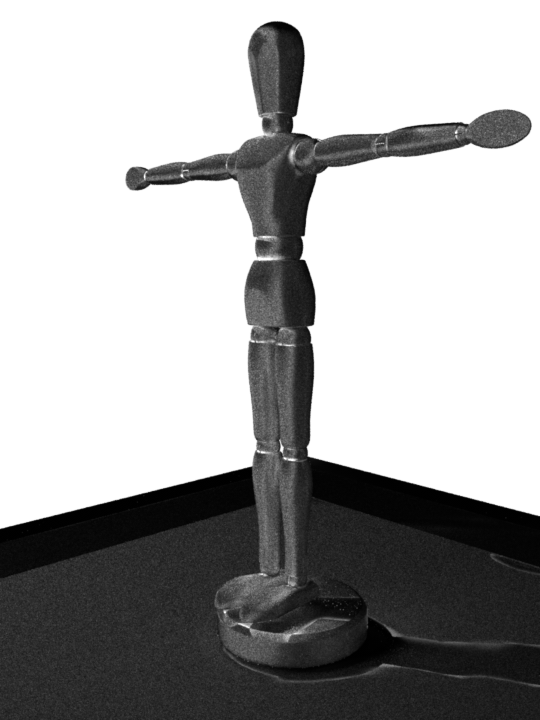}
     \includegraphics[width=\adamakwidth\textwidth]{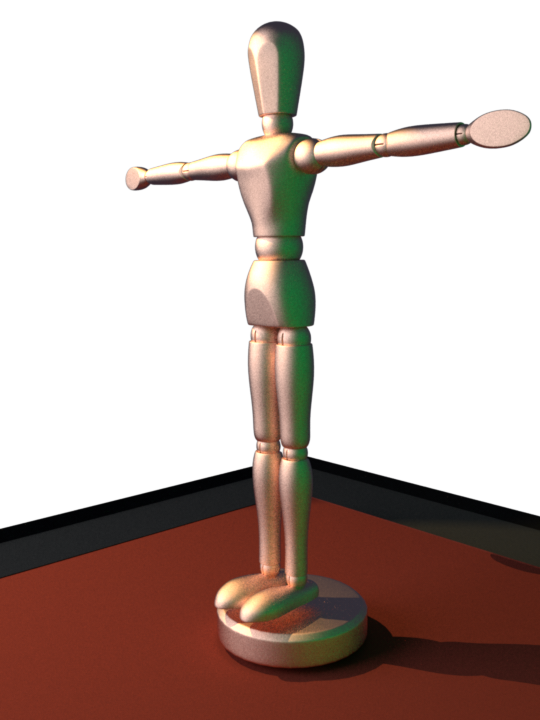}

\hspace{\mapeoffsethere}
\makebox[0pt]{{MAPE:}}
\hspace{-\mapeoffsethere}
\makebox[\adamakwidth\textwidth]{{0.257}}
\makebox[\adamakwidth\textwidth]{{0.044}}
\makebox[\adamakwidth\textwidth]{{}}
\makebox[\adamakwidth\textwidth]{\textbf{0.025}}
\makebox[\adamakwidth\textwidth]{\textbf{0.014}}
\makebox[\adamakwidth\textwidth]{{}}
\makebox[\adamakwidth\textwidth]{{}}

    \vspace{5pt}
    
     \makebox[0pt]{\rotatebox{90}{\hspace{35pt} Speak}}
     \includegraphics[width=\adamakwidth\textwidth]{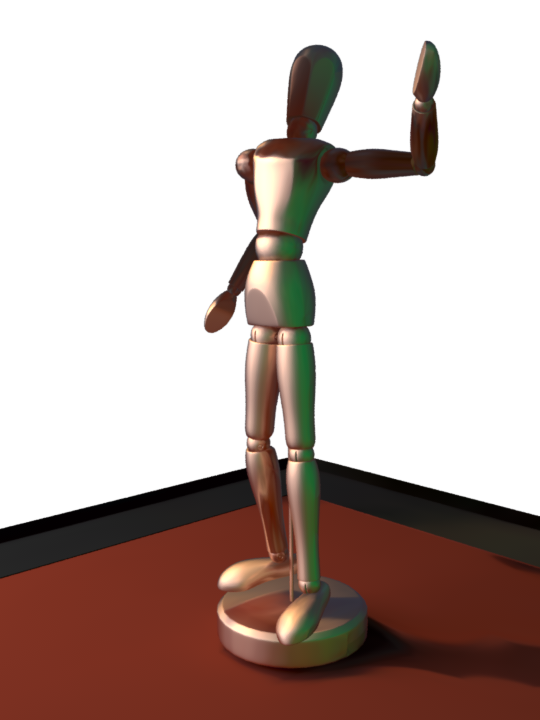}
     \includegraphics[width=\adamakwidth\textwidth]{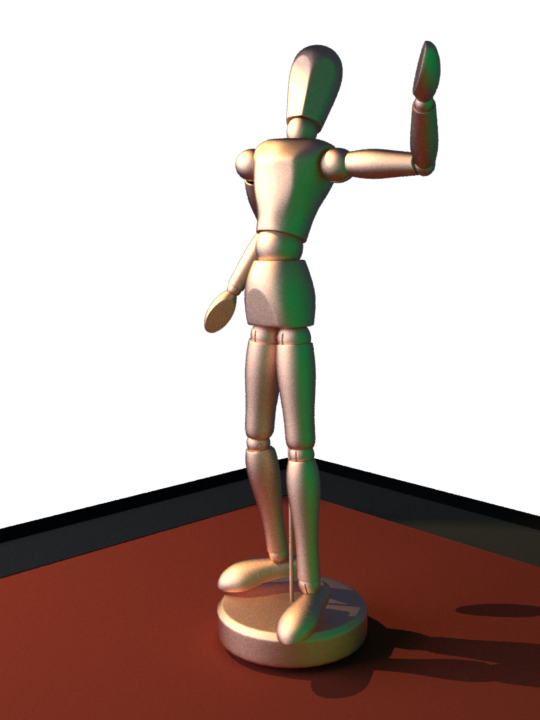}
     \includegraphics[width=\adamakwidth\textwidth]{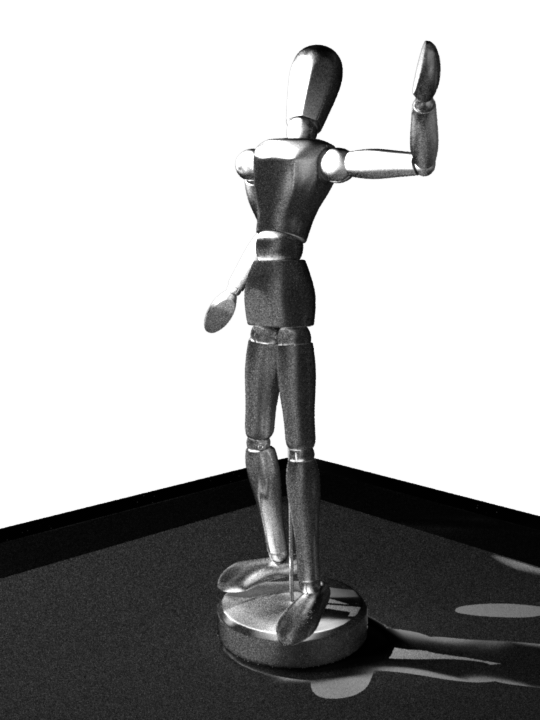}
     \includegraphics[width=\adamakwidth\textwidth]{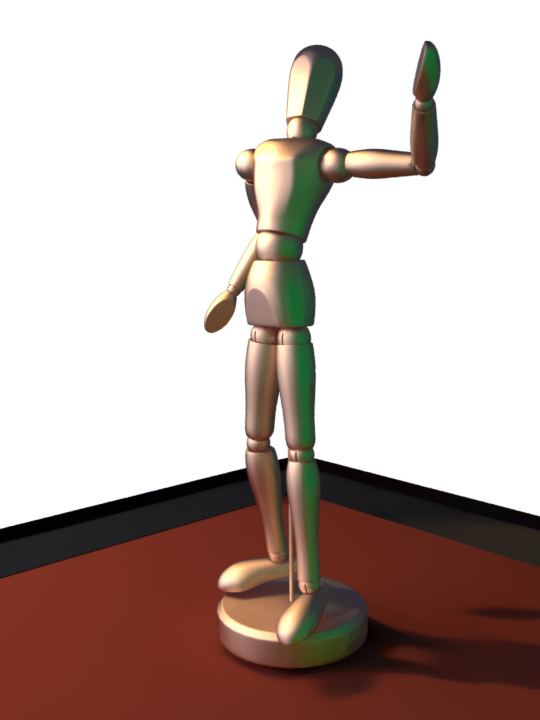}
     \includegraphics[width=\adamakwidth\textwidth]{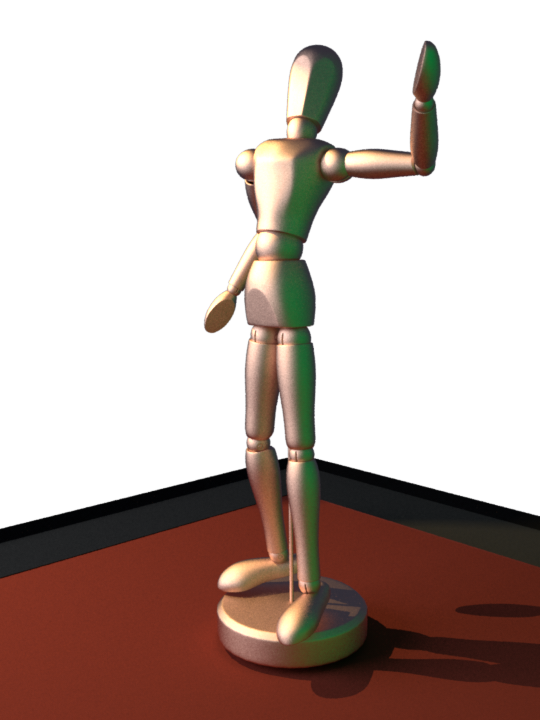}
     \includegraphics[width=\adamakwidth\textwidth]{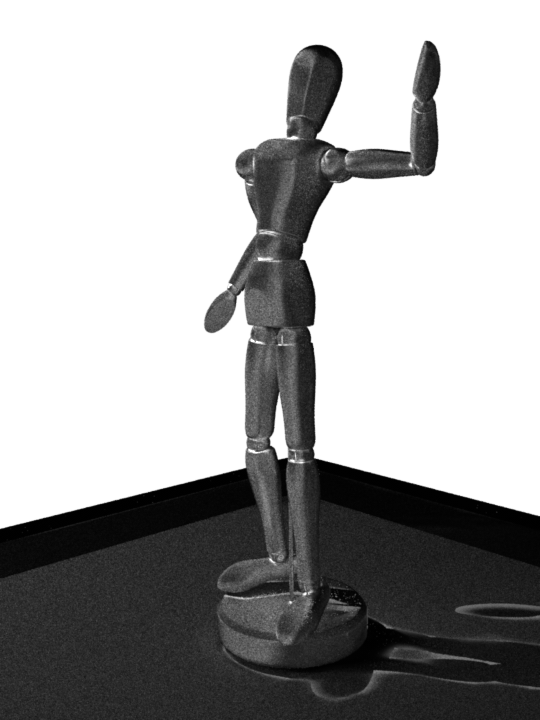}
     \includegraphics[width=\adamakwidth\textwidth]{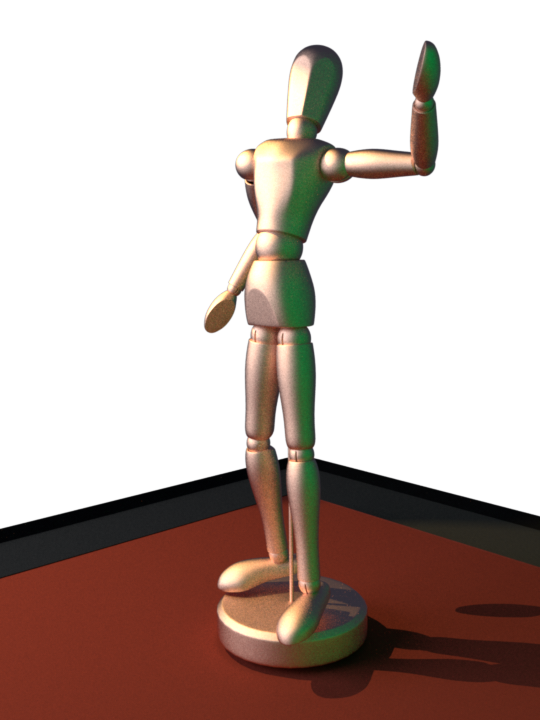}

\hspace{\mapeoffsethere}
\makebox[0pt]{{MAPE:}}
\hspace{-\mapeoffsethere}
\makebox[\adamakwidth\textwidth]{{0.060}}
\makebox[\adamakwidth\textwidth]{{0.020}}
\makebox[\adamakwidth\textwidth]{{}}
\makebox[\adamakwidth\textwidth]{\textbf{0.029}}
\makebox[\adamakwidth\textwidth]{\textbf{0.016}}
\makebox[\adamakwidth\textwidth]{{}}
\makebox[\adamakwidth\textwidth]{{}}


%


    \rule{\adamakwidth\textwidth}{1pt}
    \rule{\adamakwidth\textwidth}{1pt}
    \rule{\adamakwidth\textwidth}{1pt}
    \rule{\adamakwidth\textwidth}{1pt}
    \rule{\adamakwidth\textwidth}{1pt}
    \rule{\adamakwidth\textwidth}{1pt}
    \rule{\adamakwidth\textwidth}{1pt}

\makebox[\adamakwidth\textwidth]{{LHS (initial)}}
\makebox[\adamakwidth\textwidth]{{RHS (initial)}}
\makebox[\adamakwidth\textwidth]{{Residual (initial)}}
\makebox[\adamakwidth\textwidth]{LHS (finetuned)}
\makebox[\adamakwidth\textwidth]{RHS (finetuned)}
\makebox[\adamakwidth\textwidth]{{Residual (finetuned)}}
\makebox[\adamakwidth\textwidth]{Ground Truth}

    \caption{Dynamic scenes with fine-tuning. The Rest Pose renderings are generated by standard training of our Neural Radiosity solver. We then use the trained neural network weights as the initialization for solving the Speak pose. The network is able to very quickly adapt to the new scene configuration through fine-tuning. See also Figure~\ref{fig:retraining_charts} for a comparison of training convergence.
    }
    \label{fig:Animation}
\end{figure*}

%% file: all_scenes.tex
\newcommand\chairwidth{0.24}

\newcommand\mapeoffset{12pt}
\newcommand\mapeoffsett{26pt}

\begin{figure*}[htb]
    \centering

    \hspace{5pt}
    \makebox[\chairwidth\textwidth]{{LHS}}
    \makebox[\chairwidth\textwidth]{{RHS}}
    \makebox[\chairwidth\textwidth]{{Ground Truth}}
    \makebox[\chairwidth\textwidth]{{Relative Residual}}
    \vspace{5pt}

    
     \makebox[5pt]{\rotatebox{90}{\hspace{33pt} Chair}}
     \includegraphics[trim=0 40 0 150,clip,width=\chairwidth\textwidth]{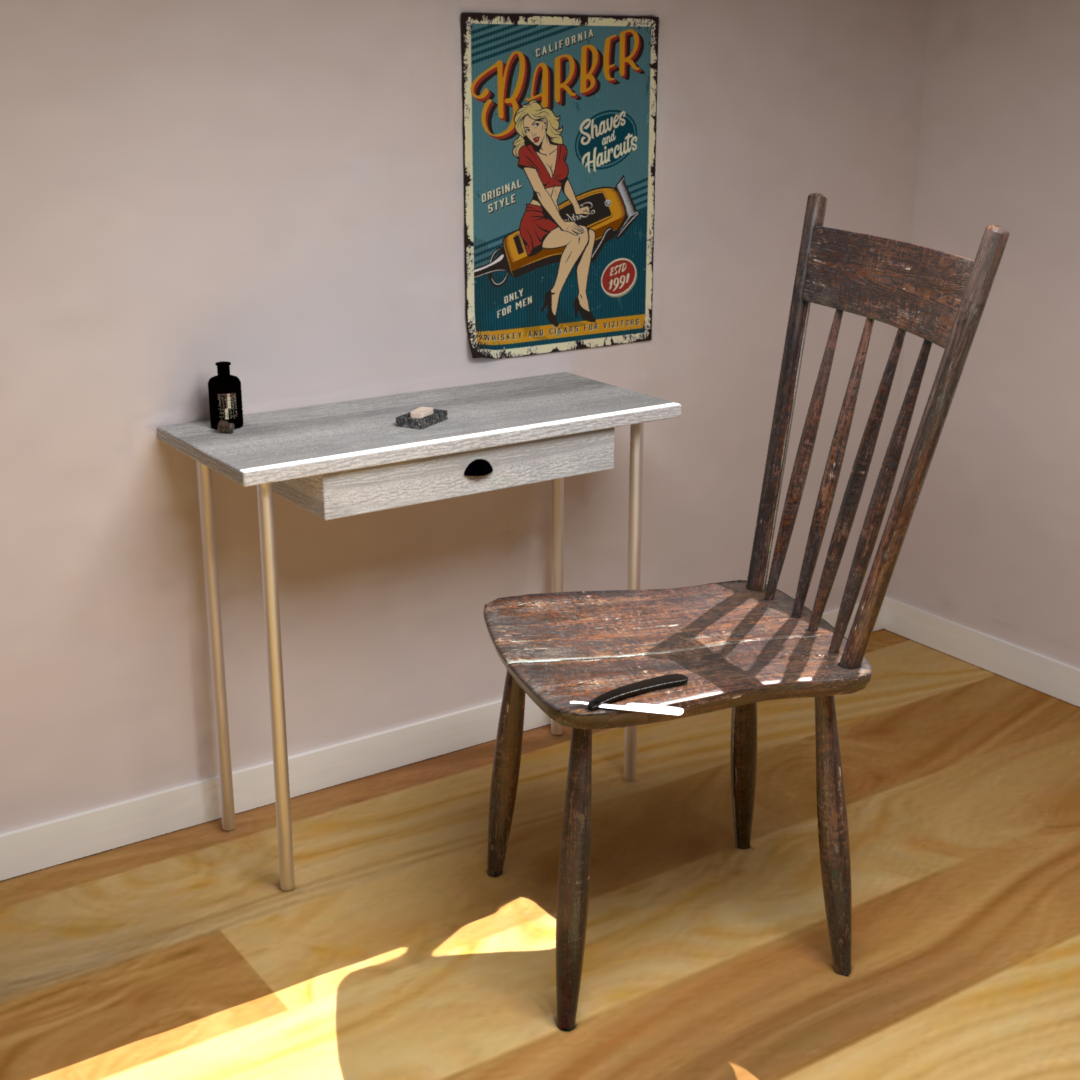}
     \includegraphics[trim=0 40 0 150,clip,width=\chairwidth\textwidth]{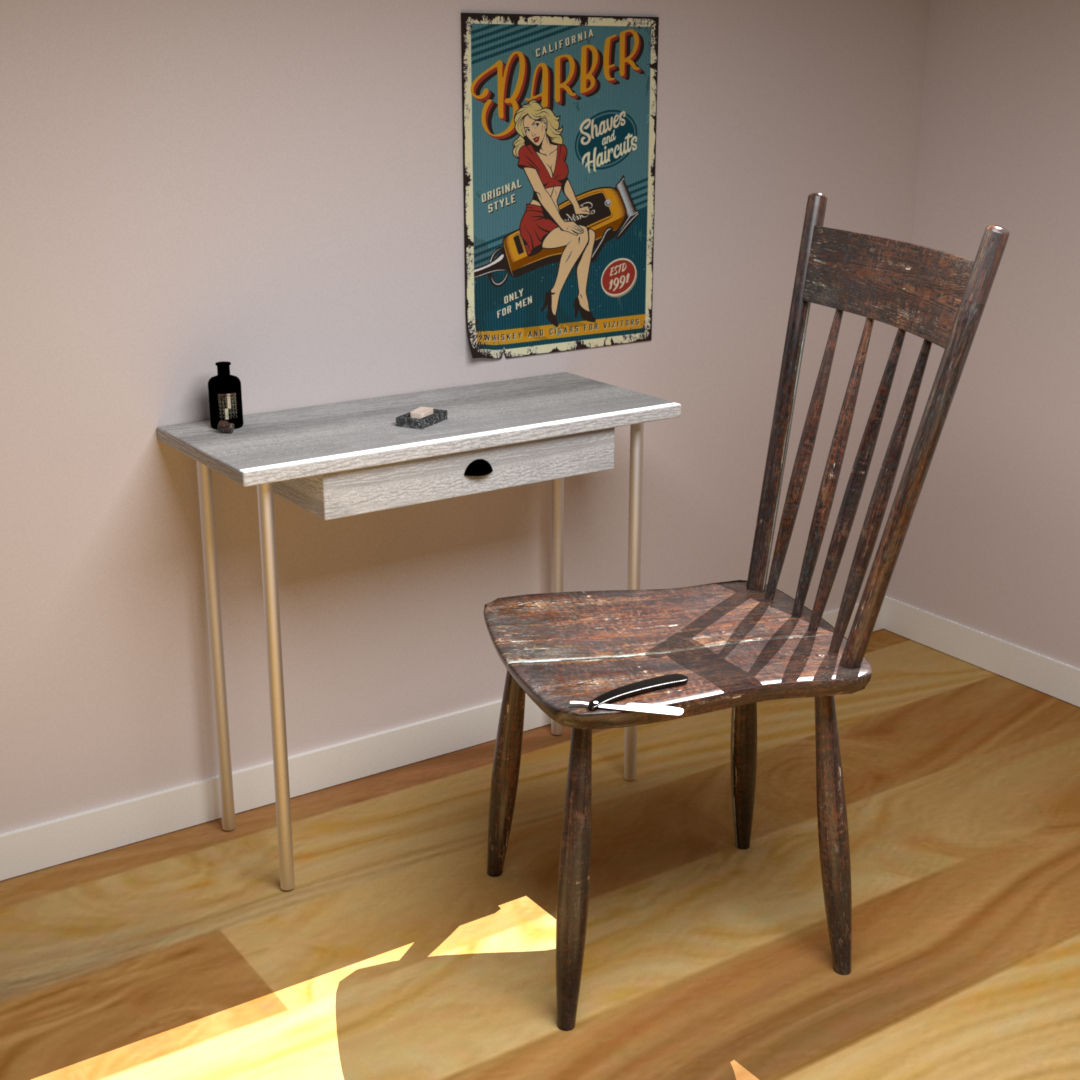}
     \includegraphics[trim=0 40 0 150,clip,width=\chairwidth\textwidth]{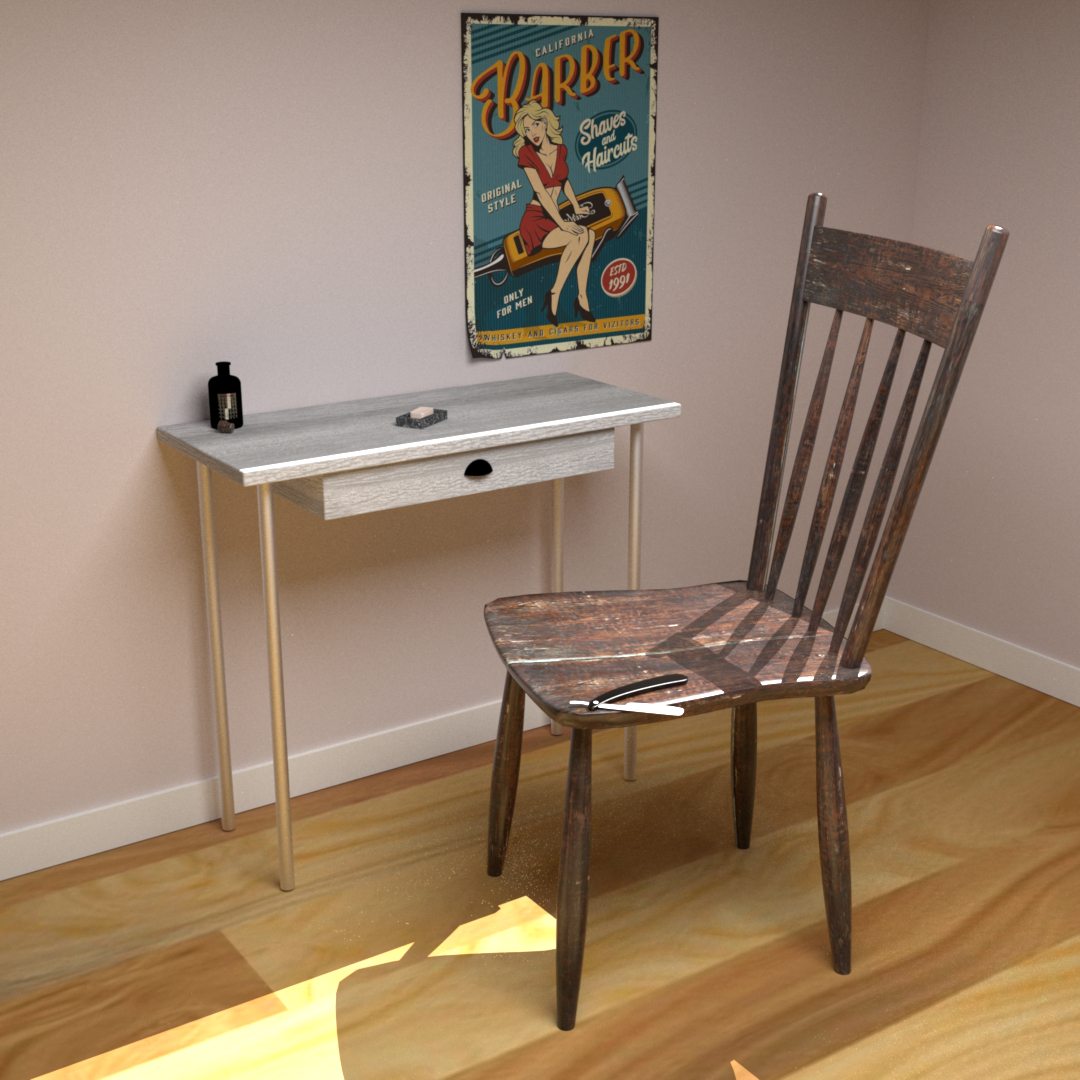}
     \includegraphics[trim=0 20 0 85,clip,width=\chairwidth\textwidth]{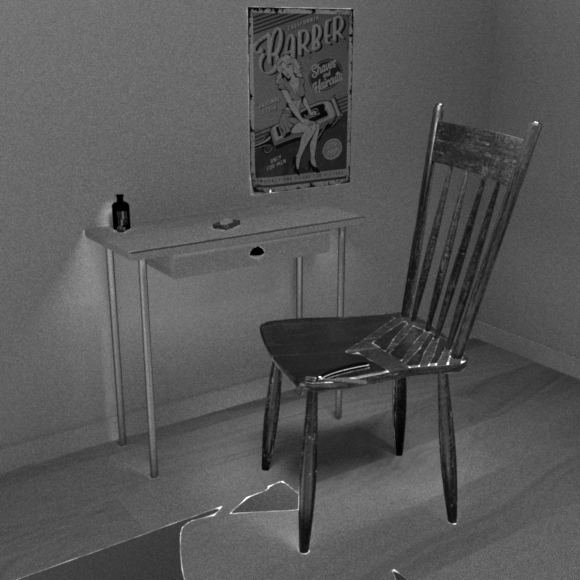}

    \vspace{-4pt}
     
    \hspace{\mapeoffsett}
    \makebox[0pt]{{spp-MAPE:}}
    \hspace{-\mapeoffsett}
    \makebox[\chairwidth\textwidth]{16-0.054}
    \makebox[\chairwidth\textwidth]{64$\times$32-0.030}
    \makebox[\chairwidth\textwidth]{1024}
    \makebox[\chairwidth\textwidth]{128$\times$8}

     \makebox[5pt]{\rotatebox{90}{\hspace{23pt} Mirror Chair}}
     \includegraphics[trim=0 20 0 85,clip,width=\chairwidth\textwidth]{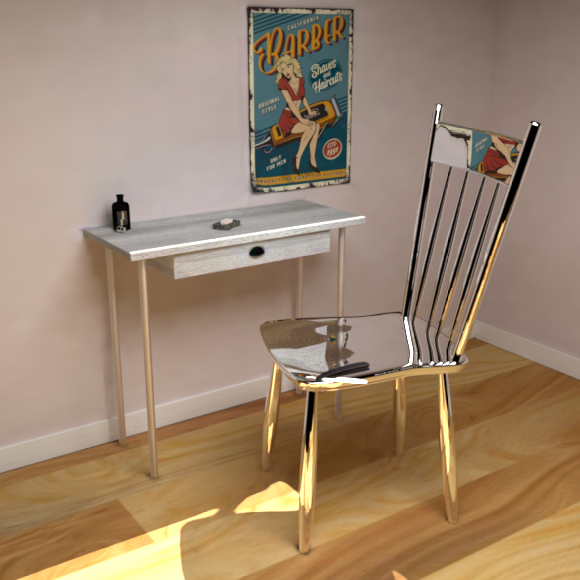}
     \includegraphics[trim=0 20 0 85,clip,width=\chairwidth\textwidth]{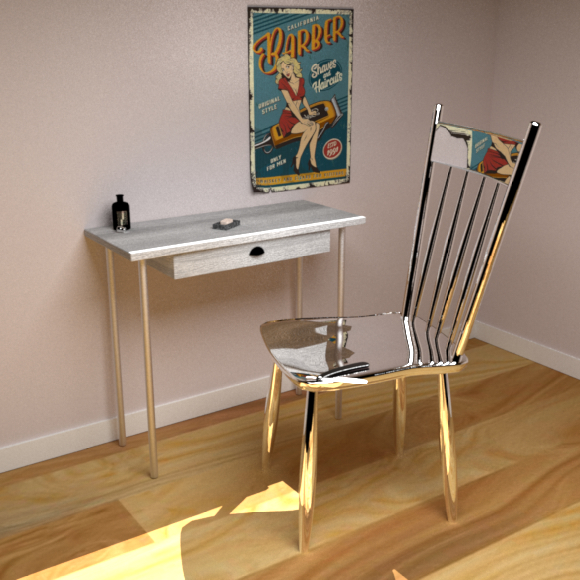}
     \includegraphics[trim=0 20 0 85,clip,width=\chairwidth\textwidth]{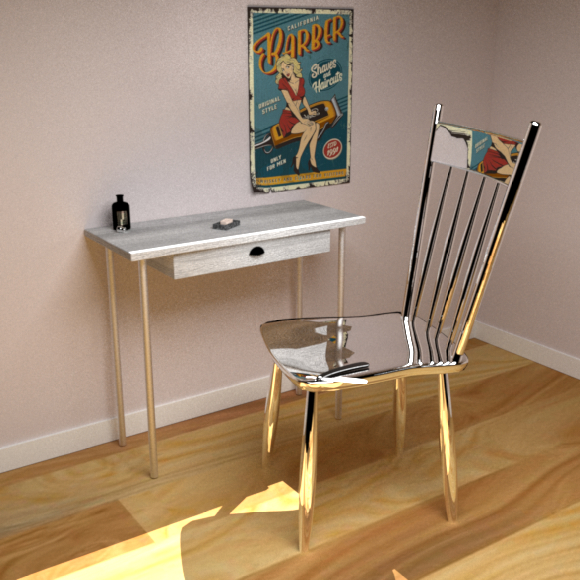}
     \includegraphics[trim=0 20 0 85,clip,width=\chairwidth\textwidth]{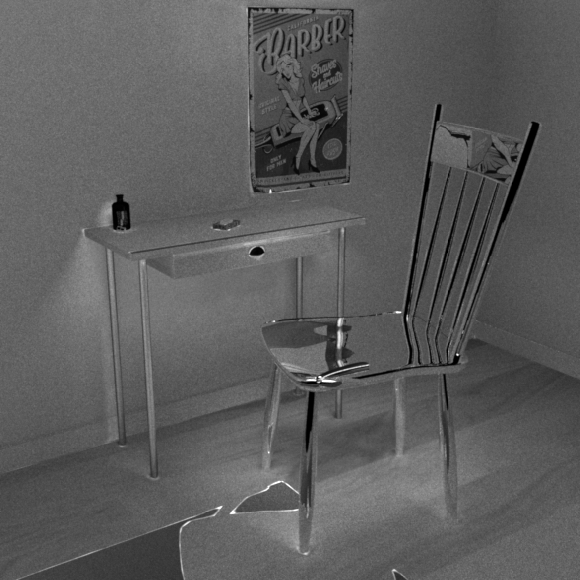}

    \vspace{-4pt}
     
    \hspace{\mapeoffsett}
    \makebox[0pt]{{spp-MAPE:}}
    \hspace{-\mapeoffsett}
    \makebox[\chairwidth\textwidth]{128-0.075}
    \makebox[\chairwidth\textwidth]{128$\times$8-0.044}
    \makebox[\chairwidth\textwidth]{1024}
    \makebox[\chairwidth\textwidth]{128$\times$8}



     \makebox[5pt]{\rotatebox{90}{\hspace{10pt} Veach Door}}
     \includegraphics[width=\chairwidth\textwidth]{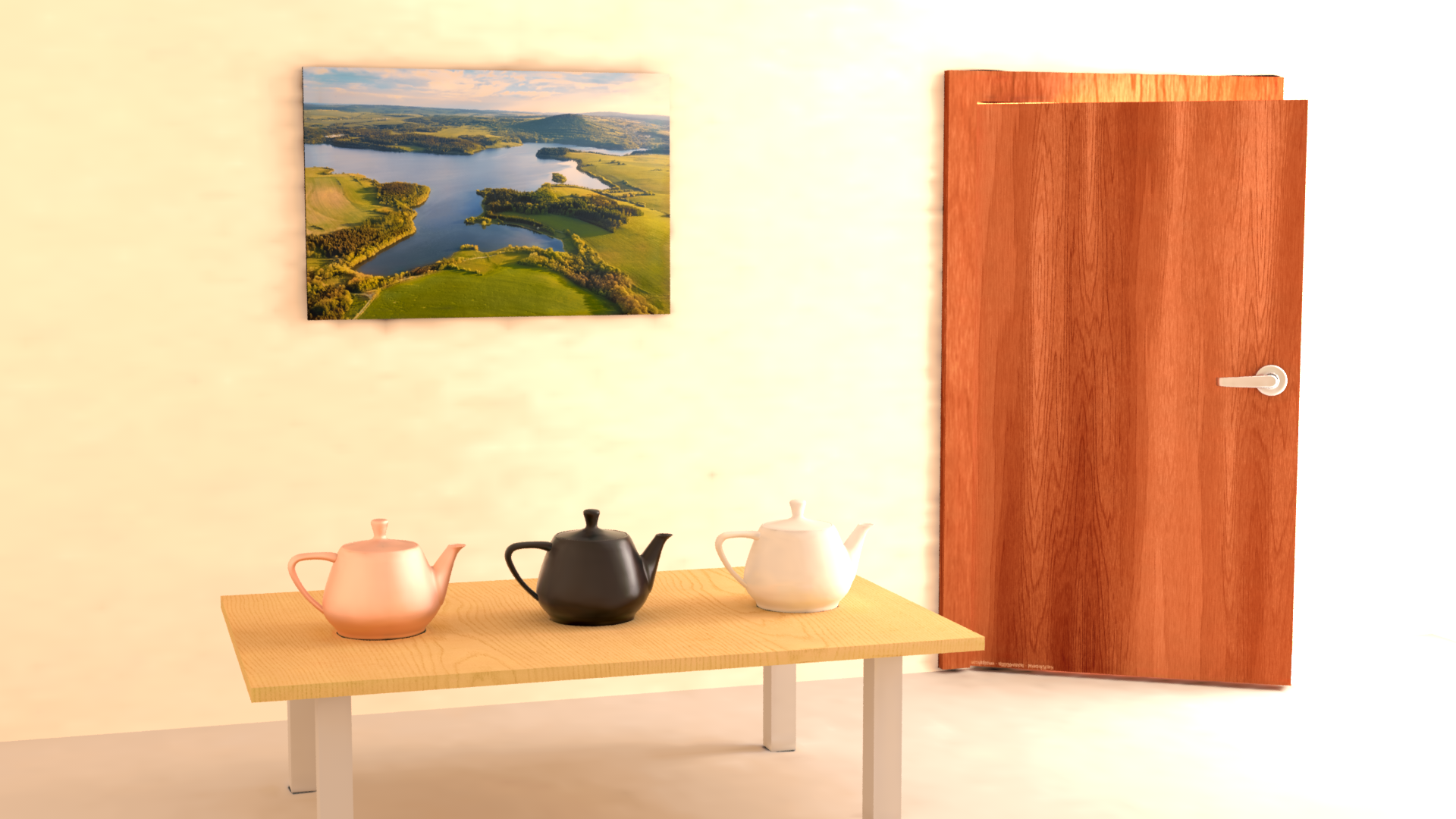}
     \includegraphics[width=\chairwidth\textwidth]{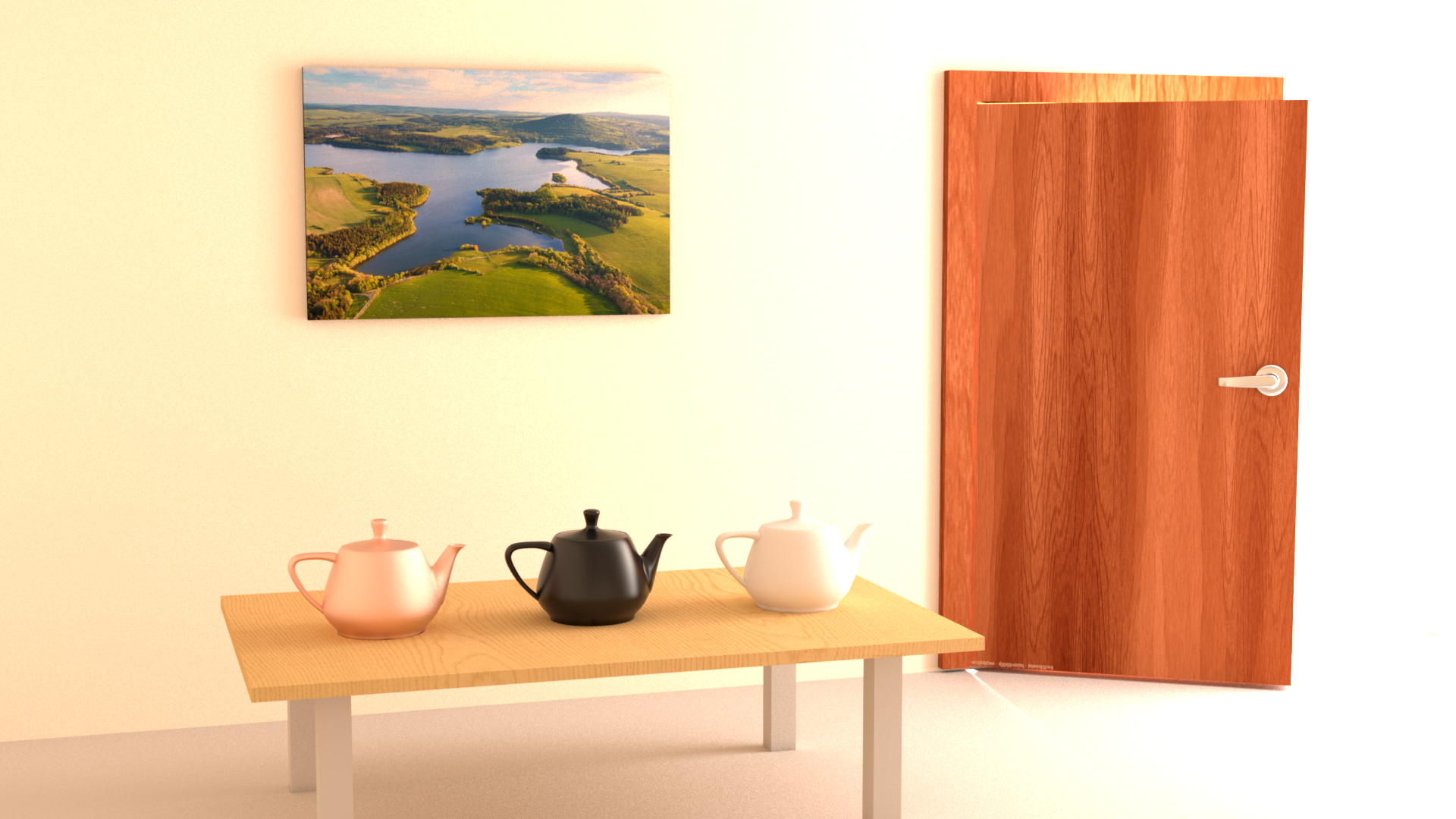}
     \includegraphics[width=\chairwidth\textwidth]{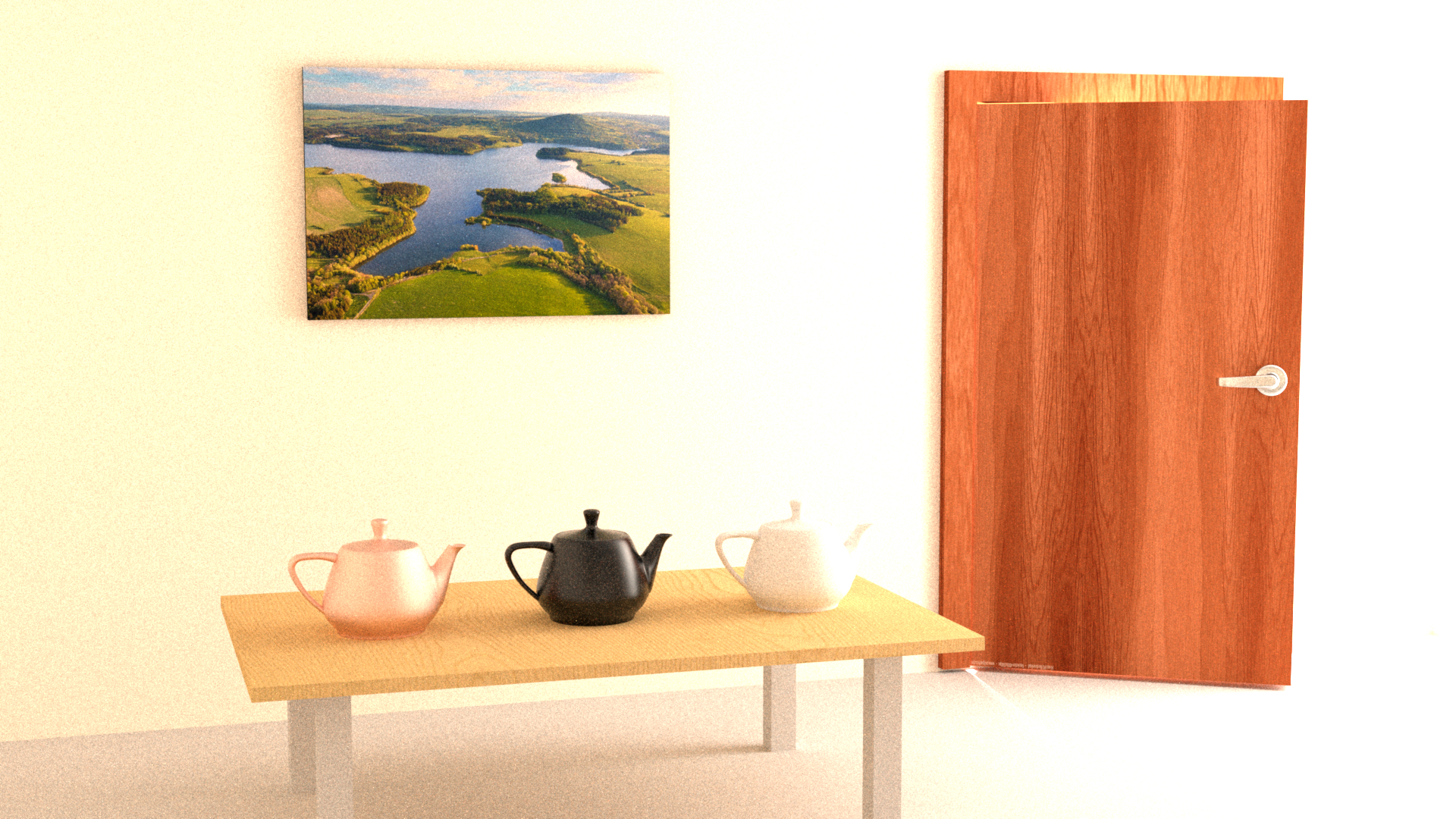}
     \includegraphics[width=\chairwidth\textwidth]{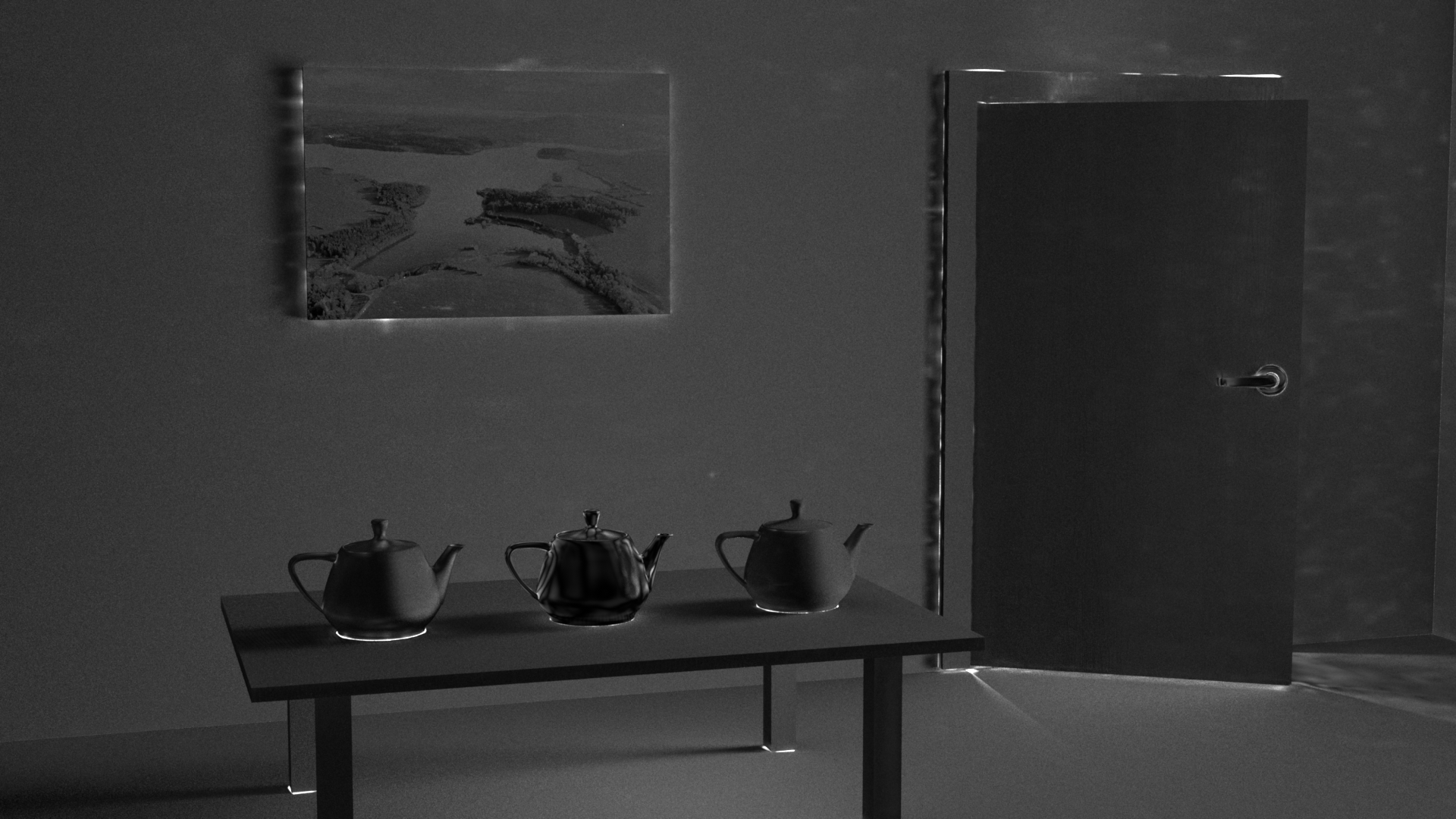}
     
    \hspace{\mapeoffsett}
    \makebox[0pt]{{spp-MAPE:}}
    \hspace{-\mapeoffsett}
    \makebox[\chairwidth\textwidth]{8-0.196}
    \makebox[\chairwidth\textwidth]{64$\times$16-0.141}
    \makebox[\chairwidth\textwidth]{1024}
    \makebox[\chairwidth\textwidth]{64$\times$16}


     \makebox[5pt]{\rotatebox{90}{\hspace{7pt} Dining Room}}
     \includegraphics[width=\chairwidth\textwidth]{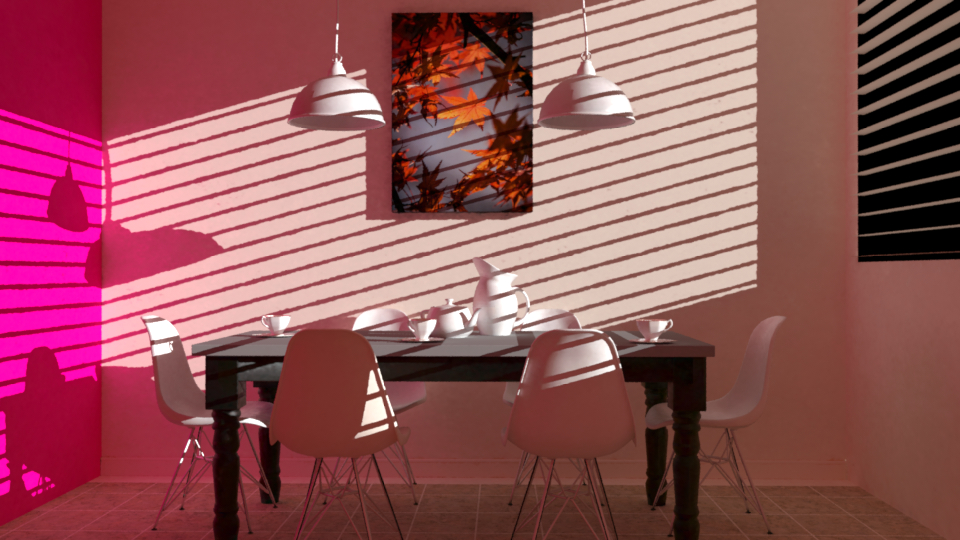}
     \includegraphics[width=\chairwidth\textwidth]{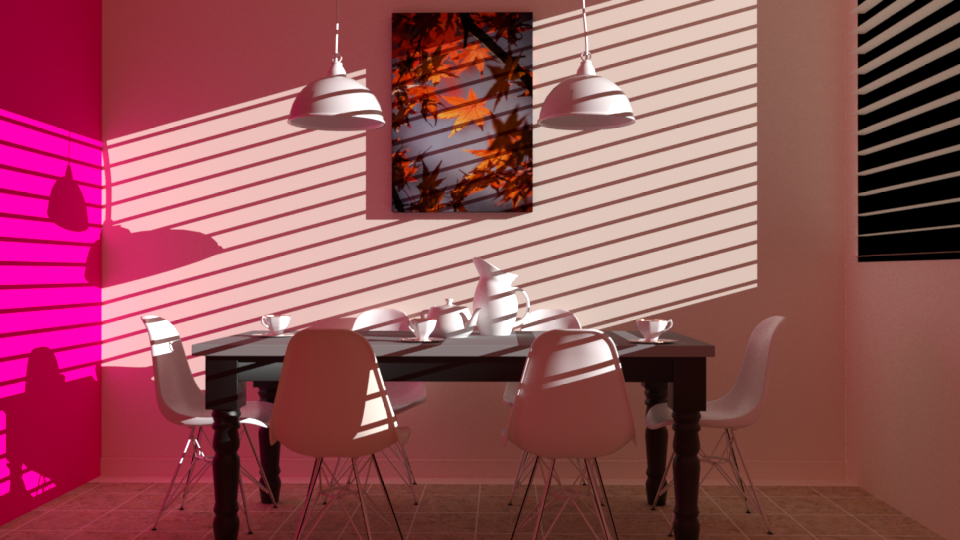}
     \includegraphics[width=\chairwidth\textwidth]{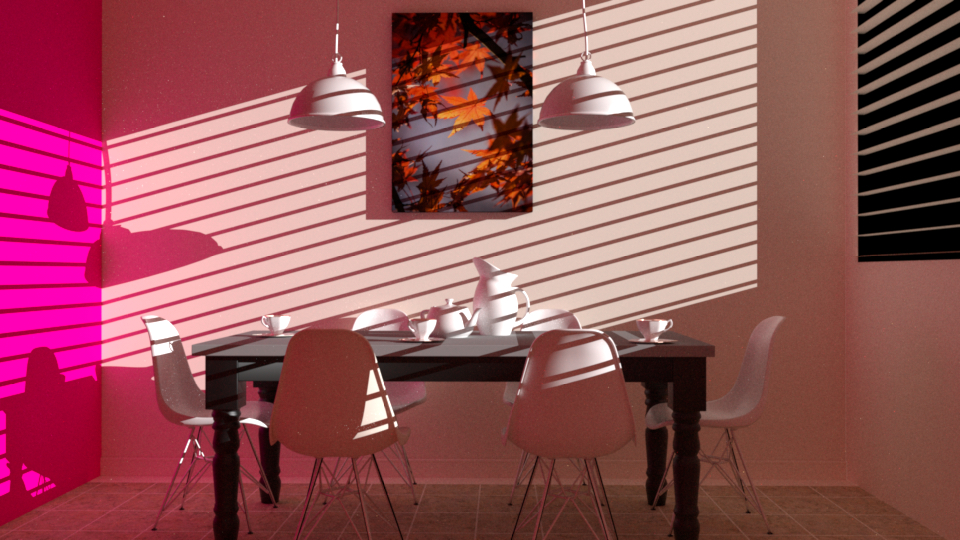}
     \includegraphics[width=\chairwidth\textwidth]{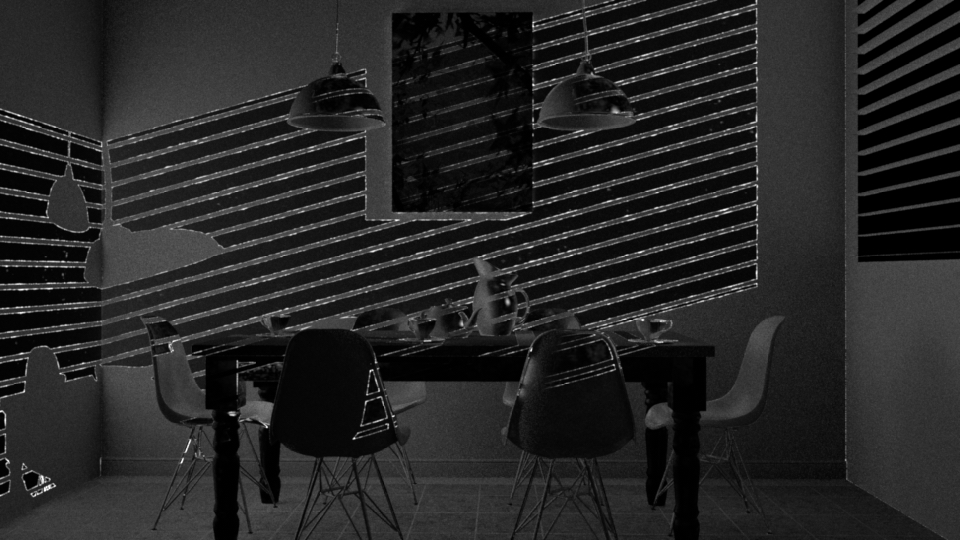}

    \hspace{\mapeoffsett}
    \makebox[0pt]{{spp-MAPE:}}
    \hspace{-\mapeoffsett}
    \makebox[\chairwidth\textwidth]{16-0.057}
    \makebox[\chairwidth\textwidth]{64$\times$16-0.030}
    \makebox[\chairwidth\textwidth]{1024}
    \makebox[\chairwidth\textwidth]{64$\times$16}


     \makebox[5pt]{\rotatebox{90}{\hspace{17pt} Bedroom}}
     \includegraphics[width=\chairwidth\textwidth]{NNMitsuba Paper/RenderingsV2/Bedroom/bedroommlp-spp16-m1.jpg}
     \includegraphics[width=\chairwidth\textwidth]{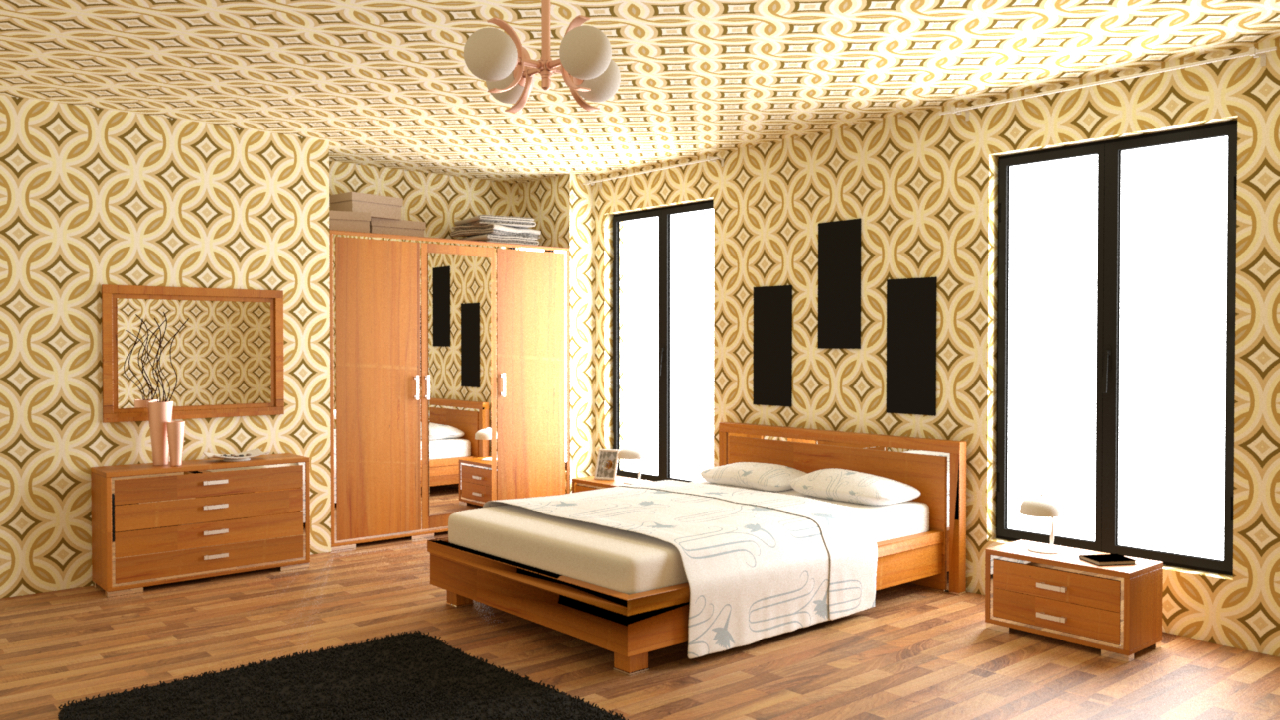}
     \includegraphics[width=\chairwidth\textwidth]{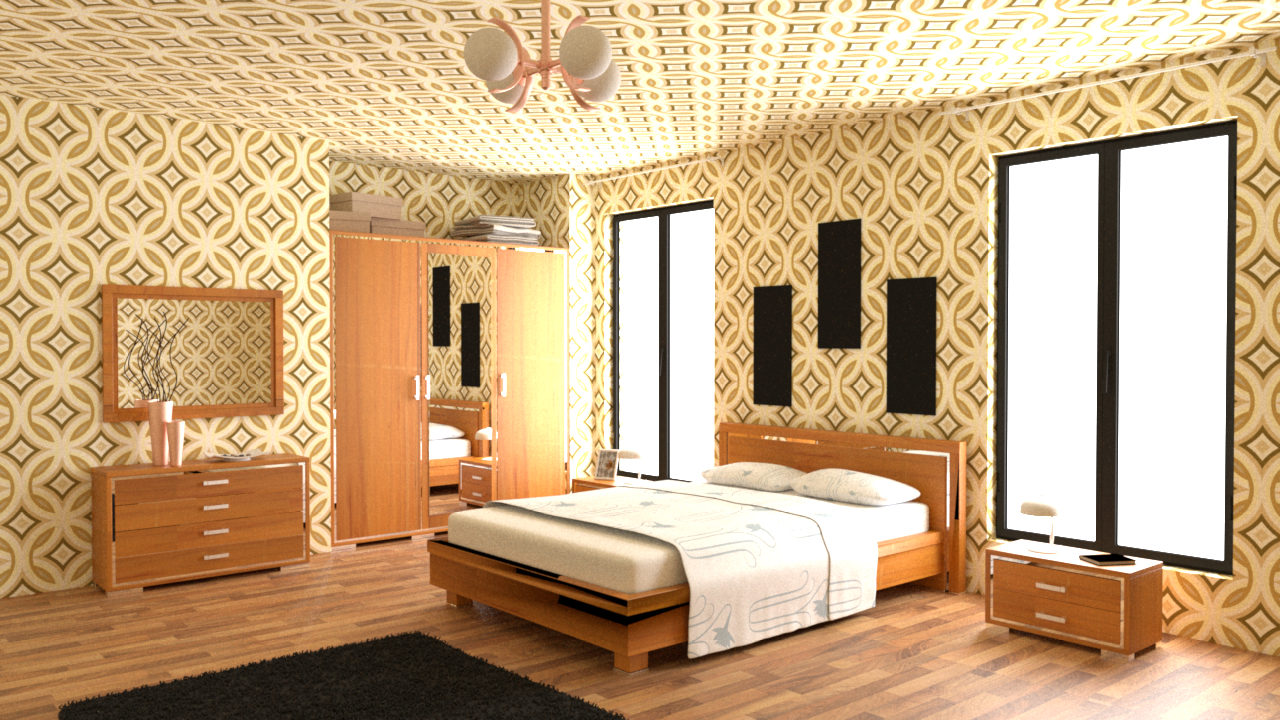}
     \includegraphics[width=\chairwidth\textwidth]{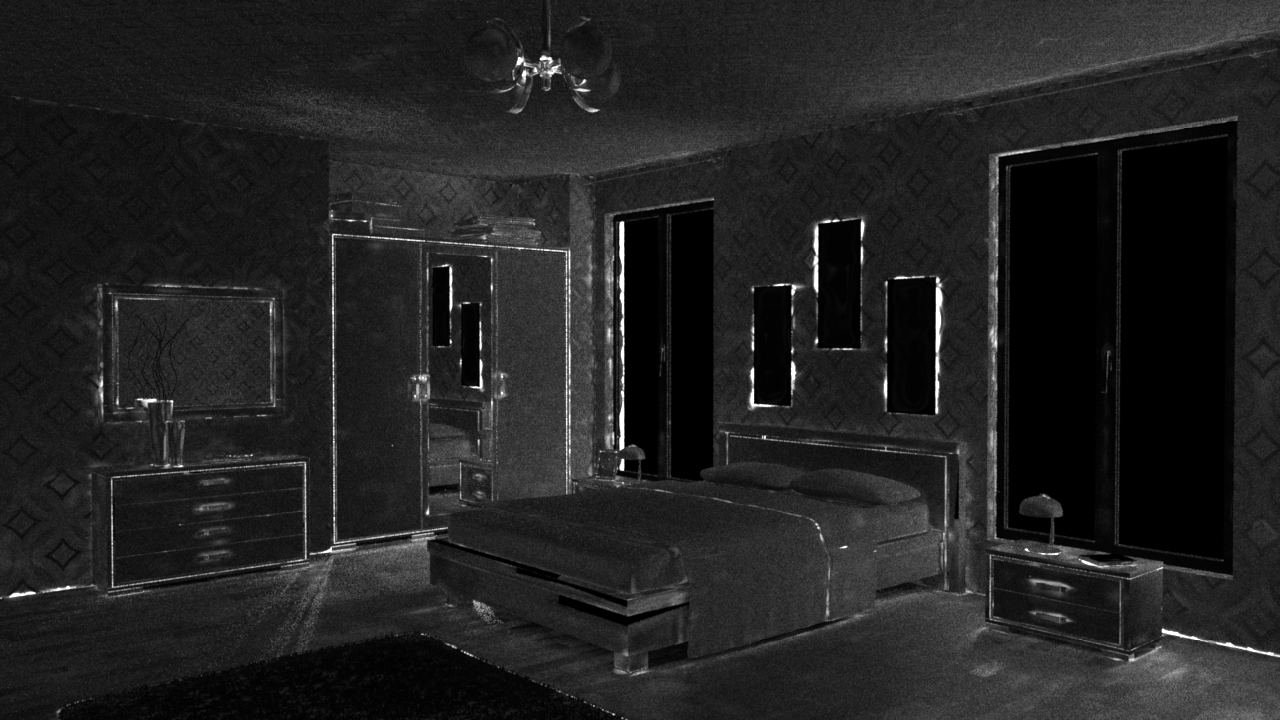}
     
    \hspace{\mapeoffsett}
    \makebox[0pt]{{spp-MAPE:}}
    \hspace{-\mapeoffsett}
    \makebox[\chairwidth\textwidth]{16-0.111}
    \makebox[\chairwidth\textwidth]{16$\times$32-0.065}
    \makebox[\chairwidth\textwidth]{512}
    \makebox[\chairwidth\textwidth]{16$\times$32}


     \makebox[5pt]{\rotatebox{90}{\hspace{7pt} Living Room}}
     \includegraphics[width=\chairwidth\textwidth]{NNMitsuba Paper/RenderingsV2/LivingRoom/livingroom-4800mlp-spp128-m1.png}
     \includegraphics[width=\chairwidth\textwidth]{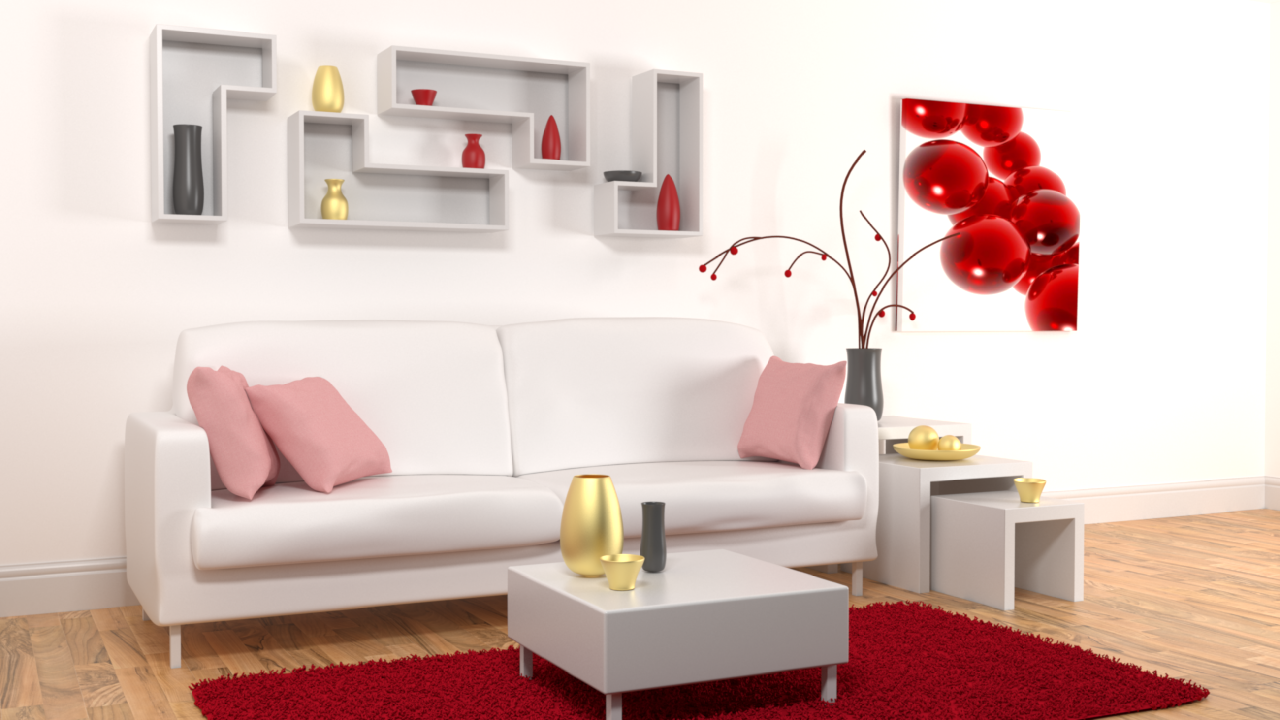}
     \includegraphics[width=\chairwidth\textwidth]{NNMitsuba Paper/RenderingsV2/LivingRoom/living-whitesofaint-scene-spp1024-m1.png}
     \includegraphics[width=\chairwidth\textwidth]{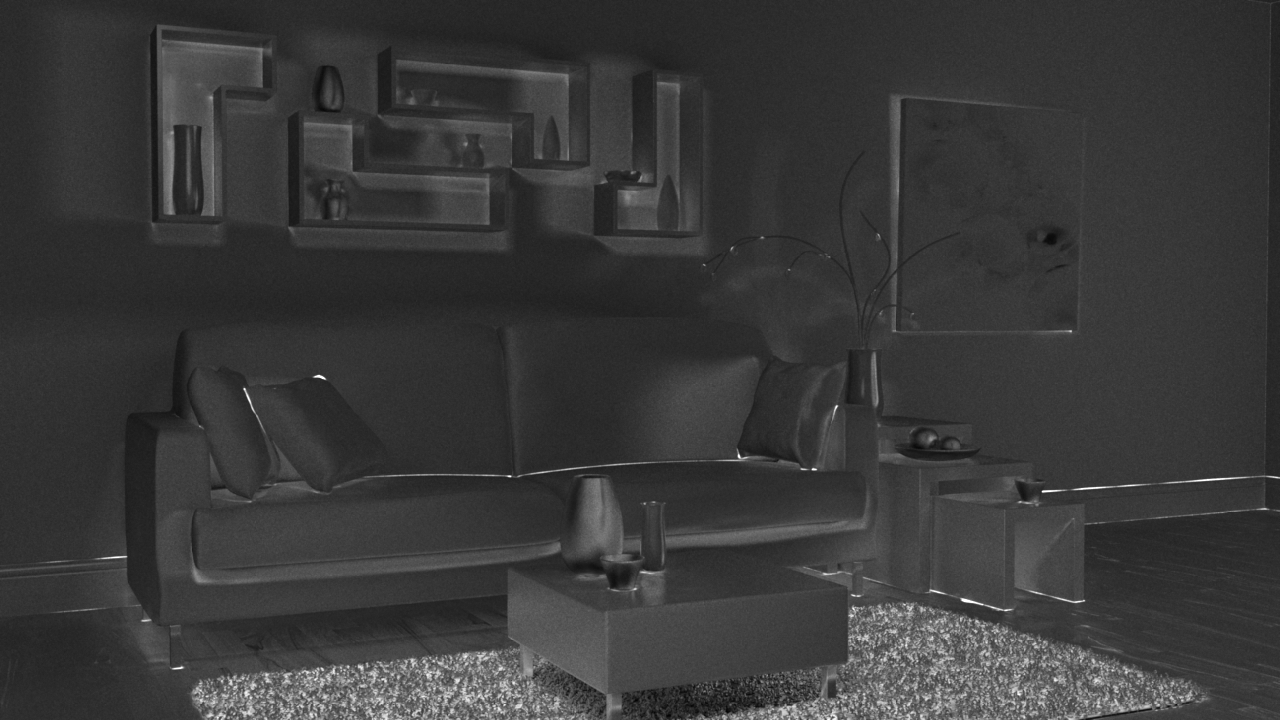}
     
    \hspace{\mapeoffsett}
    \makebox[0pt]{{spp-MAPE:}}
    \hspace{-\mapeoffsett}
    \makebox[\chairwidth\textwidth]{128-0.048}
    \makebox[\chairwidth\textwidth]{512$\times$8-0.020}
    \makebox[\chairwidth\textwidth]{1024}
    \makebox[\chairwidth\textwidth]{128$\times$8}


    \caption{Results for example scenes. The number of sample rays shot per camera pixel is denoted by ``spp'', multiplied by our RHS hyperparameter $M$, denoting the number of rays cast to compute the scattering integral. \emph{Dining Room}, \emph{Bedroom}, \emph{Veach Room}, \emph{Living Room} are provided by Bitterli~\shortcite{rendering_resources}. \emph{Chair} and \emph{Mirror Chair} are a modified versions of \cite{chair}. \label{fig:examplescenes}
    }
\end{figure*}